\documentclass{jpp}
\usepackage{graphicx}
\usepackage{epstopdf, epsfig}
\usepackage{amssymb}
\usepackage{amsmath}
\usepackage{xcolor}
\allowdisplaybreaks

\renewcommand\vec\mathbf

\newcommand\vecbar[1]{\bar{\vec{#1}}}

\newcommand\mat[1]{\mathsf{#1}}
\newcommand\uvec[1]{\hat{\vec{#1}}}

\shorttitle{Adjoint method for sensitivity of island size to magnetic field variations}
\shortauthor{A. Geraldini, M. Landreman, E. Paul}

\title{An adjoint method for determining the sensitivity of island size to magnetic field variations}

\author{Alessandro Geraldini\aff{1,2}
  \corresp{\email{ale.gerald@gmail.com}},
  M. Landreman\aff{1},
  E. Paul\aff{3}}

\affiliation{\aff{1}Institute for Research in Electronics and Applied Physics, University of Maryland, College Park, MD 20742, USA
\aff{2}Swiss Plasma Center, \'Ecole Polytechnique F\'ed\'erale de Lausanne, CH-1015 Lausanne, Switzerland
\aff{3} Department of Astrophysical Sciences, Princeton University, Princeton, NJ 08544, USA}

\begin{document}

\maketitle

\begin{abstract}
An adjoint method to calculate the gradient of island width in stellarators is presented and applied to a set of magnetic field configurations.
The underlying method of calculation of the island width is that of \cite{Cary-Hanson-1991} (with a minor modification), and requires that the residue of the island centre be small.
Therefore, the gradient of the residue is calculated in addition.
Both the island width and the gradient calculations are verified using an analytical magnetic field configuration introduced in \cite{Reiman-1986}.
The method is also applied to the calculation of the shape gradient of the width of a magnetic island in an NCSX vacuum configuration with respect to positions on a coil.
A gradient-based optimization is applied to a magnetic field configuration studied in \cite{Hanson-Cary-1984} to minimize stochasticity by adding perturbations to a pair of helical coils.
Although only vacuum magnetic fields and an analytical magnetic field model are considered in this work, the adjoint calculation of the island width gradient could also be applied to a magnetohydrodynamic (MHD) equilibrium if the derivative of the magnetic field with respect to the equilibrium parameters was known.
Using the island width gradient calculation presented here, more general gradient-based optimization methods can be applied to design stellarators with small magnetic islands.
Moreoever, the sensitivity of the island size may itself be optimized to ensure that coil tolerances with respect to island size are kept as high as possible.
\end{abstract}

\section{Introduction}

Stellarators \citep{Spitzer-1958} are promising candidates for a nuclear fusion device whose main advantage is to operate in an intrinsically steady state \citep{Helander-2014}.
In order to avoid the need of a toroidal plasma current to produce a poloidal magnetic field, stellarators lack the continuous toroidal symmetry of the magnetic field vector which is a characteristic of the tokamak.
Contrary to the tokamaks, however, stellarators have magnetic fields that tend to be non-integrable and to develop magnetic islands which break the otherwise nested toroidal magnetic surfaces \citep{Rosenbluth-1966, Cary-Hanson-1986}.
This decreases energy and particle confinement in the device. 
Thus, minimizing the number and size of magnetic islands is one of the most basic properties of a good stellarator configuration \citep{Yamazaki-1972}.

Nowadays, stellarator configurations can be produced with an extraordinarily high degree of accuracy \citep{Pedersen-2016}.
Unfortunately, due to the inherent tendency to possess islands, the configurations can be sensitive to the positions of the coils used to produce the magnetic field.
An instructive example of the importance of island width sensitivity is the National Compact Stellarator Experiment (NCSX).
In NCSX, a resonant flux surface was found to be particularly sensitive to the coils' positioning, contributing to the tight tolerances on the coils.
Construction of the device became economically unsustainable for several reasons --- including coil tolerances --- leading to the eventual cancellation of the experiment \citep{Strykowsky-2009, Neilson-2010}.
From this lesson, it is clear that a method for efficient evaluation of the sensitivity of island size on coil positioning is of fundamental importance. 

Magnetic islands tend to occur at rational flux surfaces, and especially at low order rational surfaces, due to the fact that perturbations to the intended magnetic field configuration, called error fields, can resonate with the rotational tranform of the magnetic field \citep{Helander-2014}.
The effect of error fields on stellarator configurations has been a subject of study since the measurement of magnetic islands in Wendelsteain 7-AS in \cite{Jaenicke-1993} highlighted that the assumption of flux surfaces in a stellarator experiment is incorrect.
Error fields have been studied in the CNT stellarator configuration \citep{Hammond-2016}, including those corresponding to the largest effect on the island divertor in Wendelstein 7-X (W7-X).
\cite{Lazerson-2018} conclude that tools to gauge the sensitivity of island size are important as they allow trim coils to be used to reduce the volume of magnetic islands.
A recent paper by \cite{Zhu-2019} addresses the issue of the identification and removal of the error fields responsible for island size using a Hessian matrix approach and making the simplifying assumptions that first derivatives are zero and that variations in the magnetic coordinates can be ignored.

There are several methods to calculate the width of a magnetic island.
The most basic approach relies on making a detailed scatter plot (Poincar\'e plot) of the position at which a magnetic field line intersects a given poloidal plane, and then measuring the island size from the plot.
This, however, is extremely time consuming and noisy, and is therefore especially inadequate if one wants to calculate the island width of a very large number of configurations in a short time, let alone if one wants to obtain gradient information.
Variations of this method employ automated algorithms to detect islands and calculate the island width from integration of several magnetic field lines in an island \citep{Pedersen-2006} but this is still not a viable approach to obtain accurate gradient information.
Another approach to calculate island size was developed by \cite{Lee-1990} and exploits a Fourier decomposition of the magnetic field vector.
In this work, we consider a measure of island width derived by \cite{Cary-Hanson-1991} that allows an efficient computation of the width of a magnetic island and also allows for the direct calculation of its gradient.
This approach exploits the small-island approximation to calculate a measure of width that depends only on the magnetic field line corresponding to the island centre and on the equations for linearized displacements from the island centre.

Adjoint methods have recently been applied in stellarators to obtain derivatives of neoclassical fluxes \citep{Paul-2019}, departure from quasisymmetry and several other quantities \citep{Antonsen-2019, Paul-2020}.
This work has highlighted the potential of adjoint methods in enabling efficient derivative computations with respect to a large number of parameters describing the coils or the outermost magnetic surface. 
In this work, we present an adjoint method to calculate the gradient of the Cary-Hanson measure of island width with respect to any parameter describing the magnetic field. 
The calculations of residues, island widths and gradients of these quantities are applied to three example magnetic field configurations: 
\begin{enumerate}
\item an analytical configuration (not curl-free) studied in \cite{Reiman-1986}, which we refer to as the \emph{Reiman model}; 
\item the magnetic field in NCSX, specified by the position of a set of discrete points on a set of filamentary coils and by the current through each coil;
\item a magnetic field produced by a pair of helical coils that was optimized in \cite{Hanson-Cary-1984} and \cite{Cary-Hanson-1986}.
\end{enumerate}

A potential application of the gradient calculations is the fast calculation of coil tolerances with respect to island size.
Another important application of the gradient calculations developed here is optimization of stellarator surfaces.
In order to minimize stochasticity and island size in stellarator vacuum magnetic fields, methods employed so far often minimize the magnitude of a quantity known as the \emph{residue} \citep{Greene-1968} of periodic field lines \citep{Hanson-Cary-1984, Cary-Hanson-1986}.
This quantity is calculated by linearising the equations for the magnetic field line about the island centre and calculating a matrix known as the full orbit tangent map.
This is a linear map relating the displacement of a magnetic field line from a nearby periodic field line, after a full magnetic field line period, to the initial displacement. 
In general, this map is a two-dimensional matrix with unit determinant, and therefore has three degrees of freedom.
The residue is related to the trace of this map, and provides a criterion for determining whether the closed field line is an O point (island centre) or an X point.
If the residue is zero there is no island chain (an unbroken rational flux surface), and if the residue is small the size of the island chain is small compared to the length scale of the magnetic configuration.
Therefore, the residue constitutes an extremely useful degree of freedom of the map.
Minimizing the absolute value of the residue of a periodic field line amounts to reducing the stochasticity in the magnetic field configuration and eventually also reducing the volume occupied by the corresponding island chain in the magnetic field configuration.
The gradient of the residue is used to find an optimal magnetic field configuration with small islands for a helical coil configuration previously optimized in \cite{Cary-Hanson-1986}.

An aspect of the problem that is not considered in this work is the application to magnetohydrodynamic (MHD) equilibrium configurations \citep{Hegna-2012, Hudson-2001}.
The problem of calculating the gradient of island width (or residue) with respect to magnetic field parameters amounts to calculating the gradient of the magnetic field with respect to the parameters of the equilibrium.
This is not addressed here and is left to future work.

This paper is structured as follows.
In section~\ref{sec-width}, we review the derivation of a method developed by \cite{Cary-Hanson-1991} to compute the small island width by integration along the island centre.
Then, in section~\ref{sec-shape} we derive equations for the variation of island width and residue as a function of the variation of the magnetic field configuration, using an adjoint method.
In section \ref{sec-numerical} we present some numerical results obtained by considering three different magnetic field configurations.
We also present results of a gradient-based optimization of residues in the helical coil configuration. 
Finally, in section~\ref{sec-conclusion} the main results of this paper are summarized and discussed.

\section{Calculation of island size from periodic magnetic field trajectory} \label{sec-width}

In this section, we review the calculation method of island widths introduced in \cite{Cary-Hanson-1991}.
In section \ref{subsec-width-flux} we assume the existence of toroidal flux surfaces and, upon considering an island-producing perturbation, we derive the equations for magnetic field lines in an island chain in the magnetic coordinates of the unperturbed system.
The linearized motion near the island centre (O point) is analyzed in section \ref{subsec-width-relating} and the equation for the displacement from the island centre is expressed in a frame rotating with the island centre poloidally around the magnetic axis.
The equations describing magnetic field line trajectories in cylindrical coordinates are obtained in section \ref{subsec-width-lab}.
Using the results of sections \ref{subsec-width-relating} and \ref{subsec-width-lab}, in section \ref{subsec-width-width} an expression for the island width is obtained.

\subsection{Magnetic coordinates} \label{subsec-width-flux}

It is often convenient to use magnetic coordinates \citep{Helander-2014} when describing the position along a magnetic field in systems with nested flux surfaces, such as the ideal magnetic configuration in fusion devices. 
Flux surfaces are closed toroidal surfaces where the enclosed toroidal magnetic flux $2\pi \psi$ through a surface of constant toroidal angle $\varphi$ and the enclosed poloidal magnetic flux $2\pi \chi$ through a surface of constant poloidal angle $\theta$ are constant.
The choice of magnetic coordinates is not unique: here we choose $\varphi$ to be the geometric toroidal angle which also constrains the poloidal angle $\theta$ (not generally a geometric angle).
The magnetic field is given by
\begin{align} \label{B-flux}
\vec{B} = \nabla \psi \times \nabla \theta + \nabla \varphi \times \nabla \chi (\psi, \theta, \varphi) \rm .
\end{align}
With nested flux surfaces, the poloidal flux $\chi$ is always a function of the toroidal flux $\psi$ only, since both these quantities are constant on each flux surface.
Then, $\nabla \chi$ and $\nabla \psi$ are parallel to one another and the magnetic field line trajectory never crosses the flux surfaces since $\vec{B} \cdot \nabla \chi = \vec{B} \cdot \nabla \psi = 0$.

To study the small departure from the ideal configuration with nested flux surfaces, we use magnetic coordinates $\psi$, $\theta$ and $\varphi$ which correspond to the toroidal flux, poloidal angle and geometric toroidal angle of the unperturbed system. 
Equation (\ref{B-flux}) still describes the magnetic field in such a system, but the function $\chi$ is comprised of a large piece which is equal to the poloidal flux of the unperturbed nested flux surfaces, $ \chi_0 (\psi) $, and a small perturbation that breaks the flux surfaces,  $\chi_1 (\psi, \theta, \varphi)$,
\begin{align} \label{chi-perturbed}
\chi (\psi, \theta, \varphi )= \chi_0 (\psi) +  \chi_1 (\psi, \theta, \varphi) \text{.}
\end{align}

As shown in appendix \ref{app-Ham}, the magnetic field line trajectory is a Hamiltonian system where the canonical co-ordinate $q$ is $\theta$, the canonical momentum $p$ is $\psi$, the Hamiltonian $H$ is $\chi$ and the time $t$ is $\varphi$ \citep{Cary-Littlejohn-1983}.
Hamilton's equations are therefore given by $d\psi / d \varphi = - \partial \chi / \partial \theta$ and $d \theta / d\varphi =  \partial \chi / \partial \psi $.
Since $\chi_1$ is small, to lowest order in the perturbation the magnetic field line trajectory lies in the unperturbed nested flux surfaces,
\begin{align} \label{dpsidvarphi-lowest}
\frac{d\psi}{d\varphi} \simeq 0 \text{,}
\end{align}
\begin{align} \label{dthetadvarphi-lowest}
\frac{d\theta}{d\varphi} \simeq  \chi_0' (\psi) \equiv \iota_0 (\psi ) \text{.}
\end{align}
Here, we have defined the rotational transform $\iota_0 ( \psi)$, which corresponds to the average number of poloidal turns of a magnetic field line around the magnetic axis divided by the average number of toroidal turns.
The magnetic field trajectory in magnetic coordinates is therefore approximately given by $\psi \simeq \psi_0$ (where $\psi_0$ is a constant) and $\theta \simeq \iota_0 (\psi_0 ) \varphi + \theta_{\text{i}}$, where $\theta_{\text{i}}$ is a constant.

To calculate the effect of the perturbation to the function $\chi$ on the magnetic field line trajectory, we re-express $\chi_1$ as a sum of its Fourier components,
\begin{align}
\chi_1 (\psi, \theta, \varphi ) =   \sum_{m,n} \chi_{m,n}(\psi) \exp \left( im\theta - in \varphi \right) \text{,}
\end{align}
where we have used that $\chi_1$ is periodic in $\theta$ and $\varphi$.
Here, $n$ is the toroidal mode number and $m$ is the poloidal mode number of the perturbation.
As shown in appendix \ref{app-resonant}, for any unperturbed flux surface $\chi_0 (\psi)$, the effect of $\chi_1$ is dominated by the pair of Fourier modes that resonate with the rotational transform of the unperturbed flux surface, $ n/m = \iota_0 (\psi ) $ \citep{Helander-2014, Cary-Littlejohn-1983}.
Thus, the effect of the perturbation is largest at rational flux surfaces, where the rotational transform is a rational number.
In what follows, we assume that resonances at different rational flux surfaces do not overlap and interact with each other and that higher order harmonics of the resonances have a smaller amplitude and can be neglected.
Consider the rational flux surface where 
\begin{align}
\iota_0 (\psi_0) = \frac{N}{M} \rm .
\end{align}
Here, $N$ is the toroidal mode number and $M$ is the poloidal mode number of the island-producing perturbation, which is readily re-expressed to
\begin{align}
\chi_1 (\psi, \theta, \varphi) = \epsilon (\psi) \cos \left( M \zeta (\psi) +  M \theta - N\varphi  \right) \text{.}
\end{align}
An amplitude, $\epsilon (\psi) > 0$, and a phase factor, $\zeta (\psi) $, of the resonant perturbation have been introduced to replace the pair of complex amplitudes $\chi_{M,N}$ and $\chi_{-M, -N}$ corresponding to the terms in $\chi_1$ that resonate with $\iota_0(\psi_0)$.
The subscripts $M$ and $N$ on $\epsilon$ and $\zeta$ have been omitted for brevity.
Introducing a new variable $\Theta$,
\begin{align} \label{Theta-def}
 \Theta = \theta  - \frac{N \varphi}{M} \rm ,
\end{align}
$\chi_1$ is re-expressed to
\begin{align}
\chi_1 (\psi, \Theta) = \epsilon (\psi) \cos \left( M \zeta (\psi) +  M \Theta \right) \text{.}
\end{align}
With this change of variable, the dependence on $\varphi$ of the phase of $\chi_1$ has dropped.
However, the function $\chi$ is no longer a Hamiltonian in the variables $(\psi, \Theta)$.
As shown in appendix \ref{subapp-Ham}, the Hamiltonian in the new variables is 
\begin{align} \label{K-Ham}
K (\psi, \Theta ) =  \chi (\psi, \Theta ) - \iota_0 (\psi_0)  \psi \text{,}
\end{align}
and Hamilton's equations are $d\Theta / d\varphi = \partial K / \partial \psi$ and $d\psi / d \varphi = - \partial K / \partial \Theta$.
The Hamiltonian $K$ is independent of $\varphi$ and is therefore conserved following the perturbed magnetic field lines. 

In what follows, $K(\psi, \Theta )$ is expanded close to the rational flux surface.
The perturbation in $\psi$ near the rational flux surface is 
\begin{align}
\psi_1 = \psi - \psi_0 \text{.}
\end{align}
The function $\chi$ is expanded to
\begin{align} \label{chi-expanded}
\chi (\psi_1, \Theta )& = \chi_0 (\psi_0 ) + \chi_0' (\psi_0 )  \psi_1  + \frac{1}{2} \chi_0'' (\psi_0 ) \psi_1^2    +  \epsilon (\psi_0) \cos \left( M \zeta (\psi_0) +  M \Theta  \right) \nonumber \\
 & +   \epsilon'(\psi_0) \psi_1 \cos \left( M \zeta (\psi_0) +  M \Theta  \right)  -  M \epsilon (\psi_0) \zeta'(\psi_0 ) \psi_1 \sin \left( M \zeta (\psi_0) +  M \Theta  \right)  \nonumber \\ & + O(\epsilon \hat{\psi}_1^2, \psi \hat{\psi}_1^2) \text{,}
\end{align}
where the normalized $\psi$ perturbation is $\hat{\psi}_1 = \psi_1 \iota_0'(\psi_0) $ (equivalent to the $\psi$ perturbation divided by a measure of typical variations of $\psi$ and $\chi$ across the toroidal configuration).
Using equations (\ref{chi-expanded}), (\ref{K-Ham}), and $\iota_0 = \chi_0' (\psi_0)$, the Hamiltonian $K$ becomes
\begin{align} \label{K-island}
K ( \psi_1 , \Theta )  & = \chi_0 ( \psi_0 ) - \iota_0 ( \psi_0 ) \psi_0 +   \frac{1}{2} \iota_0'(\psi_0) \psi_1^2 +  \epsilon(\psi_0) \cos \left( M \zeta (\psi_0) +  M \Theta \right) \nonumber \\
 &
+   \epsilon'(\psi_0) \psi_1 \cos \left( M \zeta (\psi_0) +  M \Theta  \right)    -  M \epsilon(\psi_0) \zeta'(\psi_0 ) \psi_1 \sin \left( M \zeta (\psi_0) +  M \Theta  \right)  \nonumber \\
 & + O(\epsilon \hat{\psi}_1^2, \psi \hat{\psi}_1^2) \text{.}
\end{align}
Hamilton's equations for the magnetic field line sufficiently close to the rational flux surface are thus
\begin{align} \label{dThetadphi}
\frac{d \Theta}{d\varphi} = \frac{\partial K}{\partial \psi_1}  = & \iota_0'(\psi_0) \psi_1+  \epsilon'(\psi_0) \cos \left( M \zeta (\psi_0) +  M \Theta \right) \nonumber  \\
& -  \epsilon(\psi_0) M \zeta'(\psi_0) \sin \left( M \zeta (\psi_0) +  M \Theta \right) + O(\hat{\epsilon} \hat{\psi}_1, \psi_1 \hat{\psi}_1 ) \text{,}
\end{align}
and
\begin{align} \label{dpsi1dphi}
\frac{d\psi_1}{d\varphi} = - \frac{\partial K}{\partial \Theta }  =  \epsilon(\psi_0) M \sin \left( M \zeta (\psi_0) +  M \Theta \right) + O(\epsilon \hat{\psi}_1) \text{,}
\end{align}
where $\hat{\epsilon} = \iota_0'(\psi_0)\epsilon (\psi_0)$.

The fixed points of the magnetic field line flow in the $(\psi_1, \Theta)$ coordinates, which represent closed magnetic field lines, occur at $d\Theta / d\varphi = d\psi_1 / d\varphi = 0$.
This corresponds to $(\psi_1, \Theta ) = (\bar{\psi}_1, \bar{\Theta})$ such that $\sin \left( M \zeta (\psi_0) +  M \bar{\Theta} \right) = 0$ and $\bar{\psi}_1 = \pm \epsilon'(\psi_0)/ \iota_0'(\psi_0) = O(\epsilon )$, where we have used $\cos \left( M \zeta (\psi_0) +  M \bar{\Theta} \right) = \mp 1$.
Introducing magnetic coordinates relative to a closed magnetic field line, $\delta \psi = \psi - \psi_0 - \bar{\psi}_1$ and $\delta \Theta = \Theta - \bar{\Theta}$, the equations (\ref{dThetadphi}) and (\ref{dpsi1dphi}) linearized near the closed magnetic field give
\begin{align} \label{near-fixed}
\frac{d}{d\varphi} \begin{pmatrix}
\delta \psi \\
\delta \Theta
\end{pmatrix} = \begin{pmatrix}
O(\hat{\epsilon}) &  \mp \epsilon(\psi_0) M^2 + O(\epsilon  \hat{\psi}_1)  \\
\iota_0'(\psi_0) + O(\iota_0'(\psi_0)\hat{\epsilon} )  & O(\hat{\epsilon})
\end{pmatrix}  \begin{pmatrix}
\delta \psi \\
\delta \Theta
\end{pmatrix}  \text{.}
\end{align}
As will be shown explicitly by expanding the Hamiltonian near the closed field line, the error terms in (\ref{near-fixed}) arising from the diagonal elements of the matrix are negligible because the only self-consistent ordering relating the characteristic sizes of $\delta \psi$ and $\delta \Theta$ is $|\iota_0'| \delta \psi \sim  M \sqrt{\hat{\epsilon}} \delta \Theta$.
Trajectories neighbouring the island centre (O point) are described when the signs of the off-diagonal elements in the matrix in (\ref{near-fixed}) are opposite, since in this case the eigenvalues are purely imaginary and trajectories are periodic.
The other sign choice corresponds to the trajectories passing close to the crossing point of the island separatrices (X point).
Thus, we can replace $\pm$ everywhere with $\pm \text{sgn} \left( \iota'_0 (\psi_0) \right)$, with the top sign and bottom signs understood to correspond to O and X points respectively.
The motion sufficiently close to the centre is thus always described by
\begin{align} \label{EOM-flux}
\frac{d}{d\varphi} \begin{pmatrix}
\delta \psi \\
\delta \Theta
\end{pmatrix} =  \begin{pmatrix}
O(\hat{\epsilon}) & - \text{sgn} \left( \iota'_0 (\psi_0) \right) \epsilon(\psi_0) M^2 ( 1 + O( \hat{\epsilon} ))  \\
 \iota_0'(\psi_0)  (1 + O(\hat{\epsilon}))   & O(\hat{\epsilon})
\end{pmatrix}  \begin{pmatrix}
\delta \psi \\
\delta \Theta
\end{pmatrix}
 \text{,}
\end{align}
where we have set $\cos \left( M \zeta (\psi_0) +  M \bar{\Theta} \right) =- \text{sgn} \left( \iota'_0 (\psi_0) \right)$ at an O point.
Solving this linear system gives
\begin{align} \label{deltau-deltaTheta}
 \begin{pmatrix}
\delta \psi (\varphi)\\
\delta \Theta (\varphi)
\end{pmatrix} = \mat{T}  \begin{pmatrix}
\delta \psi(0) \\
\delta \Theta(0)
\end{pmatrix}
 \text{,}
\end{align}
where the tangent map $ \mat{T} $ is given by
\begin{align} \label{T-u-Theta}
\mat{T} \simeq  \begin{pmatrix}
\cos(\omega \varphi )  & -  \left( \omega  / \iota_0'(\psi_0) \right)  \sin (\omega \varphi )    \\
\left( \iota_0'(\psi_0) / \omega \right) \sin (\omega \varphi )  & \cos(\omega \varphi )
\end{pmatrix} 
 \text{,}
\end{align}
and the frequency at which neighbouring points rotate around the O point is
\begin{align} \label{omega}
\omega \simeq M \sqrt{|\iota'_0(\psi_0)| \epsilon(\psi_0)}  \rm .
\end{align}
Note that, from equation (\ref{EOM-flux}), each element of the tangent map in (\ref{T-u-Theta}) has an error of $O(\hat{\epsilon})$ and the the frequency $\omega$ in (\ref{omega}) has an error of $O\left( \hat{\epsilon} \sqrt{|\iota'_0(\psi_0)| \epsilon(\psi_0)} \right)$ \citep{Cary-Hanson-1991}.

In order to relate the linearized motion along a field line neighbouring the island centre described by equations (\ref{deltau-deltaTheta})-(\ref{omega}) to the island width, the motion along the island separatrix is studied.
The Hamiltonian on the separatrix is constant and equal to its value at the X point,
\begin{align} \label{H-separatrix}
K (\bar{\psi}, \bar{\Theta} ) = \chi_0 ( \psi ) - \iota_0 ( \psi ) \psi_0 + \rm sgn \left( \iota_0'(\psi_0) \right) \epsilon(\psi_0) + O(\hat{\epsilon}\epsilon) \text{,}
\end{align}
where we inserted $\bar{\psi} = - \epsilon'(\psi_0)/ |\iota_0'(\psi_0)| $ (X point), $\sin \left( M \zeta (\psi_0) +  M \bar{\Theta} \right) = 0$ (fixed point) and $\cos \left( M \zeta (\psi_0) +  M \bar{\Theta} \right) = \text{sgn} \left( \iota_0'(\psi_0) \right) $ (X point) in equation (\ref{K-island}).
From equation (\ref{K-island}) and $\epsilon (\psi_0) > 0$, the value of $\psi_1^2$ is largest when $\cos \left( M \zeta (\psi_0) +  M \Theta \right) = -\text{sgn} \left( \iota_0'(\psi_0) \right)$ on the separatrix, and so by evaluating $K (\psi, \Theta )$ at this point and equating it to equation (\ref{H-separatrix}) we obtain
\begin{align}
 \frac{1}{2} | \iota_0'(\psi_0) | \psi_1^2 = 2  \epsilon(\psi_0) + O(\epsilon \hat{\psi}_1, \hat{\epsilon}\epsilon )  \text{.}
\end{align}
Hence, the values of $\psi_1$ at the separatrix at the point where the island is largest are $\psi_1  = \pm 2 \sqrt{    \epsilon(\psi_0) / | \iota_0'(\psi_0) | } +  O(\epsilon ) \sim \sqrt{\hat{\epsilon}} / \iota_0'(\psi_0)$, and the full island width, denoted $\Upsilon$, is
\begin{align} \label{width-flux}
\Upsilon  =  4 \sqrt{ \frac{   \epsilon(\psi_0) }{\left| \iota_0'(\psi_0) \right| } }   \rm .
\end{align}
Note that $\Upsilon$ is approximately equal to the full-island width in the magnetic coordinate $\psi$ with an absolute error of $O(\epsilon )$, equivalent to a relative error of $O\left( \hat{\epsilon}^{1/2}\right)$.
Note also that $|\bar{\psi} | \ll \Upsilon$, and so the island centre is equidistant from the two branches of the separatrix to lowest order in $\hat{\epsilon}$.
From (\ref{omega}) and (\ref{width-flux}), the matrix $\mat{T}$ in (\ref{T-u-Theta}) can be rewritten with $\Upsilon$ in place of $\iota_0'$ to obtain
\begin{align} \label{T-width}
\mat{T} \simeq  \begin{pmatrix}
\cos(\omega \varphi ) & -  \text{sgn} \left(  \iota_0'(\psi_0) \right)  \left( M \Upsilon  / 4 \right)  \sin (\omega \varphi )    \\
\text{sgn} \left(  \iota_0'(\psi_0) \right)  \left( 4 / M \Upsilon \right) \sin (\omega \varphi )  & \cos(\omega \varphi )
\end{pmatrix} 
 \text{.}
\end{align}

When linearising the equations for the magnetic field line near the island centre, the associated Hamiltonian is equation (\ref{K-Ham}) expanded near the island centre up to quadratic terms in $\delta \psi$ and $\delta \Theta$.
Retaining only the lowest order terms in $\epsilon$ gives
\begin{align} \label{K-expanded}
K = &  \text{const}   +  \frac{1}{2} \iota_0'(\psi_0) \delta \psi^2 + \frac{1}{2} \text{sgn} \left( \iota_0' (\psi_0) \right) M^2\epsilon \delta \Theta^2 \nonumber \\ & + O ( \hat{\epsilon} \delta \psi^2 \iota_0'(\psi_0 ), M^2 \hat{\epsilon} \delta \psi \delta \Theta, M^2\hat{\epsilon} \epsilon \delta \Theta , M^2\hat{\epsilon} \epsilon \delta \Theta^2  ) \text{,}
\end{align}
where we have deduced that the only self-consistent ordering is $\iota_0'(\psi_0) \delta \psi  \sim M \sqrt{\hat{\epsilon}}  \delta \Theta$.
Note that equation (\ref{K-expanded}) can be made to be exactly quadratic: the linear term in the error in (\ref{K-expanded}) could be cancelled by calculating the first order correction in $\hat{\epsilon}$ of the value of $\Theta$ at the island centre and redefining $\delta \Theta$ relative to this more accurate coordinate.
The neglected $O(M^2 \hat{\epsilon} \delta \psi \delta \Theta)$ cross-term in (\ref{K-expanded}) gives rise to the diagonal terms in the matrix in equation (\ref{near-fixed}), which have a negligible contribution once the ordering $\iota_0'(\psi_0) \delta \psi  \sim M \sqrt{\hat{\epsilon}}  \delta \Theta$ is taken into account.
The magnitude of the quadratic perturbation to the Hamiltonian, 
\begin{align} \label{deltaK}
|\delta K | = \delta \vec{u} \cdot \mat{K} \cdot \delta \vec{u} = \begin{pmatrix}
\delta \psi , & 
\delta \Theta
\end{pmatrix}   \begin{pmatrix}
 \frac{1}{2} | \iota_0'(\psi_0) | + O ( \hat{\epsilon} \iota_0'(\psi_0) )  & O(M^2\hat{\epsilon} )  \\
O(M^2\hat{\epsilon} )   & \frac{1}{2} M^2\epsilon + O (M^2\hat{\epsilon} \epsilon )
\end{pmatrix} \begin{pmatrix}
\delta \psi \\ 
\delta \Theta
\end{pmatrix}   \text{,}
\end{align}
is a scalar invariant, as it is conserved following the field lines neighbouring the O point. 
To lowest order in $\hat{\epsilon}$, $\mat{K}$ is diagonal and thus the trajectories infinitesimally close to the island centre in magnetic coordinates are ellipses that are approximately aligned with the magnetic coordinate directions and elongated in the $\Theta$ direction.  
The angle between the characteristic directions of the ellipses and the magnetic coordinate axes is small, $O(\hat{\epsilon})$.
Note that if we diagonalized the matrix $\mat{K}$ exactly, we would obtain additional error terms of the same order in the diagonal terms.

\subsection{Relating magnetic coordinates to lengths at the island centre} \label{subsec-width-relating}

The relationship between the displacement from the island centre measured as a length in the poloidal cross-section at a given $\varphi$ and the same displacement measured in magnetic coordinates must be linear if the displacement is infinitesimally small (as the relationship between the two sets of coordinates must be locally described by a Taylor expansion).
Thus, we define the local linear change of variables $\delta \vec{u} =  (\delta \psi, \delta \Theta ) \rightarrow \delta \boldsymbol{\xi} = ( \delta \xi_{\perp}, \delta \xi_{\parallel} )$, where $\delta \xi_{\perp} (\varphi)$ and $\delta \xi_{\parallel}(\varphi)$ are displacements from the island centre in two orthogonal directions (yet unspecified) in the poloidal plane with toroidal angle $\varphi$. 
The two sets of coordinates are related by $\delta \vec{u} (\varphi ) = \mat{Q} (\varphi ) \cdot \delta \boldsymbol{\xi} (\varphi)$ for any $\varphi$, where 
\begin{align} \label{Qmat-def}
\mat{Q}(\varphi ) = \begin{pmatrix}
\frac{\partial \psi }{\partial \xi_{\perp}}(\varphi )  & \frac{\partial \psi }{\partial \xi_{\parallel}}(\varphi )  \\
\frac{\partial \Theta }{\partial \xi_{\perp}}(\varphi )  & \frac{\partial \Theta }{\partial \xi_{\parallel}}(\varphi )  
\end{pmatrix} \rm .
\end{align} 
For each value of $\varphi$, the scalar invariant can be cast in the new coordinates, $|\delta K | = \delta \boldsymbol{\xi} (\varphi) \cdot \mat{Q}^\intercal (\varphi)\cdot \mat{K} \cdot \mat{Q}(\varphi) \cdot \delta \boldsymbol{\xi} (\varphi) $, where $\mat{Q}^{\intercal}$ denotes the transpose of $\mat Q$ and the local laboratory invariant matrix $ \mat{Q}^\intercal (\varphi) \cdot \mat{K} \cdot \mat{Q} (\varphi )$ is symmetric since $\mat{K}$ is symmetric.
Thus, the eigenvectors of the local laboratory invariant matrix are orthogonal and the points of intersection of a trajectory, infinitesimally close to the island centre, with any given poloidal plane $\varphi$ lie on an ellipse.

The dependence of the scalar invariant on $\varphi$ expresses the fact that, when viewing the motion continuously in different $\varphi$-planes, the poloidal displacement of a field line infinitesimally close to the island centre lies on a \emph{continuously varying ellipse}.
This is a consequence of the fact that in a stellarator the poloidal cross-section of flux surfaces taken at different toroidal angles is generally different, and thus the shape of the island continuously changes and rotates poloidally as the island centre is followed around.
Nonetheless, a set of equivalent flux surface sections always exists for values of $\varphi$ that differ by an integer multiple of a field period, $2\pi / n_0$, where $n_0 \geqslant 1$.
The flux surface sections are, in general, only equivalent to lowest order in the island-producing perturbation, $\epsilon$, because an island chain may break the field periodicity ($n_0 = 1$ is a special case where the field periodicity is never broken). 
The closed field line intersects the set of approximately equivalent poloidal planes a finite number of times, $L$, before returning to the original position. 
Therefore, when snapshots of the position along a field line neighbouring the island centre are taken at $\varphi = \varphi_k + 2\pi Q L / n_0$ for any positive integer $Q$ and initial toroidal angle $\varphi_k$, the motion appears to be around \emph{the same ellipse}.
In this work, we choose one set of an infinite number of possible sets of symmetry planes by specifying the toroidal plane corresponding to $\varphi = 0$ and considering the set of symmetry planes given by $\varphi_ k = 2\pi k / n_0$, where $k$ is a positive integer. 
For stellarators with stellarator symmetry, the magnetic configuration in the plane $\varphi = 0$ is chosen to be up-down symmetric. 
Henceforth, any quantity that is a function of $\varphi_k$ will be denoted with a subscript $k$, e.g. $\mat{Q}_k = \mat{Q} (\varphi_k ) $.

As the matrix $\mat{K}$ is (approximately) diagonal, the diagonalization of the invariant matrix $ \mat{Q}_k^\intercal \cdot \mat{K} \cdot \mat{Q}_k$ is achieved (approximately) by choosing the coordinates $\delta \xi_{\parallel, k}$ and $\delta \xi_{\perp, k}$ such that $\partial \psi / \partial \xi_{\parallel, k} = \partial \Theta / \partial \xi_{\perp, k} = 0$ for all $k$ and thus
\begin{align}
\delta \psi_k  = \frac{\partial  \psi}{\partial  \xi_{\perp,k}}  \delta \xi_{\perp, k} \rm ,
\end{align}
\begin{align}
\delta \Theta_k = \frac{\partial  \Theta}{\partial  \xi_{\parallel, k}}  \delta \xi_{\parallel, k}  \rm .
\end{align}
With this choice, the coordinates $\delta \xi_{\perp, k}$ and $\delta \xi_{\parallel, k}$ quantify the displacement (as a length) from the island centre in the poloidal plane $\varphi = 2\pi k / n_0$, measured in the directions associated with $\delta \psi$ (across the flux surface) and $\delta \Theta$ (along the flux surface), respectively. 
In the new coordinates, the displacement from the fixed point satisfies the equation
\begin{align} \label{deltaxiperp-deltaxipar}
 \delta \boldsymbol{\xi}_{k+q} (\varphi) = \mat{S}_{R,k}^q  \cdot \delta \boldsymbol{\xi}_k \rm ,
\end{align}
where the tangent map in the rotating frame, $ \mat{S}_{R,k}^q =  \mat{Q}_{k+q}^{-1}  \cdot \mat{T}_{k+q} \cdot \mat{Q}_k $, is given by
\begin{align} \label{S-width-1}
\mat{S}_{R,k}^q  \simeq  \begin{pmatrix}
 \frac{ \partial \xi_{\perp, k+q} }{  \partial \psi }  \frac{ \partial \psi }{ \partial \xi_{\perp, k} }  \cos\left( \frac{2\pi \omega q }{ n_0} \right) & - \frac{ \partial \xi_{\perp, k+q} }{  \partial \psi } \frac{ \partial \Theta }{ \partial \xi_{\parallel, k} }  \frac{ M\Upsilon }{ 4}  \sin\left( \frac{2\pi \omega q }{ n_0} \right)    \\
 \frac{ \partial \xi_{\parallel, k+q} }{  \partial \Theta } \frac{ \partial \psi }{ \partial \xi_{\perp, k} }  \frac{ 4}{ M\Upsilon }  \sin \left( \frac{2\pi \omega q }{ n_0} \right) &  \frac{ \partial \xi_{\parallel, k+q} }{  \partial \Theta }  \frac{ \partial \Theta }{ \partial \xi_{\parallel, k} }  \cos\left( \frac{2\pi \omega q }{ n_0} \right)
\end{pmatrix} 
 \text{.}
\end{align}
Note that
\begin{align} \label{wperpbar-def}
\bar{w}_{\perp, k} = \Upsilon \frac{ \partial \xi_{\perp, k} }{  \partial \psi } \rm 
\end{align}
is approximately equal to the island width measured at $\varphi = 2\pi k / n_0$, with an absolute error of $O( \Upsilon^2 \partial^2 \xi_{\perp, k} / \partial \psi^2 ) $ giving rise to a relative error of $ O(\hat{\epsilon}^{1/2}  )$.
This relative error is to be added to the equally large error in approximating the island width in the coordinate $\psi$ as $\Upsilon$ in equation (\ref{width-flux}).
Using equation (\ref{wperpbar-def}), the off-diagonal elements of $\mat{S}_{R,k}^q$ explicitly include the island width. 
Upon imposing $\delta \xi_{\parallel, k} = 0$ as an initial condition, we obtain the equation
\begin{align} \label{width-intermediate}
\frac{\delta \xi_{\parallel, k+q} }{ \delta \xi_{\perp, k}} \simeq  \frac{ \partial \xi_{\parallel, k+q} }{  \partial \Theta }   \frac{4}{ M \bar{w}_{\perp, k}  } \sin \left( \frac{2\pi \omega q}{n_0}   \right)   \rm 
 \text{,}
\end{align} 
where only the bottom-left element of (\ref{S-width-1}) appears.

Since the derivative in $ \partial \xi_{\parallel, k+q}  /  \partial \Theta  $ is taken at fixed $\varphi$, the definition of $\Theta$ in (\ref{Theta-def}) implies that
\begin{align}
\frac{\partial \xi_{\parallel, k+q}}{\partial \theta}  = \frac{\partial \xi_{\parallel, k+q}}{\partial \Theta}  \rm .
\end{align}
Integrating in $\theta$ at fixed $\varphi = \varphi_{k+q}$ gives the circumference $\bar C$ around the unperturbed flux surface, 
\begin{align} \label{C-dxidTheta}
\bar C = \int_{-\pi}^{\pi}  \frac{\partial \xi_{\parallel, k+q}}{\partial \theta}  d\theta  \rm .
\end{align}
Since each plane under consideration, given by $\varphi = 2\pi ( k + q) / n_0$, has the same unperturbed flux surface indepently of the value of $k+q$, the circumference is independent of the index and thus $\bar C = \bar C_{k+q}$.
Consider the definition of $\Theta$ in equation (\ref{Theta-def}). 
Following the island centre $\Theta_{k+q} = \Theta_k =  \theta_k - N \varphi_k / M$ is conserved and therefore the poloidal angle changes according to the equation $\theta_{k+q} -  \theta_k = 2\pi Nq / (n_0 M)$.
Since we are only considering the set of equivalent planes separated by intervals of $2\pi / n_0$ in $\varphi$, the poloidal angles $\theta_{k+q}$ correspond to the same flux surface poloidal cross-section. 
The field line first returns to the same poloidal location in an equivalent plane when $N L / (n_0 M) = \bar{n}$, where $L$ and $\bar{n}$ are the smallest possible integers satisfying this relation.
Thus, the number of distinct islands crossed by a unique island centre magnetic field line in an equivalent poloidal plane is
\begin{align}
L = \frac{\bar{n} n_0}{N} M \rm .
\end{align}
The interval in poloidal angle $\theta$, when defined between $-\pi$ and $\pi$, between the fixed points in a given plane is $2\pi / L$, as there are $L$ equally spaced fixed points.
Thus, if $L$ is sufficiently large, $L \gg 1$, the integral in (\ref{C-dxidTheta}) can be replaced by a sum and the circumference can be approximated by
\begin{align} \label{C-intermediate}
\bar C  \simeq \frac{2\pi}{L} \sum_{q=q_0 }^{q_0+L-1} \frac{\partial \xi_{\parallel, k+q}}{\partial \theta} \simeq \sum_{q=q_0}^{q_0 + L-1}   \frac{ \pi ( \delta \xi_{\parallel, k+q } /  \delta \xi_{\perp, k}  )  M \bar{w}_{\perp, k} }{2L  \sin \left(   2\pi \omega q  / n_0 \ \right)  } \rm .
\end{align}
The error made in approximating the integral over the periodic function $\theta$ as a sum with equally spaced integration points is exponentially small in $1/L$, $O\left( \exp (-L) \right)$.

The integer $q_0$ can be chosen arbitrarily, although a convenient choice is made as follows. 
If $\sin \left( 2\pi \omega (k+q) / n_0 \right) = 0$, the denominator in the summand with index $q$ is zero.
However, no matter how precise, a numerical calculation of $\delta \xi_{\parallel, k+q}$ will not give exactly zero, due to the fact that the elliptical motion described using the tangent map (\ref{T-width}) has small errors both in the aspect ratio and in the axes directions (the higher order terms in $\hat{\epsilon}$ that were neglected).
To avoid the resulting divergence of the error, the first index of the summand is chosen such that the the sum is centred around $\sin \left( 2\pi \omega q / n_0 \right) = 1$, $2\pi \omega (q_0 + L/2 ) / n_0 = \pi / 2$, giving
\begin{align} \label{q0}
 q_0   = \text{int}\left( -  \frac{L}{2} +  \frac{n_0 }{ 4 \omega } \right) \rm ,
\end{align}
where int denotes a function that rounds to the nearest integer.
This amounts to following the linearized trajectory for a large number, $q_0 / L$, of closed magnetic field line periods.
With this choice, the linearized trajectory is followed until it has rotated by as close as possible to $\pi / 2$ in magnetic coordinates for values of $q$ in the middle of the summation interval.
Moreover, we assume that the frequency of rotation, $\omega$ in equation (\ref{omega}), is small enough that the change in the quantity $2\pi \omega (k+q) / n_0$ in the summation interval $q_0 \leqslant q < q_0 + L$ is also small, $2\pi \omega L / n_0 \ll 1$, giving $\sin \left( 2\pi \omega q / n_0 \right) \simeq 1$ for all values of $q$ in the sum and thus
\begin{align} \label{C-intermediate-2}
C  \simeq \frac{M\pi}{2L} \sum_{q=q_0}^{q_0 + L-1}   \frac{ \delta \xi_{\parallel, k+q } }{  \delta \xi_{\perp, k} } \bar{w}_{\perp, k}    \rm .
\end{align}
In \cite{Cary-Hanson-1991}, this assumption is not made and the sine function in equation (\ref{C-intermediate}) is kept. 
Using equations (\ref{omega}) and the pessimistic ordering in which $L/n_0$ is largest, $L \sim n_0 M$, this assumption requires the modestly stricter ordering $2\pi M^2 \sqrt{\iota'_0 \epsilon} \sim 2\pi M^2 \sqrt{\hat{\epsilon}} \ll 1$. \footnote{\cite{Cary-Hanson-1991} point out that --- according to \cite{Greene-1979} --- the rotation angle of points near the stable fixed point of a standard map (that resembles the full-orbit map $\mat{M}_k$ considered here) has a threshold of about $60^{\circ}$ above which chaos ensues.
Therefore, if $2\pi \omega L / n_0$ is not small, it is likely that the concept of island width has broken down.
The footnote in page 299 of \cite{Rosenbluth-1966} argues that $\epsilon$ must decrease with $M$ more rapidly than $M^{-3}$ in order to have non-overlapping islands, further justifying the stricter ordering.}

\subsection{Cylindrical coordinates} \label{subsec-width-lab}

Since stellarators are toroidal devices, we choose the right-handed cylindrical coordinates $(R,  \varphi, Z)$ as a convenient fixed coordinate system.
Here $\varphi$ is the toroidal coordinate, $R$ is the smallest distance of a point from the axis through the centre of the torus and $Z$ is the smallest distance of a point from the mid-plane of the device.

In what follows, we study the equations describing the magnetic field line poloidal position, $\vec{X} = (R, Z)$, as a function of toroidal angle $\varphi$.
Considering the magnetic field line as a streamline of the flow field $\vec{B}$, the streamline trajectory satisfies 
\begin{align}
\frac{Rd\varphi}{B_{\varphi}} = \frac{dR}{B_R} = \frac{dZ}{B_Z} \rm ,
\end{align}
and thus
\begin{align} \label{EOM}
\frac{d\vec{X}}{d\varphi} = \vec{V}(\vec{X}, \varphi) \equiv  \frac{R \vec{B}_p (\vec{X} , \varphi )}{ B_{\varphi}(\vec{X}, \varphi)}  \rm ,
\end{align}
where we have denoted the component of the magnetic field vector in the poloidal plane as $\vec{B}_{\rm p} = (B_R, B_Z)$.
To obtain the position of a magnetic field line at $\varphi = \varphi_{k+q}$, we integrate equation (\ref{EOM}) in $\varphi$ from an initial poloidal position $\vec{X}_k = \vec{X} ( \varphi_k )$,
\begin{align}
\vec{X}_{k+q} = F^q_k \left(  \vec{X}_{k} \right)   \equiv \vec{X}_k + \int_{\varphi_k}^{\varphi_{k+q}} \vec{V}(\vec{X}(\varphi), \varphi) d\varphi \text{.}
\end{align}
A closed magnetic field line $\vecbar{X}$, such as the island centre, satisfies
\begin{align} \label{periodic-field-line}
 \vecbar{X}_{k+L} = F^L_k (\vec{X}_k) = \vecbar{X}_k    \text{.}
\end{align}

For an initial condition that is infinitesimally close to the island centre, $\vec{X}(\varphi_k ) = \vecbar{X}_{k} + \delta \vec{X} (\varphi_k) $, the displacement from the closed field line as a function of toroidal angle satisfies the linearized equation
\begin{align} \label{EOM-delta}
\frac{ d \delta \vec{X}}{d\varphi}  = \left[ \nabla_{\vec X} \vec{V} \left( \vecbar{X}, \varphi \right) \right]^{\intercal} \cdot \delta \vec{X}   \rm ,
\end{align}
where the Jacobian of the field-line-following equation is
\begin{align} \label{Jacobian}
\nabla_{\vec X} \vec{V} = \frac{ \uvec{e}_R  \vec{B}_p (\vecbar{X} , \varphi )}{ B_{\varphi}(\vecbar{X}, \varphi)}  +  \frac{R \nabla_{\vec X} \vec{B}_p (\vecbar{X} , \varphi )}{ B_{\varphi}(\vecbar{X}, \varphi)}  -  \frac{R ( \nabla_{\vec X} B_{\varphi}(\vecbar{X}, \varphi) )  \vec{B}_p (\vecbar{X} , \varphi )  }{ B_{\varphi}(\vecbar{X}, \varphi)^2} \rm .
\end{align}
We have denoted $ \left[  \nabla_{\vec X} \vec{V} \right]^{\intercal} \cdot \delta \vecbar{X} =  \delta \bar{R}  \partial \vec{V} / \partial R +  \delta \bar{Z}  \partial \vec{V} / \partial Z$, and $\uvec{e}_R = \nabla_{\vec X} R$.
It is useful to introduce a 2x2 matrix $\mat{S}_k(\varphi)$ that solves a similar equation to (\ref{EOM-delta}), 
\begin{align}  \label{EOM-U}
\frac{d \mat S_k }{d\varphi}  = \left[ \nabla_{\vec X} \vec{V} \left( \vecbar{X}, \varphi \right) \right]^{\intercal} \cdot  \mat S_k (\varphi)  \rm ,
\end{align}
with initial condition $\mat{S}_k(\varphi_k) = \mat{I}$.
Then any solution to (\ref{EOM-delta}) can be found from $\delta \vec{X}(\varphi) = \mat{S}_k(\varphi) \delta \vec{X}(\varphi_k)$, as can be verified by substituting this expression into (\ref{EOM-delta}). 
Denoting $\mat{S}_k(\varphi_{k+q}) = \mat{S}_k^q$ gives the equation
\begin{align} \label{U-def}
\delta \vec{X}_{k+q} =  \textsf{S}_k^q \delta \vec{X}_k \text{,}
\end{align}
so we can identify $\mat{S}_k^q$ with $\mat{S}_{R,k}^q$ in equations (\ref{deltaxiperp-deltaxipar})-(\ref{S-width-1}).
The tangent map $\mat{S}_k^q$ is obtained from the integral
\begin{align}
 \textsf{S}_k^q  \equiv \textsf{I} +  \int_{\varphi_k}^{\varphi_{k+q}}  \left[ \nabla_{\vec X} \vec{V} \left( \vecbar{X}, \varphi \right) \right]^{\intercal} \cdot  \mat S_k (\varphi)  d\varphi  \text{.}
\end{align} 
In order to carry out this integration, the function $\vecbar{X}(\varphi )$ must be calculated separately from equation (\ref{EOM}).
As the tangent map is linear in $\delta \vec{X}$, it satisfies the property 
\begin{align} \label{map-comptan}
 \mat{S}_k^q  \equiv \mat{S}^1_{k+q-1}  \mat{S}_{k+q-2}^1  \ldots \mat{S}_{k+1}^1 \mat{S}_{k}^1 \rm .
\end{align}
The full-orbit tangent map is denoted
\begin{align}
 \textsf{M}_k = \mathsf{S}_k^{L}  \text{.} 
\end{align}

An important property of the full-orbit tangent map is that it has exactly unit determinant, $\det (\mathsf{M}_k ) = 1$ for all $k$ \citep{Cary-Hanson-1986}.
This follows from the underlying Hamiltonian nature of the magnetic field line trajectory \citep{Meiss-1992}. 
Thus, the characteristic equation for the eigenvalues of $  \mathsf{M}_k $ is $\lambda^2 -  \lambda \text{Tr} (  \mathsf{M}_k ) + 1 = 0$, giving
\begin{align}
 \lambda =   \frac{1}{2} \text{Tr} (  \mathsf{M}_k ) \pm \sqrt{   \left(  \frac{1}{2} \text{Tr} (  \mathsf{M}_k )   \right)^2 - 1  } \text{.}
\end{align}
Hence, for $ | \text{Tr} (  \mathsf{M}_k ) | < 2 $ the eigenvalues are complex numbers on the unit circle, $\lambda_{\pm} =  \exp \left( \pm i \alpha \right) $, with
\begin{align} \label{alpha-rotation}
\alpha = \arccos \left( \frac{1}{2} \text{Tr} (\textsf{M}_k ) \right) \text{.}
\end{align} 
The angle $\alpha$ is the average angle of rotation around the island centre of a neighbouring trajectory after the island centre comes back to its original poloidal position.
For this reason, it takes the same value irrespective of the fixed point $k$ used to calculate it.
From equation (\ref{T-u-Theta}), the average angle of rotation after following around the island centre is $2\pi \omega L / n_0$.
Hence, we obtain an expression for the average frequency of rotation of linearized trajectories about the island centre,
\begin{align} \label{omega-alpha}
\omega = \frac{n_0}{2\pi L} \arccos \left( \frac{1}{2} \text{Tr} (\textsf{M}_k ) \right) \text{.}
\end{align}

It is useful to recall that a closed magnetic field line is not necessarily an island centre.
Since the relation $\det (\mat M_k) = 1$ always holds, the closed field line can be elliptic or hyperholic: it can be an O point or an X point respectively.
When the closed field line is not an O point, the concept of a rotation frequency breaks down.
A quantity that can be used to determine whether the closed field line is a centre or an X point is the residue \citep{Greene-1968} of the full orbit tangent map,
\begin{align} \label{Residue}
\mathcal{R} = \frac{1}{2} - \frac{1}{4} \text{Tr} (\textsf{M}_k )\rm . 
\end{align}
For $0 < \mathcal{R} < 1$, a rotation frequency can be calculated from (\ref{omega-alpha}) and thus the closed field line is a centre.
If $\mathcal{R} < 0$ or $\mathcal{R} > 1$ the closed field line is an X point, as the magnitude of the trace is greater than unity and (\ref{omega-alpha}) does not have a real solution for $\omega$.
In the island width calculation of this section, it is not only assumed that $0 < \mathcal{R} < 1$, but also that $\mathcal{R} \ll 1$ such that $2\pi \omega L / n_0 \ll 1$.

Two-dimensional matrices with unit determinant, such as $\mathsf{M}_k$, satisfy the equation 
\begin{align} \label{symplectic-property}
\mat{M}_k^{\mathrm{T}} \mat{\sigma} \mathsf{M}_k  = \mat{M}_k \mat{\sigma} \mat{M}_k^{\mathrm{T}} =   \mat{\sigma} \rm ,
\end{align}
with
\begin{align}
\sigma = \begin{pmatrix}
0 & 1 \\ 
-1 & 0 \\
\end{pmatrix} \text{.}
\end{align}
From equation (\ref{symplectic-property}) we get $\mat{M}_k^{\mathrm{T}} ( \mat{\sigma} \mat{M}_k ) \mat{M}_k   =  \mat{\sigma} \mat{M}_k$. 
Therefore, the matrix $\mat{\sigma} \mat{M}_k $ is an invariant of the full-orbit tangent map.
The symmetrized matrix 
\begin{align} \label{W-matrix}
 \mat{W}_k = \frac{1}{2} \left(  \sigma  \mat{M}_k +  \mat{M}_k^\intercal  \sigma^\intercal \right)
\end{align}
satisfies the same property that the unsymmetrized counterpart $ \mat{\sigma} \mat{M}_k $ satisfies,
\begin{align}
\mat{M}_k^{\mathrm{T}} \cdot  \mat{W}_k \cdot  \mat{M}_k   =  \mat{W}_k \text{.}
\end{align}

Consider the scalar invariant discussed after equation (\ref{Qmat-def}) calculated at $\varphi = \varphi_k$, $|\delta K | = \delta \boldsymbol{\xi}_k^\intercal \cdot \mat{Q}_k^\intercal \cdot \mat{K} \cdot \mat{Q}_k \cdot \delta \boldsymbol{\xi}_k $.
This quantity is constant on a trajectory crossing the planes $\varphi = 2\pi (k+LQ) / n_0$, where $Q$ is an integer, as it directly follows from the quadratic perturbation to the Hamiltonian (in the magnetic coordinate analysis) about the island centre, $|\delta K | = \delta \vec{u}^\intercal \cdot \mat{K} \cdot \delta \vec{u}$.
Similarly, the quantity $\delta \vec{X}_k^\intercal \cdot \mat{W}_k \cdot \delta \vec{X} $ is constant on a trajectory crossing the plane $\varphi = 2\pi (k + LQ) / n_0$, since $\delta \vec{X}_k^\intercal \cdot \mat{M}_k^\intercal \cdot  \mat{W}_k \cdot  \mat{M}_k \cdot \delta \vec{X}_k   =  \delta \vec{X}_k^\intercal \cdot \mat{W}_k \cdot \delta \vec{X}_k  $.
The vectors $\delta \vec{X}_k$ and $\delta \boldsymbol{\xi}_k$ are equivalent displacement vectors expressed in two different coordinate systems that are rotated with respect to one another. 
Since the scalar invariant remains unchanged after a rotation, the two invariants can only be related by an overall constant,  $\delta \vec{X}_k^\intercal \cdot \mat{W}_k \cdot \delta \vec{X}_k = \gamma_k | \delta K |$ \citep{Cary-Hanson-1991}.
Hence, the symmetric invariant matrix $\textsf{W}_k$ has the same unit eigenvectors as the (approximately diagonal) matrix 
\begin{align} \label{QTKQ-invariant}
\mat{Q}_k^\intercal \cdot \mat{K} \cdot \mat{Q}_k =  \begin{pmatrix}
 \frac{1}{2} \iota_0'(\psi_0) \left(\frac{\partial \psi}{ \partial \xi_{\perp} } \right)^2 + O \left( \frac{ \epsilon }{ \bar{C}^2}  \right)  & O \left( \frac{ M^2 \epsilon }{ \bar{C}^2}  \right)  \\
O \left( \frac{ M^2 \epsilon }{  \bar{C}^2}  \right)  & \frac{1}{2} M^2\epsilon \left(\frac{ \partial \xi_{\parallel} }{\partial \Theta} \right)^2  + O \left( \frac{ M^2 \hat{\epsilon}\epsilon }{\bar{C}^2}  \right)
\end{pmatrix} \rm ,
\end{align}
where in the error terms we have assumed that $ | \partial \boldsymbol{\xi} / \partial \Theta| \sim \bar C$ and  $ | \partial \boldsymbol{\xi} / \partial \psi| \sim \bar C  \iota'_0 (\psi_0)$.
Moreover, the eigenvalues of $\textsf{W}_k$ are equal to the eigenvalues of $\mat{Q}_k^\intercal \cdot \mat{K} \cdot \mat{Q}_k$ multiplied by a factor of $\gamma_k$. 
From equation (\ref{QTKQ-invariant}), the smallest eigenvalue of $\mat{Q}_k^\intercal \cdot \mat{K} \cdot \mat{Q}_k$ is approximately equal to the bottom-right element, associated with the eigenvector $(\delta \xi_{\perp}, \delta \xi_{\parallel}) \simeq ( O(\hat{\epsilon}), 1+  O(\hat{\epsilon}))$, and the largest eigenvalue is approximately equal to the top-left element, associated with the eigenvector $(\delta \xi_{\perp}, \delta \xi_{\parallel}) = ( 1 + O(\hat{\epsilon}),  O(\hat{\epsilon}) )$. 
Therefore, the eigenvector of $\textsf{W}_k$ corresponding to the smallest eigenvalue is approximately tangent to the flux surface, and is denoted $\uvec{e}_{k\parallel}$, and the eigenvector corresponding to the largest eigenvalue is approximately normal to the flux surface, and is denoted $\uvec{e}_{k\perp} $. 
Since $\uvec{e}_{k\parallel}$ and $\uvec{e}_{k\perp} $ are just defined as eigenvectors of $\mat{W}_k$, their sign is yet unspecified.
We impose the constraint $\uvec{e}_{\parallel, k} \cdot \mat{M}_k \cdot \uvec{e}_{\perp, k} > 0$ for all $k$ to fix the \emph{relative} sign of the eigenvectors.

\subsection{Island width} \label{subsec-width-width}

An expression for the island width at $\varphi = \varphi_k = 2\pi k / n_0$ can be obtained by rearranging (\ref{C-intermediate-2}),
\begin{align} \label{width-initial}
\bar{w}_{\perp, k}  = \frac{2L\bar{C} }{M\pi } \left( \sum_{q=q_0}^{q_0+L-1}  \frac{ \delta \xi_{\parallel, k+q }  }{ \delta \xi_{\perp, k}    } \right)^{-1}  \rm.
\end{align}
Recall that $\delta \xi_{\parallel, k} = 0$ and $q_0$ is chosen for convenience according to equation (\ref{q0}), such that $\delta \xi_{\perp, k+q} \simeq 0$ for all $q$ in the sum.
An accurate calculation of the island width thus rests upon an accurate calculation of  $\bar{C}$ and $ \delta \xi_{\parallel, k+q }  / \delta \xi_{\perp, k} $.

To calculate $ \delta \xi_{\parallel, k+q }  / \delta \xi_{\perp, k} $, we follow the linearized magnetic field with initial condition $\delta \vec{X}_k \cdot \uvec{e}_{\parallel k} = 0$, corresponding to $\delta \xi_{\parallel, k} \simeq 0$ and $\delta \xi_{\perp, k} \simeq  \delta \vec{X}_k \cdot \uvec{e}_{\perp, k} $. 
Using equation (\ref{EOM-U}), or repeated application of partial tangent maps as in (\ref{map-comptan}), we obtain $\delta\vec{X}_{k+q} = \mat{S}^q_k \cdot \delta \vec{X}_k$.
The final displacement from the fixed point in the direction parallel to the flux surface is $\delta \xi_{\parallel, k+q} \simeq  \uvec{e}_{\parallel, k+q} \cdot \delta \vec{X}_{k+q} $, giving 
\begin{align} \label{deltaxiparallel}
\frac{ \delta \xi_{\parallel, k+q} }{\delta \xi_{\perp, k}} \simeq  \uvec{e}_{\parallel,k+q} \cdot \mat{S}_k^q \cdot \uvec{e}_{\perp, k}  \rm . 
\end{align}
Equation (\ref{deltaxiparallel}) is defined to be positive for all $k$ and $q$, which provides a second constraint to fix the sign of the vectors $\uvec{e}_{\perp, k}$ and $\uvec{e}_{\parallel, k}$.
With this additional constraint, the signs of the eigenvectors $\uvec{e}_{\perp,k}$, $\uvec{e}_{\parallel,k}$, $\uvec{e}_{\perp, k+q}$, $\uvec{e}_{\parallel, k+q}$ for all $k$ and $q$ are specified with respect to one another.
Note that the two constraints can only both be self-consistently applied if the small island width approximation is valid; if attempting to apply both constraints leads to a contradiction, the islands are too wide. 
We then have 
\begin{align} \label{Sigmak-def}
\sum_{q=q_0}^{q_0+L-1}  \frac{ \delta \xi_{\parallel, k+q }  }{ \delta \xi_{\perp, k}    } \simeq \Sigma_k \equiv \sum_{q=q_0}^{q_0+L-1}  \uvec{e}_{\parallel,k+q} \cdot \mat{S}_k^q \cdot \uvec{e}_{\perp, k}  \text{,}
\end{align}
where we have defined the positive quantity $\Sigma_k$.

The circumference $\bar{C}$ can be approximated, for large enough number of fixed points, from the sum of the chords in the poloidal plane.
However, to obtain the appropriate chords, the fixed points must first be reordered to proceed monotonically around the flux surface. 
To do so, we consider a reordering function $\rho(k)$ such that $p = \rho(k) $ is the index labelling the reordered fixed points.
We define $\rho(0) = 0$ such that $p=0$ for $k =0$, i.e. the set of reordered fixed points have the same point indexed as zero as the set of fixed points ordered by following the field line.
Moreover, we reorder the fixed points to monotonically circulate the magnetic axis in the same direction that they \emph{appear} to circulate when ordered following the magnetic field.
The inverse of the reordering function is denoted as $\rho^{-1}$ and is defined such that $\rho^{-1}(\rho (k)) = k$, i.e. the field-line-following index $k$ is returned for a given reordered index $p$.
It can be shown that
\begin{align}
\rho(k) = (n_{\rm turns} k) ~\text{mod} ~L \rm 
\end{align}
and
\begin{align}
\rho^{-1}(p) = \frac{  \left(p ~\text{mod} ~n_{\text{turns}} \right)L + p }{n_{\text{turns}}} \text{ mod } L    \rm ,
\end{align}
where $n_{\rm turns}$ is the total number of poloidal turns around the magnetic axis when following the field-line-ordered fixed points $\vec{X}_k$ (i.e. the number of times that the magnetic axis \emph{appears} to have been circulated poloidally).
Here, $k \text{ mod } q$ indicates the remainder of $k/q$, equal to an integer between $0$ and $q-1$.
The circumference $\bar C$ is approximated by the sum of the chords obtained by joining the reordered fixed points,
\begin{align} \label{circ}
\bar C \simeq C \equiv \sum_{k=0}^{L-1} \left| \vec{c}_{k} \right| \text{,}
\end{align}
where 
\begin{align}
\vec{c}_{k} = \vec{X}_{k} -   \vec{X}_{k_-} \rm ,
\end{align}
and $\vec{X}_{k_-}$ is the fixed point preceding $\vec{X}_{k}$ in the \emph{reordered} set, such that
\begin{align}
k_- = \rho^{-1}(\rho(k) - 1) \rm .
\end{align}
The relative error in the chord approximation, i.e. $\bar{C} \simeq C$, is $O(1/L)$.

Using equations (\ref{width-initial}), (\ref{Sigmak-def}) and (\ref{circ}), the analytical small island width $\bar{w}_{\perp, k} $ can be approximated by $\bar{w}_{\perp, k}  \simeq w_{\perp, k} $, where 
\begin{align} \label{width}
w_{\perp, k} \equiv \frac{2LC }{M\pi \Sigma_k}   \rm.
\end{align}
The value of $M$ is obtained by counting how many of the $L$ fixed points $\vecbar X_k$ are also fixed points in the plane $\varphi = 0$, i.e. satisfy $\vec X (2\pi L / n_0 ) = \vec X(0) = \vecbar X_k$. 
The relative error in the island width $w_{\perp, k}$ compared to the true island width is $O(\hat{\epsilon}^{1/2}, L^{-1})$.
The $O(\hat{\epsilon}^{1/2})$ piece comes from the error in approximating the width in the coordinate $\psi$ using $\Upsilon$ in equation (\ref{width-flux}), and the additional error in relating changes in $\psi$ to lengths in equation (\ref{wperpbar-def}).
The $O(L^{-1})$ piece comes from the chord approximation.
Considering instead the relative error between $w_{\perp, k}$ in equation (\ref{width}) and $\bar{w}_{\perp, k}$ in equation (\ref{wperpbar-def}), this is $O(\hat{\epsilon}, L^{-1})$.
Here, the $O(\hat{\epsilon})$ piece comes from the error in equation (\ref{deltaxiparallel}), which in turn comes from the $O(\hat{\epsilon})$ error in the bottom-left matrix element of (\ref{S-width-1}) due to the matrix $\mat{K}$ in (\ref{deltaK}) only being approximately diagonal.

\section{Island size and residue variation using adjoint equations} \label{sec-shape}

Here we consider the variation of the different quantities --- including the width --- related to magnetic islands following the infinitesimally small variation of the magnetic field configuration from $\vec{B} (\vec{R}) $ to $ \vec{B} (\vec{R}) + \Delta \vec{B} (\vec{R}) $. 
From equation (\ref{width}), the variation of the island width is 
\begin{align} \label{Deltaw}
\frac{\Delta w_{\perp,k}}{w_{\perp,k}}  \simeq   \frac{ \Delta C }{C} -  \frac{ \Delta \Sigma_k  }{ \Sigma_k } \text{,}
\end{align}
where $ \Delta C$ is the variation of the circumference, approximated by the sum of chords in equation (\ref{circ}), and $\Delta \Sigma_k$ is the variation of the sum in equation (\ref{Sigmak-def}). 
Most of this section is devoted to deriving expressions for $\Delta C$ and $\Delta \Sigma_k$ in terms of $\Delta \vec{B} ( \vec{R})$, using an adjoint method.
In the final subsection, an expression for $\Delta \mathcal{R}$ is derived.
Note that the variation of the on-axis rotational transform is directly related to $\Delta \mathcal{R}$ if one considers the magnetic axis as the periodic field line instead of an island chain O or X point.

\subsection{Circumference variation}

A result of the change in magnetic configuration from $\vec{B} (\vec{R}) $ to $ \vec{B} (\vec{R}) + \Delta \vec{B} (\vec{R}) $ is that the periodic field line position changes from $\vecbar{X}(\varphi)$ to $\vecbar{X}(\varphi) + \Delta \vecbar{X} ( \varphi )$. 
For the purpose of the island width calculation, the variation of the periodic field line position affects the circumference.
This changes from $C$ in equation (\ref{circ}) to $C+ \Delta C$, where $\Delta C$ is given by
\begin{align}\label{DeltaC-int}
\Delta C(\vecbar{X}(\varphi); \Delta \vecbar{X}(\varphi)) = \sum_{k=1}^L  \left( \Delta \vecbar{X}_k  - \Delta \vecbar{X}_{k_-} \right) \cdot  \hat{\vec{c}}_k \rm ,
\end{align}
where $\uvec{c}_{k} = \vec{c}_{k} / |\vec{c}_{k} |$.
Equation (\ref{DeltaC-int}) can be re-expressed as
\begin{align} \label{DeltaC-DeltaX-int}
\Delta C(\vecbar{X}(\varphi); \Delta \vecbar{X}(\varphi)) =  \int_0^{2\pi L/n_0} d\varphi \Delta \vecbar{X} ( \varphi )  \cdot \left[  \sum_{k=0}^{L-1}  \hat{\vec{c}}_k \left( \delta (\varphi - \varphi_k ) - \delta (\varphi - \varphi_{k_-}) \right) \right] \rm .
\end{align}
Notice that the second sum in equation (\ref{DeltaC-DeltaX-int}) can be re-cast as
\begin{align} \label{sum-equivalence}
\sum_{k'=0}^{L-1}  \hat{\vec{c}}_{k'} \delta (\varphi - \varphi_{k'_-}) = \sum_{k=0}^{L-1}  \uvec{c}_{k_+} \delta (\varphi - \varphi_k ) \rm ,
\end{align}
where 
\begin{align}
k_+ = \rho^{-1}(\rho(k)+1) \rm ,
\end{align}
since the order in which the sum is taken is not important.
Re-expressing equation (\ref{DeltaC-DeltaX-int}) using equation (\ref{sum-equivalence}) gives the more convenient expression
\begin{align} \label{DeltaC-DeltaX}
\Delta C(\vecbar{X}(\varphi); \Delta \vecbar{X}(\varphi)) =  \int_0^{2\pi L/n_0} d\varphi \Delta \vecbar{X} ( \varphi )  \cdot \left[  \sum_{k=0}^{L-1}  \left( \hat{\vec{c}}_k  - \hat{\vec{c}}_{k_+}\right) \delta (\varphi - \varphi_k )   \right] \rm .
\end{align}

We impose the constraint that $\vecbar{X} (\varphi) + \Delta \vecbar{X} (\varphi)$ remain a point along a closed magnetic field by introducing a Lagrangian
\begin{align} \label{LC}
\mathcal{L}_C = C + \left\langle \boldsymbol{\lambda}, \frac{d\vec{X}}{d\varphi} - \vec{V}(\vec{X}, \varphi ) \right\rangle \rm ,
\end{align}
where we have defined an inner product such that
\begin{align}
\left\langle \vec{A}_1, \vec{A}_2 \right\rangle = \int_0^{2\pi L / n_0} d\varphi \vec{A}_1 \cdot \vec{A}_2 \rm .
\end{align}
Extremization of $\mathcal{L}$ with respect to $\boldsymbol{\lambda}$ leads to the constraint that $\vec{X}(\varphi) $ is the position along a magnetic field line, i.e. that $\vec{X} (\varphi ) $ satisfies the differential equation (\ref{EOM}).
We have already found the periodic field line $\vecbar{X} (\varphi)$ satisfying (\ref{EOM}) with periodic boundary conditions.
Now consider extrema of $\mathcal{L}$ with respect to changes in $\vec{X} (\varphi)$,
\begin{align}
\Delta \mathcal{L}_C =  \Delta C + \left\langle \boldsymbol{\lambda}, \frac{d\Delta \vec{X}}{d\varphi} - \left[ \nabla_{\vec X} \vec{V} \right]^{\intercal}\cdot \Delta \vec{X}  \right\rangle = 0 \rm,
\end{align}
which can be re-expressed as
\begin{align}
 \Delta C - \left\langle \Delta \vec{X} , \frac{d \boldsymbol{\lambda}}{d\varphi} + \nabla_{\vec X} \vec{V} \cdot \boldsymbol{\lambda} \right\rangle + \boldsymbol{\lambda}(2\pi L / n_0) \cdot \Delta \vec{X} ( 2\pi L/n_0)  -  \boldsymbol{\lambda}(0) \cdot \Delta \vec{X} ( 0 ) = 0 \rm .
\end{align}
By definition, the island centre maintains its periodicity after the perturbation, so we impose $ \Delta \vec{X} ( 2\pi L / n_0 )  = \Delta \vec{X} ( 0 )$.
Additionally imposing the periodic boundary condition $ \boldsymbol{\lambda} (0) = \boldsymbol{\lambda} (2\pi L / n_0)  $ one obtains the equation
\begin{align}
\left\langle \Delta \vec{X}  ,   \left[  \sum_{k=0}^{L-1}  \left( \hat{\vec{c}}_k  - \hat{\vec{c}}_{k_+} \right) \delta (\varphi - \varphi_k )   \right] - \frac{d \boldsymbol{\lambda}}{d\varphi} - \nabla_{\vec X} \vec{V} \cdot \boldsymbol{\lambda} \right\rangle = 0 \rm ,
\end{align}
which leads to the differential equation
\begin{align} \label{C-adjoint}
\frac{d \boldsymbol{\lambda}}{d\varphi} + \nabla_{\vec X} \vec{V} \cdot \boldsymbol{\lambda} =  \sum_{k=0}^{L-1}  \left( \hat{\vec{c}}_k  - \hat{\vec{c}}_{k_+} \right) \delta (\varphi - \varphi_k )  \rm .
\end{align}
Equation (\ref{C-adjoint}) is the adjoint equation useful for finding derivatives of the circumference.

 If $\vecbar{X}(\varphi)$ is a periodic solution of (\ref{EOM}) and $\bar{\boldsymbol \lambda} (\varphi)$ is a periodic solution of (\ref{C-adjoint}), then $\mathcal{L}_C$ is stationary with respect to changes  in $\vecbar{X} (\varphi)$ and $\bar{\boldsymbol \lambda} (\varphi)$ arising from changes in the magnetic field $\vec{B} (\vec{R})$.
 Therefore, $\mathcal{L}_C$ is only affected by terms containing changes in the magnetic field explicitly, $\Delta \mathcal{L}_C =  \left\langle \bar{\boldsymbol \lambda} (\varphi),  - \Delta \vec{V}(\vecbar{X}, \varphi)  \right\rangle $.
Furthermore, since $\vecbar{X}(\varphi)$ is a magnetic field line trajectory, the equation $d\vecbar{X} / d\varphi = \vec{V}(\vecbar{X}(\varphi), \varphi )$ is always satisfied and $\Delta C = \Delta \mathcal{L}_C$ (from equation (\ref{LC})), giving 
\begin{align} \label{DeltaL-DeltaB}
\Delta C = - \int_0^{2\pi L / n_0} d\varphi \bar{\boldsymbol \lambda} \cdot  \Delta \vec{V}(\vecbar{X}, \varphi)  \rm,
\end{align}
where
\begin{align} \label{DeltaV}
\Delta \vec{V} (\vecbar{X}, \varphi) = \frac{\bar{R} \Delta \vec{B}_{\rm p} (\vecbar{X}, \varphi)}{B_{\varphi}(\vecbar{X}, \varphi)}  - \frac{\bar{R} \vec{B}_{\rm p} (\vecbar{X}, \varphi) \Delta B_{\varphi} (\vecbar{X}, \varphi) }{B_{\varphi}(\vecbar{X}, \varphi)^2 }  \rm .
\end{align}

\subsection{Variation of $ \Sigma_k$}

Upon varying the magnetic field from $\vec{B}(\vec{R})$ to $\vec{B}(\vec{R}) + \Delta \vec{B}(\vec{R})$, the quantity $\Sigma_k$ in equation (\ref{Sigmak-def}) varies due to the variation of the eigenvectors $\uvec{e}_{\perp k}$ and $\uvec{e}_{\parallel k+q} $ and of the tangent map $\mat{S}_k^q$,
\begin{align} \label{Deltaw-includingDeltaevecs}
\Delta \Sigma_k =  \sum_{q=q_0}^{q_0+L-1}  \left( \uvec{e}_{\parallel, k+q } \cdot \Delta \mat{S}_k^q \cdot \uvec{e}_{\perp,k }  +  \Delta e_{\parallel, k+q} \uvec{e}_{\perp k+q} \cdot \mat{S}_k^q \cdot \uvec{e}_{\perp, k }  \right. \nonumber  \\ \left. +  \Delta e_{\perp, k} \uvec{e}_{\parallel, k+q} \cdot \mat{S}_k^q \cdot \uvec{e}_{\parallel , k}  \right) \text{.}
\end{align}
In equation (\ref{Deltaw-includingDeltaevecs}), we have used the fact that the variation of a unit vector is perpendicular to the unit vector itself to re-express the variation of the normalized eigenvectors of $\mat{W}_k$ as $\Delta \uvec{e}_{\perp, k} = \Delta e_{\perp, k} \uvec{e}_{\parallel, k} $ and $\Delta \uvec{e}_{\parallel, k} = \Delta e_{\parallel, k} \uvec{e}_{\perp, k} $. 
The slow rotation of nearby points around the O point gives $\sin \left( 2\pi \omega k / n_0 \right) \simeq 1$  and $\cos \left( 2\pi \omega k / n_0 \right) \simeq 0$.
Thus, the matrix elements $ \uvec{e}_{\perp, k+q} \cdot \mat{S}_k^q \cdot \uvec{e}_{\perp, k}$ and $ \uvec{e}_{\parallel, k+q} \cdot \mat{S}_k^q \cdot \uvec{e}_{\parallel , k} $, equal to the diagonal matrix elements in $\mat{S}_{R, k}^q$ (equation (\ref{S-width-1})), are both small in $2\pi \omega L / n_0$.
This gives
\begin{align} \label{DeltaSigmak-1}
\Delta \Sigma_k \simeq  \sum_{q=q_0}^{q_0+L-1}  \uvec{e}_{\parallel, k+q } \cdot \Delta \mat{S}_k^q \cdot \uvec{e}_{\perp, k }  =  \sum_{q=q_0}^{q_0+L-1}   \uvec{e}_{\parallel, k+q } \cdot \Delta \vec{s}_k^q    \text{;}
\end{align}
here, we have introduced the variable $\vec{s}_k(\varphi) = \delta \vec{X} (\varphi) / |\delta \vec{X}(2\pi k / n_0)| $ satisfying the differential equation (\ref{EOM-delta}),
\begin{align} \label{EOM-s}
\frac{d\vec{s}_k}{d\varphi} =  \vec{s}_k \cdot \nabla_{\vec X} \vec{V} \rm ,
\end{align}
with boundary condition $\vec{s}_k(2\pi k / n_0) = \vec{s}_{ k}^0 = \uvec{e}_{\perp, k}$, and defined the variation $\Delta \vec{s}_k (\varphi)$. 
We re-express (\ref{DeltaSigmak-1}) to
\begin{align} \label{DeltaSigmak}
\Delta \Sigma_k \simeq \sum_{Q=0}^{Q_0-1} \int_{2\pi ( k + LQ) / n_0}^{2\pi (k+L Q+ L) / n_0}   \Delta \vec{s}_k (\varphi )  \cdot \uvec{e}_{\parallel, k+q } \delta (\varphi - \varphi_{k+q} ) d\varphi  \text{,}
\end{align}
where the integer $Q_0 = \lfloor q_0 / L \rfloor + 2$, with the brackets $\lfloor$ and $\rfloor$ denoting the floor function, is chosen such that all values of $q$ in the sum in (\ref{DeltaSigmak-1}) are counted.
Introducing the notation
\begin{align}
\left\langle \vec{A}_1 \cdot \vec{A}_2 \right\rangle_{k,Q} =  \int_{2\pi (k+L Q) / n_0}^{2\pi (k+L Q + L)/n_0}  \vec{A}_1 \cdot \vec{A}_2 d\varphi \rm ,
\end{align}
we define the Lagrangian of $\Sigma_k$,
\begin{align} \label{LSigmak}
\mathcal{L}_{\Sigma_k} = \Sigma_k + \sum_{Q=0}^{Q_0-1} \left( \left\langle \boldsymbol{\lambda}_{k,Q},  \frac{d\vec{X}}{d\varphi} - \vec{V}(\vec{X}, \varphi)  \right\rangle_{k,Q} +  \left\langle \boldsymbol{\mu}_k, \frac{d\vec{s}_k}{d\varphi} -  \vec{s}_k \cdot \nabla_{\vec X} \vec{V} (\vec{X}, \varphi)  \right\rangle_{k,Q} \right) \rm .
\end{align}
In (\ref{LSigmak}) we have introduced two constraints using the adjoint variables $\boldsymbol{\lambda}_{k,Q}(\varphi )$ and $\boldsymbol{\mu}_k (\varphi )$. 
With these constraints, the quantities $\mathcal{L}_{\Sigma_k}$ and $\Sigma_k$ are only equal to each other if $\vec{X} (\varphi )$ is a field line trajectory, satisfying equation (\ref{EOM}), and $\vec{s}_k$ satisfies the linearized equation (\ref{EOM-s}) for small displacements about the field line.\footnote{Note that the calculation could be carried out without replacing $\mat S_k$ by $\vec s_k$ if the vector adjoint variable $\boldsymbol \mu_k$ were replaced by a matrix $\mat \mu_k$.}

Extremization with respect to variations in $\vec{X} (\varphi )$ gives
\begin{align} \label{DeltaL-Sigmak}
\Delta \mathcal{L}_{\Sigma_k} =  - \sum_{Q=0}^{Q_0-1}  \left\langle \Delta \vec{X}, \frac{d\boldsymbol{\lambda}_{k,Q}}{d\varphi} + \nabla_{\vec X} \vec{V} \cdot \boldsymbol{\lambda}_{k,Q}  + \vec{s}_k \cdot \nabla_{\vec X} \nabla_{\vec X} \vec{V} \cdot \boldsymbol{\mu}_k \right\rangle \nonumber \\ + \sum_{Q=0}^{Q_0-1} \left[  \Delta \vec{X}(2\pi(k+QL+L)/n_0) \cdot \boldsymbol{\lambda}_{k,Q} (2\pi(k+QL+L)/n_0) \right. \nonumber  \\
 \left.  - \Delta \vec{X}(2\pi (k + QL)/n_0) \cdot \boldsymbol{\lambda}_{k,Q} (2\pi (k+QL)/n_0) \right] = 0 \rm ,
\end{align}
where the Hessian of the field-line following equation (\ref{EOM}) is
\begin{align}
\nabla_{\vec X} \nabla_{\vec X} \vec{V} (\vec{X}, \varphi) = & \frac{ 2 \uvec{e}_R \nabla_{\vec X} \vec{B}_p (\vec{X}, \varphi)  }{B_{\varphi} (\vec{X}, \varphi) } -  \frac{ 2 \uvec{e}_R ( \nabla_{\vec X} B_{\varphi} (\vec{X}, \varphi)  ) \vec{B}_p (\vec{X}, \varphi) }{B_{\varphi}(\vec{X}, \varphi)^2} + \frac{ R\nabla_{\vec X} \nabla_{\vec X} \vec{B}_p (\vec{X}, \varphi) }{B_{\varphi}(\vec{X}, \varphi) }  \nonumber \\ & -  \frac{ 2 R ( \nabla_{\vec X} B_{\varphi}(\vec{X}, \varphi) ) ( \nabla_{\vec X} \vec{B}_p (\vec{X}, \varphi) )}{B_{\varphi}(\vec{X}, \varphi)^2} -  \frac{ R ( \nabla_{\vec X} \nabla_{\vec X} B_{\varphi}(\vec{X}_p, \varphi) ) \vec{B}_p(\vec{X}, \varphi) }{B_{\varphi}(\vec{X}, \varphi)^2}  \nonumber \\ &
+  \frac{ 2 R ( \nabla_{\vec X} B_{\varphi} (\vec{X}, \varphi) ) ( \nabla_{\vec X} B_{\varphi} (\vec{X}, \varphi) ) \vec{B}_p (\vec{X}, \varphi)  }{B_{\varphi}(\vec{X}, \varphi)^3}   \text{.}
\end{align}
We consider $\vec{X} (\varphi) $ to be a periodic field line $\bar{\vec{X}} (\varphi) $ (an island centre) after the variation of the magnetic field configuration, such that $ \Delta \vec{ X} ( 2\pi (k+QL)/n_0 )  =  \Delta \vec{ X}  ( 2\pi (k+QL+L)/n_0 ) $.
Therefore, all the boundary terms in equation (\ref{DeltaL-Sigmak}) vanish if $\boldsymbol{\lambda}_{k,Q} ( 2\pi (k+QL)/n_0 ) = \boldsymbol{\lambda}_{k,Q} ( 2\pi (k + QL + L)/n_0 ) $.
The adjoint equations for $\boldsymbol{\lambda}_{k,Q} (\varphi )$ are thus
\begin{align} \label{tangent-adjoint-lambda}
\frac{d\boldsymbol{\lambda}_{k,Q}}{d\varphi} + \nabla_{\vec X} \vec{V} \cdot \boldsymbol{\lambda}_{k,Q}  + \vec{s}_k \cdot \nabla_{\vec X} \nabla_{\vec X} \vec{V} \cdot \boldsymbol{\mu}_k = 0  \rm ,
\end{align}
to be solved for periodic solutions $\bar{ \boldsymbol \lambda}_{k,Q} ( \varphi )$ in the interval $2\pi QL / n_0 \leqslant \varphi - 2\pi k / n_0 \leqslant  2\pi (QL+L) / n_0$ for all $k$ and $Q$.

Extremizing $\mathcal{L}_{\Sigma_k}$ with respect to variations in $\vec{s}_k (\varphi )$ gives an equation for $\boldsymbol{\mu}_k$,
\begin{align} \label{DeltaLk-1}
\Delta \mathcal{L}_{\Sigma_k} =  \sum_{Q=0}^{Q_0-1} \left[  \left\langle  \Delta \vec{s}_k  ,  \sum_{q=q_0}^{q_0+L-1} \uvec{e}_{\parallel k+q}  \delta ( \varphi - \varphi_{k+q} ) - \frac{d \boldsymbol{\mu}_k}{d\varphi}  - \nabla_{\vec X} \vec{V} \cdot \boldsymbol{\mu}_k \right\rangle_Q \right] \nonumber \\ \left. + \boldsymbol{\mu}_k (2\pi (k+ L Q_0 )/n_0) \cdot \Delta \vec{s}_{k} (2\pi (k+L Q_0)/n_0) - \boldsymbol{\mu}_k (2\pi k / n_0) \cdot \Delta \vec{s}_{k} (2\pi k / n_0) \right. = 0 \rm ,
\end{align}
where all boundary terms at $\varphi = 2\pi (k+LQ)/n_0$ for $Q\neq 0$ and $Q\neq Q_0$ have vanished because $\boldsymbol{\mu}_k$ is a continuous variable in the interval $2\pi k / n_0 \leqslant \varphi \leqslant 2\pi (k + Q_0L ) / n_0$.
When considering variations in $\vec{s}_k$, the initial condition $\vec{s}_{k}(2\pi k / n_0 ) = \vec{s}_k^0 = \uvec{e}_{\perp k}$ can be assumed to be unchanged and therefore $\Delta \vec{s}_k (2\pi k / n_0) = 0$.
Hence, the boundary terms in (\ref{DeltaLk-1}) vanish by imposing $\boldsymbol{\mu}_k \left( 2\pi (k+LQ_0)/n_0 \right) = 0$, resulting in the differential equation 
\begin{align} \label{tangent-adjoint-mu}
\frac{d \boldsymbol{\mu}_k}{d\varphi}  = \sum_{q=q_0}^{q_0+L-1}  \uvec{e}_{\parallel k+q}  \delta ( \varphi - \varphi_{k+q} )  - \nabla_{\vec X} \vec{V} \cdot \boldsymbol{\mu}_k \rm .
\end{align}
 Note that $\boldsymbol{\mu}_k (2\pi k / n_0) $ can be obtained from $\boldsymbol{\mu}_k (2\pi (k+L Q_0) / n_0) = 0$ using equation (\ref{tangent-adjoint-mu}).
This can be solved by repeated applications of appropriate partial tangent maps and jump conditions at the angles $\varphi_{k+q}$; as shown in appendix \ref{app-DeltaX}, the linear mapping from $\boldsymbol \mu_k$ to $\boldsymbol \mu_{k+1}$ that follows from the homogeneous term on the right hand side of (\ref{tangent-adjoint-mu}) is the adjoint of the partial tangent map $\mat S_k^1$, such that $ \boldsymbol \mu_k \left( 2\pi (k+q)/L\right) = \left( \mat S_{k+q}^1 \right)^{\intercal} \cdot \boldsymbol \mu_k \left( 2\pi (k+q+1)/L\right)$. 
The adjoint variables $\boldsymbol{\lambda}_{k,Q}  (\varphi)$ can be obtained from equation (\ref{tangent-adjoint-lambda}) once the the adjoint variables $\boldsymbol{\mu}_k (\varphi)$ are obtained from (\ref{tangent-adjoint-mu}).

The magnetic field configuration is varied while considering $\bar{ \vec X} (\varphi)$ to be the magnetic field line trajectory at the island centre, satisfying equation (\ref{EOM}) with a periodic boundary condition, and while constraining $\vec{s}_k (\varphi)$ to satisfy equation (\ref{EOM-s}) for linearized trajectories about the island centre.
We thus conclude that $ \Sigma_k =  \mathcal{L}_{\Sigma_k}$ and thus $\Delta \Sigma_k = \Delta \mathcal{L}_{\Sigma_k}$.
Therefore, the variation of $\Sigma_k$ is given by
\begin{align}
 \Delta \Sigma_k = - \sum_{Q=0}^{Q_0-1} \int_{2\pi k / n_0}^{2\pi (k + QL)/n_0} d\varphi \left(  \bar{ \boldsymbol \lambda}_{k,Q} \cdot \Delta \vec{V} (\vecbar{X}, \varphi)  +   \vec{s}_{ k} \cdot \nabla_{\vec X} \Delta \vec{V} (\vecbar{X}, \varphi) \cdot  \boldsymbol{\mu}_k  \right) \rm ,
\end{align}
where $\bar{\boldsymbol \lambda}_{k,Q}$ is a periodic solution of (\ref{tangent-adjoint-lambda}) and $\boldsymbol{\mu}_k$ is the solution of (\ref{tangent-adjoint-mu}) satisfying the boundary condition $\boldsymbol{\mu}_k (2\pi (k+L Q_0) / n_0) = 0$.
The function $\nabla_{\vec X} \Delta \vec{V} (\vecbar{X}, \varphi) $ can be written explicitly in terms of variations of the magnetic field and its gradients by differentiating (\ref{DeltaV}),
\begin{align}
 \nabla_{\vec X} \Delta \vec{V} (\vecbar{X}, \varphi)  = &  \frac{\uvec{e}_R \Delta \vec{B}_p (\vecbar{X}, \varphi) }{B_{\varphi} (\vecbar{X}, \varphi) } 
- \frac{\uvec{e}_R \vec{B}_p(\vecbar{X}, \varphi)  \Delta B_{\varphi} (\vecbar{X}, \varphi) }{B_{\varphi} (\vecbar{X}, \varphi)^2 } 
+ \frac{R \nabla_{\vec X} \Delta \vec{B}_p (\vecbar{X}, \varphi)  }{B_{\varphi} (\vecbar{X}, \varphi) }  \nonumber \\
& - \frac{R (\nabla_{\vec X} \vec{B}_p (\vecbar{X}, \varphi)  ) \Delta B_{\varphi} (\vecbar{X}, \varphi) }{B_{\varphi}(\vecbar{X}, \varphi)^2} 
 - \frac{R \nabla_{\vec X} B_{\varphi} (\vecbar{X}, \varphi)  \Delta \vec{B}_p (\vecbar{X}, \varphi)  }{B_{\varphi}(\vecbar{X}, \varphi)^2}   \nonumber \\
&  - \frac{R \nabla_{\vec X} (\Delta B_{\varphi } (\vecbar{X}, \varphi)  ) \vec{B}_p (\vecbar{X}, \varphi)  }{B_{\varphi}(\vecbar{X}, \varphi)^2}  
   +  \frac{2 R (\nabla_{\vec X}  B_{\varphi } ) \vec{B}_p \Delta B_{\varphi} }{B_{\varphi}^3}  \rm .   
\end{align}

\subsection{Residue variation}

The residue $\mathcal{R}$ varies due to the variation of the full orbit tangent map $\mat{M}_k$.
We re-express the trace of the full orbit tangent map as
\begin{align} \label{Trace-new}
\rm Tr (\mat M_0) = \mat{I} : \mat M_0 \rm,
\end{align}
where the operation denoted by the colon $:$ corresponds to multiplying each matrix element of one matrix with the corresponding matrix element of the other matrix and adding the results together.
Note that $k=0$ was chosen in (\ref{Trace-new}), as the residue is independent of $k$.
From equation (\ref{Residue}) and (\ref{Trace-new}) we obtain
\begin{align} \label{DeltaR-DeltaTr}
\Delta \mathcal R  = - \frac{1}{4}  \mat{I} : \Delta \mat M_0   \text{.}
\end{align}
Recall that $\mat M_0 = \mat S_0 ( 2\pi L / n_0 )$ and that the tangent map $\mat S_0$ satisfies the differential equation (\ref{EOM-U}) with boundary condition $\mat{S}_0(0) = \mat I$.
We define the Lagrangian of $\mathcal R$,
\begin{align} \label{LRes}
\mathcal{L}_{\mathcal R} = \mathcal{R} +  \int_0^{2\pi L / n_0} d\varphi \left[ \boldsymbol{\lambda}_{\mathcal R} \cdot \left(  \frac{d\vec{X}}{d\varphi} - \vec{V}(\vec{X}, \varphi)  \right)  \right. \nonumber \\ \left. +  \mu_{\mathcal{R}} : \left(  \frac{d\mat{S}_0}{d\varphi} -   \left( \nabla_{\vec X} \vec{V} (\vec{X}, \varphi) \right)^\intercal \cdot  \mat{S}_0  \right) \right] \rm .
\end{align}
In (\ref{LRes}) we have introduced two constraints using the adjoint variables $\boldsymbol{\lambda}_{\mathcal R}(\varphi )$ (a vector) and $\mu_{\mathcal R} (\varphi )$ (a matrix). 
With these constraints, the quantities $\mathcal{L}_{\mathcal R}$ and $\mathcal R$ are equal to each other if $\vec{X} (\varphi )$ is a field line trajectory, satisfying equation (\ref{EOM}), and $\mat S_0$ satisfies equation (\ref{EOM-U}) for the tangent map.

Extremization with respect to variations in $\vec{X} (\varphi )$ gives
\begin{align} \label{DeltaL-Residue}
\Delta \mathcal{L}_{\mathcal R}  =  -   \left\langle \Delta \vec{X}, \frac{d\boldsymbol{\lambda}_{\mathcal R}}{d\varphi} + \nabla_{\vec X} \vec{V} \cdot \boldsymbol{\lambda}_{\mathcal R}  + \left( \nabla_{\vec X} \nabla_{\vec X} \vec{V} \cdot \mat \mat{\mu}_{\mathcal R} \right) :  \mat S_0 \right\rangle \nonumber \\ +   \Delta \vec{X}(2\pi L/n_0) \cdot \boldsymbol{\lambda}_{\mathcal R} (2\pi L/n_0)  - \Delta \vec{X}(0) \cdot \boldsymbol{\lambda}_{\mathcal R} (0) = 0 \rm ,
\end{align}
Again, all the boundary terms in equation (\ref{DeltaL-Residue}) vanish if $ \boldsymbol{\lambda}_{\mathcal R} ( 2\pi L/n_0 ) = \boldsymbol{\lambda}_{\mathcal R} ( 0 ) $. 
The adjoint equation for $\boldsymbol{\lambda}_{\mathcal R} (\varphi )$ is thus
\begin{align} \label{Res-adjoint-lambda}
\frac{d\boldsymbol{\lambda}_{\mathcal R}}{d\varphi} + \nabla_{\vec X} \vec{V} \cdot \boldsymbol{\lambda}_{\mathcal R}  + \left( \nabla_{\vec X} \nabla_{\vec X} \vec{V} \cdot \mat \mat{\mu}_{\mathcal R} \right) :  \mat S_0 = 0  \rm ,
\end{align}
to be solved in the interval $0 \leqslant \varphi  \leqslant  2\pi L / n_0$ for periodic solutions, denoted $\bar{ \boldsymbol \lambda}_{\mathcal R} (\varphi )$.

Extremizing $\mathcal{L}_{\mathcal R} $ with respect to variations in $\mat{S}_0 (\varphi )$ gives an equation for $\mat \mu_{\mathcal R}$,
\begin{align} \label{DeltaLRes-1}
\Delta \mathcal{L}_{\mathcal R}  = - \frac{1}{4} \mat{I} : \Delta \mat M_0 + \left\langle \mat{\mu}_{\mathcal R} ,  \frac{d \Delta \mat S_0 }{d\varphi}  - (\nabla_{\vec X} \vec{V})^\intercal \cdot \Delta \mat S_0   \right\rangle = 0 \rm .
\end{align}
Note that we have used the following definition for the inner product of two matrix quantities,
\begin{align}
\left\langle \mat{A}_1, \mat{A}_2 \right\rangle = \int_0^{2\pi L / n_0} d\varphi \mat{A}_1 : \mat{A}_2 \rm .
\end{align}
We re-express (\ref{DeltaLRes-1}) to
\begin{align} \label{DeltaLRes-2}
-\frac{1}{4} \mat{I} : \Delta \mat M_0 - \left\langle  \Delta \mat S_0 ,  \frac{d  \mat{\mu}_{\mathcal R} }{d\varphi}  + \nabla_{\vec X} \vec{V} \cdot  \mat \mu_{\mathcal R}  \right\rangle + \mat \mu_{\mathcal R} (2\pi L / n_0 ) : \Delta S_0 ( 2\pi L / n_0 ) \nonumber \\ -  \mat \mu_{\mathcal R} (0 ) : \Delta S_0 ( 0 ) = 0 \rm .
\end{align}
Noting that $\Delta \mat{S}_0 (0) = 0$ and $\Delta S_0 (2\pi L / n_0 ) = \Delta \mat M_0$ both hold true by definition, the boundary terms are zero provided $\mat \mu_{\mathcal R} ( 2\pi L / n_0 ) = \frac{1}{4} \mat{I}$ is chosen. 
Thus, the adjoint equation for $\mu_{\mathcal R}$ is  
\begin{align}
 \frac{d  \mat{\mu}_{\mathcal R} }{d\varphi}  = - \nabla_{\vec X} \vec{V} \cdot \mat \mu_{\mathcal R} \rm ,
\end{align}
with $\mat \mu_{\mathcal R} ( 2\pi L / n_0 ) =  \frac{1}{4} \mat{I}$ as a boundary condition.
From appendix~\ref{app-DeltaX}, an equivalent form of this boundary condition is $\mat \mu_{\mathcal R} ( 0 ) = \frac{1}{4} \mat{M}_0^{\intercal}$.

As before, the variation of the residue is given by
\begin{align} \label{DeltaR}
\Delta \mathcal{R} =-  \int_0^{2\pi L /n_0} \left( \bar{\boldsymbol \lambda}_{\mathcal R} \cdot  \Delta \vec{V} (\vecbar{X}, \varphi)  +   ( \nabla_{\vec X} \Delta \vec{V} (\vecbar{X}, \varphi) \cdot \mu_{\mathcal R} ) : \mat S_0  \right) d\varphi \rm ,
\end{align}
where $\bar{ \boldsymbol \lambda}_{\mathcal R}$ and $\mu_{\mathcal R}$ are the solutions of the adjoint equations (\ref{tangent-adjoint-lambda}) and (\ref{tangent-adjoint-mu}) with the appropriate boundary conditions.

To conclude this section, we briefly discuss a subsidiary result of the preceeding analysis. 
The trace of the full orbit tangent map calculated at the magnetic axis is related the rotation angle of nearby trajectories about the magnetic axis.
By definition, this is proportional to the on-axis rotational transform. 
Equating (\ref{alpha-rotation}) (with $k=0$) to the product of the interval in toroidal angle, $2\pi  / n_0$, and the on-axis rotational transform, $\bar{\iota}$, and re-arranging for $\bar{\iota}$ gives
\begin{align} \label{iotabar}
\bar{\iota} = \frac{n_0}{2\pi} \arccos \left( \frac{1}{2} \rm Tr \left(  \mat M_0 \right) \right) \rm .
\end{align}
Note that this equation is correct provided $\bar{\iota} < n_0$, a condition which is satisfied in most stellarators.
Remembering the definition (\ref{Residue}) of the residue, the variation of the on-axis rotational tranform is given by
\begin{align} \label{Deltaiota}
\Delta \bar{\iota}  = \frac{n_0}{\pi \sin \left( 2\pi \bar{\iota} / n_0 \right)} \Delta  \mathcal R  \rm .
\end{align}

\section{Numerical results} \label{sec-numerical}

In this section, we present the numerical results obtained for the gradients of the island width and of other properties of the periodic field line.
In section~\ref{subsec-num-scheme}, we briefly explain the numerical scheme used to obtain our results.
We then present, in section~\ref{subsec-num-Reiman} numerical results for the island width, its gradient and the gradient of other island-related quantities in an analytical magnetic configuration studied in \cite{Reiman-1986}.
In section~\ref{subsec-num-NCSX}, we present results for the shape gradient of the width of a magnetic island in NCSX with respect to the positions on a type A modular coil.
Finally, in section~\ref{subsec-num-heli}, we apply the gradient of the residue of a periodic field line to optimize a helical magnetic configuration of the kind studied in \cite{Cary-Hanson-1986} and \cite{Hanson-Cary-1984}.

\subsection{Numerical scheme} \label{subsec-num-scheme}

A Runge-Kutta 4th order explicit scheme is used to integrate equations (\ref{EOM}), (\ref{EOM-U}), (\ref{C-adjoint}), (\ref{tangent-adjoint-lambda}) and (\ref{tangent-adjoint-mu}) in toroidal angle $\varphi$.
The number of grid points in $\varphi$ per field period is denoted $N_{\varphi}$, such that the number of grid points per toroidal turn is $n_0 N_{\varphi}$.

A Newton method is used to search for periodic solutions of the magnetic field line such as the magnetic axis and a magnetic island centre.
The search proceeds as follows.
The position of the periodic field line is initially guessed as $\vec{X}^0 (0)$ and equation (\ref{EOM}) is integrated in $\varphi$ from $\varphi = 0$ to $\varphi = 2\pi L / n_0$, where $L$ is an integer ($=1$ for the magnetic axis).
The tangent map is also integrated following the magnetic field line.
The final position $\vec{X}(  2\pi L / n_0  )$ and tangent map $\mat{S} (  2\pi L / n_0  )$ are used to evaluate a next guess for the magnetic axis as follows.
The step $\vec{X}_{\rm step} (0)$ in position necessary to move closer to the periodic solution at the next iteration, $\vec{X}^{i+1} (0) = \vec{X}^{i} (0) + \vec{X}^i_{\rm step} (0)$, is calculated by imposing $\vec{X}^i(0) + \vec{X}^i_{\rm step} (0) = \vec{X}^i ( 2\pi L / n_0 ) + \vec{X}^i_{\rm step}  ( 2\pi L / n_0 )$ on the linearized equations.
This leads to $\vec{X}^i(0) +  \vec{X}_{\rm step}^i(0) = \vec{X}^i( 2\pi L / n_0 ) + \mat{S}^i ( 2\pi L / n_0  ) \cdot  \vec{X}^i_{\rm step}(0) $ and, upon rearranging, to
\begin{align} \label{step-to-periodic}
\vec{X}_{\rm step}^i (0) = \left( \mat{I} - \mat{S}^i ( 2\pi L / n_0 ) \right)^{-1} \cdot \left( \vec{X}^i ( 2\pi L / n_0 ) - \vec{X}^i(0) \right)   \rm .
\end{align}
The error is calculated from
\begin{align}
\mathcal E_i = \frac{\left| \vec{X}^i ( 2\pi L / n_0  ) -  \vec{X}^i (0) \right|}{\left| \vec{X}^i (0) \right|}
\end{align}
where the magnitude of the poloidal vector $\vec{X} = (R, Z)$ is defined as $|\vec{X} | = \sqrt{R^2 + Z^2}$.
A periodic field line solution is found if $\mathcal E_i $ falls below a threshold value $\mathcal E_{\rm thresh} = 10^{-13}$; we then consider $\vecbar{X}(\varphi) = \vec{X}^i (\varphi)$.

Once a periodic field line is found, the values of the magnetic field and its first and second derivatives on the toroidal grid points and in the intermediate Runge Kutta steps are stored to accelerate subsequent parts of the code such as the solutions of the adjoint equations and the calculations of the gradients.

\subsection{Island width and gradient calculation: Reiman model} \label{subsec-num-Reiman}

\begin{figure} 
\includegraphics[width = 1.0\textwidth]{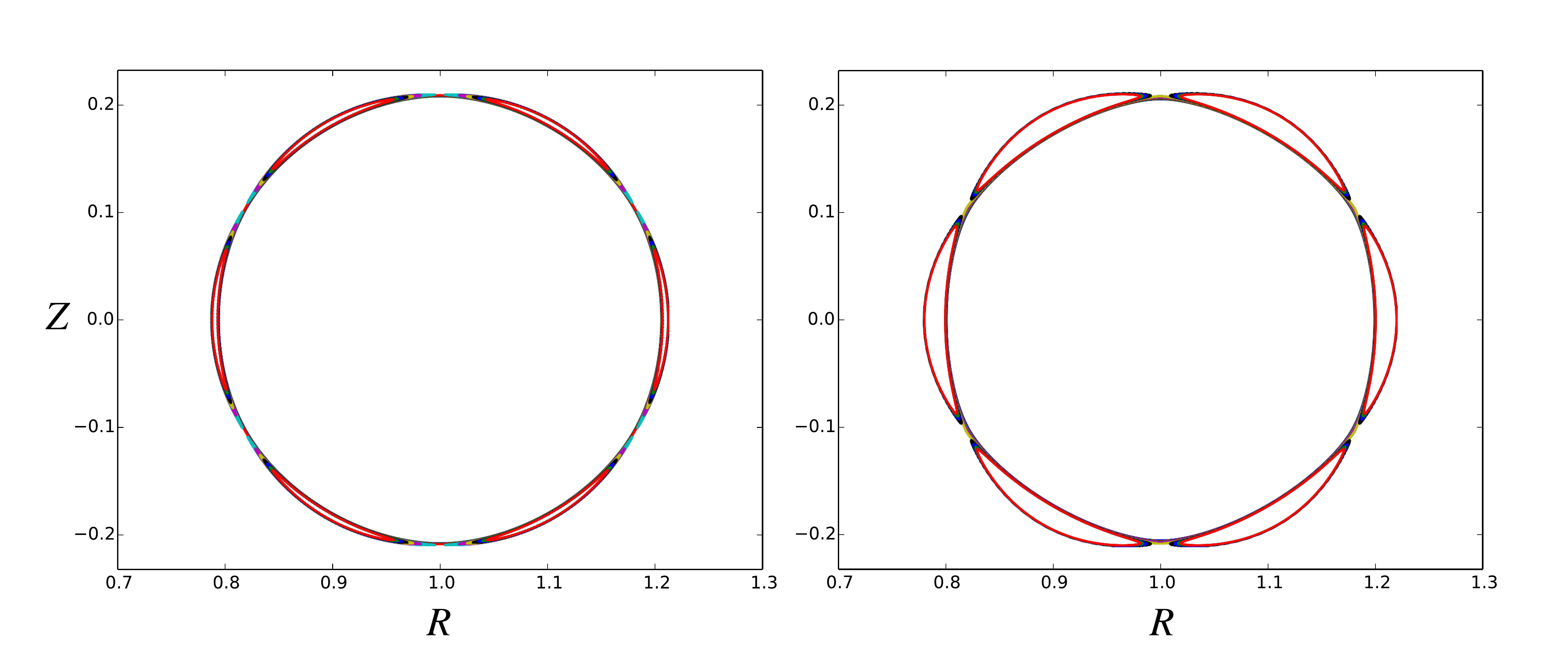}
\caption{Poincar\'e plots showing magnetic field lines near the island separatrix in the Reiman magnetic field configuration with $\iota_{\rm ax} = 0.15$, $\iota'_{\rm ax} = 0.38$ and $\varepsilon_i = 0$ for $i\neq 6 $, for $\varepsilon_6 = 0.001$ (left) and $\varepsilon_6 = 0.01$ (right).}
\label{fig-surfaces-Reiman}
\end{figure}

We study magnetic configurations of a form similar to the model field used in section~5 of \cite{Reiman-1986}, given by equation (\ref{B-flux}) with
\begin{align} \label{psi-Reiman}
\psi = \frac{1}{2} r^2 \rm ,
\end{align}
\begin{align} \label{chi-Reiman}
\chi = \iota_{\rm ax} \psi + \iota'_{\rm ax} \psi^2 - \sum_{k=1}^{k_{\rm max}}  \varepsilon_k (2\psi)^{k/2} \cos \left( k \theta - \varphi \right) \rm ,
\end{align}
where $k_{\rm max}$ is the largest value of $k$ for which $\varepsilon_k \neq 0$.
Here, $r$ and $\theta$ are chosen to be the poloidal minor radius and the geometric poloidal angle, such that
\begin{align} \label{r-Reiman}
r = \sqrt{ \left( R - 1 \right)^2 + Z^2 } \rm ,
\end{align}
and
\begin{align} \label{theta-Reiman}
\tan \theta = \frac{Z}{ R - 1 } \rm .
\end{align}
The explicit expressions for the magnetic field and its derivatives are derived from equations (\ref{B-flux}) and (\ref{psi-Reiman})-(\ref{theta-Reiman}), and are given in Appendix~\ref{app-Reiman}.
The unperturbed configuration ($\varepsilon_k = 0$ for all $k$) is symmetric in the geometric poloidal and toroidal angles (it is not curl-free and is therefore not a vacuum magnetic configuration).
For the scope of this paper, we focus on the set of parameters $\iota_{\rm ax} = 0.15$, $\iota'_{\rm ax} = 0.38$ and $\varepsilon_k = 0$ for $k\neq 6$.
In figure~\ref{fig-surfaces-Reiman}, we show a Poincar\'e plot of the island chain in the plane $\varphi = 0$ resulting from the parameters $\varepsilon_6 = 0.01$ and $\varepsilon_6 = 0.001$.

\begin{figure} 
\includegraphics[width = 1.0\textwidth]{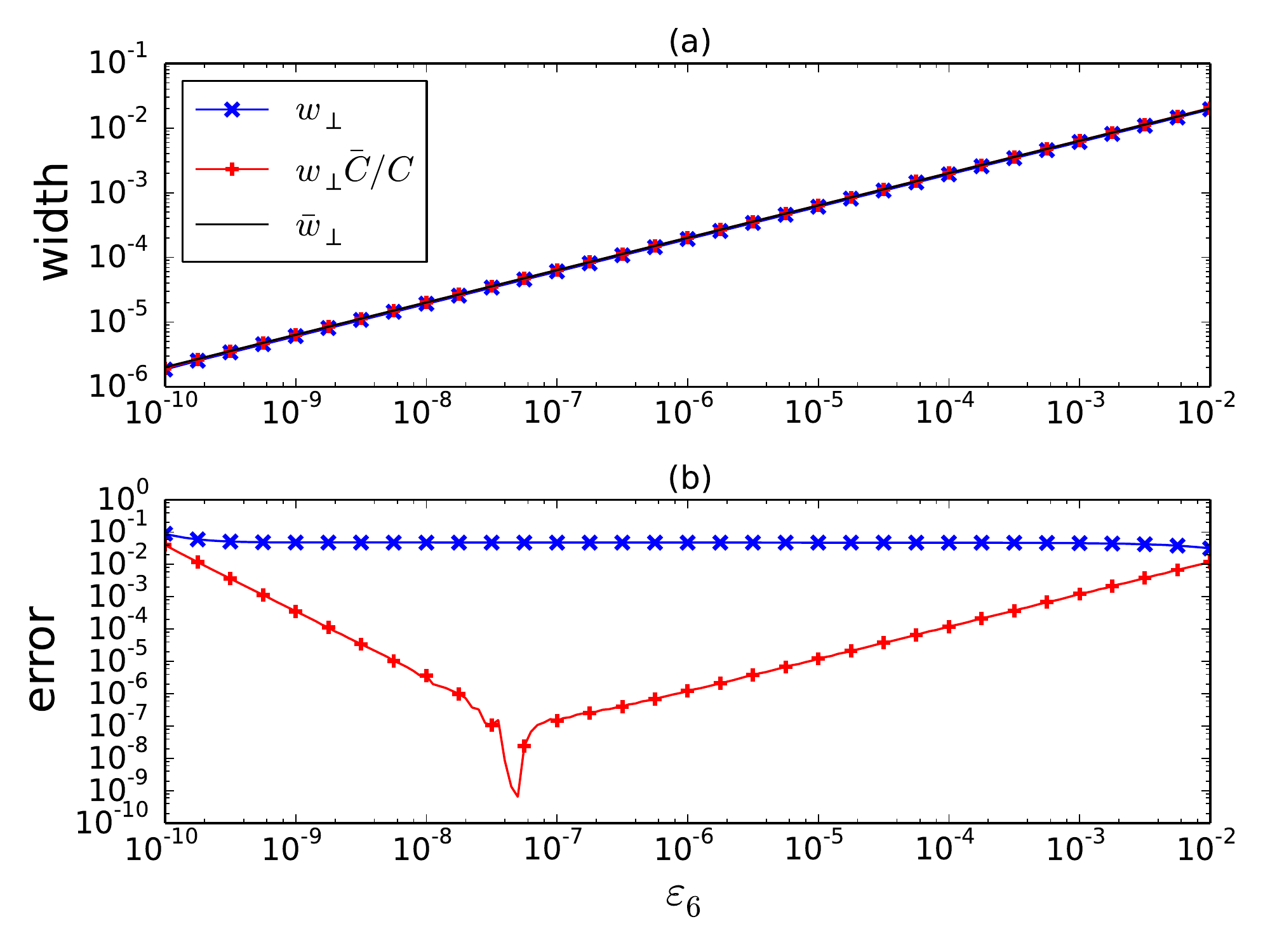}
\caption{Width of magnetic islands at the resonant flux surface with rotational transform $\iota_{\rm res} = 1/6$ calculated for the Reiman model magnetic field with $\iota_{\rm ax} = 0.15$, $\iota'_{\rm ax} = 0.38$ and $\varepsilon_i = 0$ for $i\neq 6$, shown as a function of $\varepsilon_6$. (a) Uncorrected $w_{\perp}$ ($\times$) and corrected $w_{\perp} \bar{C} / C$ (+) computed values of width are compared with the analytical value $\bar{w}_{\perp}$ in equation (\ref{width-Reiman}) (solid line). Here, $w_{\perp}$ is calculated from equation (\ref{width}), $C$ from equation (\ref{circ}) and $\bar{C}$ from equation (\ref{C-Reiman}). (b) Normalized error $| 1 - w_{\perp} / \bar{w}_{\perp}|$ ($\times$) and $| 1 - w_{\perp} \bar{C} / (\bar{w}_{\perp} C)|$ (+). For $\varepsilon_6 > 10^{-7}$, the error in the corrected width decreases linearly with $\varepsilon_6$, as expected from the discussion at the end of section~\ref{sec-width}. For smaller values of $\varepsilon_6$ this error changes sign and increases with $\varepsilon_6^{-2}$, most likely due to roundoff error propagation. One in five markers are shown in both plots. The toroidal angle resolution was $N_{\varphi} = 80$.}
\label{fig-width-Reiman}
\end{figure}

Although the Reiman magnetic field configuration is an experimentally unrealistic one, it is extremely useful and convenient to test the island width calculation presented in section~\ref{sec-width}.
Taking $\varepsilon_k = 0$ for $k \neq 6$ in equation (\ref{chi-Reiman}) results in equation (\ref{chi-perturbed}) with $\chi_0 = \iota_{\rm ax} \psi + \iota'_{\rm ax} \psi^2$ and $\chi_1 =  \varepsilon_6 (2\psi)^{3} \cos \left( k \theta - \varphi + \pi \right) $.
The (unperturbed) minor radius of the island chain is obtained by calculating $r = \bar{r}$ at the resonance, $\iota_{\rm res}= 1/6 = \iota_{\rm ax} + \iota'_{\rm ax} \bar{r}^2 $, giving $\bar{r} = \sqrt{(\iota_{\rm res} - \iota_{\rm ax} )/\iota'_{\rm ax}}$ and $\psi_0 =  (\iota_{\rm res} - \iota_{\rm ax} )/(2\iota'_{\rm ax})$.
Therefore, the circumference of the unperturbed resonant flux surface section in the Reiman model is
\begin{align} \label{C-Reiman}
\bar{C} = 2\pi \sqrt{\frac{ \iota_{\rm res} - \iota_{\rm ax} }{\iota'_{\rm ax}} } \rm .
\end{align}
The width of the island chain expressed in the magnetic coordinate $\psi$, obtained by inserting $\epsilon (\psi_0)  =  \varepsilon_6 \left( (\iota_{\rm res} - \iota_{\rm ax} )/\iota'_{\rm ax} \right)^3$ and $\iota_0'(\psi_0) = 2\iota'_{\rm ax}$ into equation (\ref{width-flux}), is
\begin{align} \label{Upsilon-Reiman}
\Upsilon = 4 \sqrt{  \frac{ \varepsilon_6  }{ 2\iota'_{\rm ax} } }  \left( \frac{\iota_{\rm res} - \iota_{\rm ax} }{\iota'_{\rm ax}} \right)^{3/2} \rm .
\end{align} 
From equation (\ref{wperpbar-def}) and the equality $\partial \xi_{\perp} / \partial \psi = d r / d \psi = 1/\bar{r}$, the expression for the island width in terms of the parameters of the Reiman model is
\begin{align} \label{width-Reiman}
\bar{w}_{\perp} = 4 \sqrt{  \frac{ \varepsilon_6  }{ 2\iota'_{\rm ax} } } \frac{\iota_{\rm res} - \iota_{\rm ax} }{\iota'_{\rm ax}}  \rm .
\end{align} 
Note that the subscript $k$ in $w_{\perp, k}$ is unnecessary here, since the poloidal symmetry of the Reiman model implies that the width of all islands in the chain is identical.

In figure~\ref{fig-width-Reiman} we plot the island width calculated using equation (\ref{width}) and compare it with that calculated using equation (\ref{width-Reiman}).
The discrepancy between the two equations is almost entirely due to the chord approximation for the circumference in equation (\ref{circ}).\footnote{Note that the other approximation (\ref{C-intermediate}) is exact in the Reiman model due to the unperturbed configuration being poloidally symmetric.}
This can be seen by comparing the island width calculated using the two methods with the quantity $\bar{C}  w_{\perp} / C$, a corrected island width where the sum of chords in (\ref{circ}) is replaced with the more accurate measure of circumference in (\ref{C-Reiman}).
The corrected island width has a near-perfect overlap with the width calculated from (\ref{width-Reiman}). The error between the two quantities decreases with $\varepsilon_6$ --- consistent with the discussion in the final paragraph of section \ref{sec-width} (recall that (\ref{width-Reiman}) comes from (\ref{wperpbar-def}))--- and is thus limited by the accuracy of the small-island analysis.
Conversely, the uncorrected island width $w_{\perp}$ is a less accurate approximation, as seen by the saturation of the error with decreasing $\varepsilon_6$.
This is due to the chord approximation of the circumference limiting the accuracy of the island width evaluation, as the $O(L^{-1})$ error is dominant at small enough $\varepsilon_6 \propto \hat{\epsilon}$.

The gradient of the island width with respect to the parameters $\kappa \in \lbrace \iota_{\rm ax}, \iota'_{\rm ax}, \varepsilon_6\rbrace$ is calculated using the method derived in this paper.
When one such parameter is varied infinitesimally, the infinitesimal magnetic field variation can be expressed as $\Delta \vec{B} = \Delta \kappa \partial \vec{B} / \partial \kappa$, where $\partial \vec{B} / \partial \kappa$ is the gradient of the magnetic field with respect to $\kappa$.
The infinitesimal variation of the magnetic field gradient can be expressed as $\Delta \nabla_{\vec X} \vec{B} = \Delta \kappa  \partial (\nabla_{\vec X} \vec{B}) / \partial \kappa$.
Both $\partial \vec{B} / \partial \kappa$ and $ \partial (\nabla_{\vec X} \vec{B}) / \partial \kappa$ are straightforwardly obtained from the equations in appendix \ref{app-Reiman}.
The variation of the circumference and of the tangent map can also be expressed as $\Delta C = \Delta \kappa \partial C / \partial \kappa$ and $\Delta \Sigma_k =  \Delta \kappa \partial \Sigma_k / \partial \kappa$.
Thus, the gradient of the circumference with respect to $\kappa$ is
\begin{align}  \label{gradC}
\frac{\partial C }{\partial \kappa } = - \int_0^{2\pi L / n_0} d\varphi  \frac{\partial \vec{V}}{\partial \kappa} \cdot \boldsymbol{\lambda} \rm,
\end{align}
and the gradient of $\Sigma_k$ is 
\begin{align} \label{gradSigma}
\frac{\partial \Sigma_k}{\partial \kappa} = -  \sum_{Q=0}^{Q_0-1} \int_{2\pi (k + LQ)/n_0}^{2\pi (k+ LQ+L)/n_0} \left( \frac{\partial \vec{V}}{\partial \kappa} \cdot \boldsymbol{\lambda}_Q   + \vec{s}_{\perp k} \cdot \left(    \frac{\partial  \nabla_{\vec X} \vec{V}}{\partial \kappa}\right) \cdot \boldsymbol{\mu} \right)  \rm ,
\end{align}
where
\begin{align}
\frac{\partial \vec{V}}{\partial \kappa} = \frac{R }{B_{\varphi}} \frac{\partial \vec{B}_{\rm p}}{\partial \kappa} - \frac{R \vec{B}_{\rm p} }{B_{\varphi}^2 } \frac{ \partial B_{\varphi}  }{\partial \kappa }  \rm ,
\end{align}
\begin{align}
\nabla_{\vec X} \frac{\partial \vec{V}}{\partial \kappa } = & \frac{\uvec{e}_R }{B_{\varphi}} \frac{\partial \vec{B}}{\partial \kappa }
- \frac{\uvec{e}_R \vec{B}_{\rm p}}{B_{\varphi}^2} \frac{\partial B_{\varphi}}{\partial \kappa }
+ \frac{R \nabla_{\vec X} }{B_{\varphi}} \frac{\partial \vec{B}_{\rm p}}{\partial \kappa }
- \frac{R (\nabla_{\vec X} \vec{B}_{\rm p} ) }{B_{\varphi}^2} \frac{\partial B_{\varphi}}{\partial \kappa }
 - \frac{R (\nabla_{\vec X} B_{\varphi}) }{B_{\varphi}^2} \frac{\partial \vec{B}}{\partial \kappa }  \nonumber \\
&  
 -  \left( \nabla_{\vec X} \frac{\partial B_{\varphi}}{\partial \kappa }  \right) \frac{R  \vec{B}_{\rm p} }{B_{\varphi}^2} 
 +  \frac{2 R (\nabla_{\vec X} B_{\varphi } ) \vec{B}_{\rm p}  }{B_{\varphi}^3} \frac{\partial B_{\varphi} }{\partial \kappa }  \rm .   
\end{align}
Hence, and using equation (\ref{Deltaw}), the gradient of the island width is given by
\begin{align} \label{gradwidth}
\frac{\partial \ln w_k }{\partial \kappa }  = \frac{ \partial \ln C }{\partial \kappa} - \frac{\partial  \ln \Sigma_k }{\partial \kappa }  \text{,}
\end{align}
with $\partial C / \partial \kappa $ and $ \partial \Sigma_k / \partial \kappa $ given by equations (\ref{gradC}) and (\ref{gradSigma}) respectively.
The gradient of the residue of a periodic field line follows from (\ref{DeltaR}),
\begin{align} \label{gradRes}
\frac{\partial \mathcal R}{\partial \kappa}  =-  \int_0^{2\pi L /n_0} \left( \boldsymbol{\lambda}_{\mathcal R} \cdot  \frac{\partial \vec{V}}{\partial \kappa}   +   \left( \frac{\partial  \nabla_{\vec X}  \vec{V} }{\partial \kappa }  \cdot \mu \right) : \mat S_0  \right) d\varphi \rm .
\end{align}
Considering the residue at the magnetic axis, the gradient of the on-axis rotational transform follows from (\ref{Deltaiota}),
\begin{align} \label{gradiota}
\frac{\partial \bar{\iota}}{\partial \kappa} = \frac{n_0}{\pi \sin \left( 2\pi \bar{\iota} / n_0 \right)} \frac{\partial  \mathcal{R}}{\partial \kappa}  \rm .
\end{align}
where $\bar{\iota}$ is obtained from (\ref{iotabar}) and is found to be equal to $\iota_{\rm ax}$, as expected.

\begin{figure} 
\includegraphics[width = 1.0\textwidth]{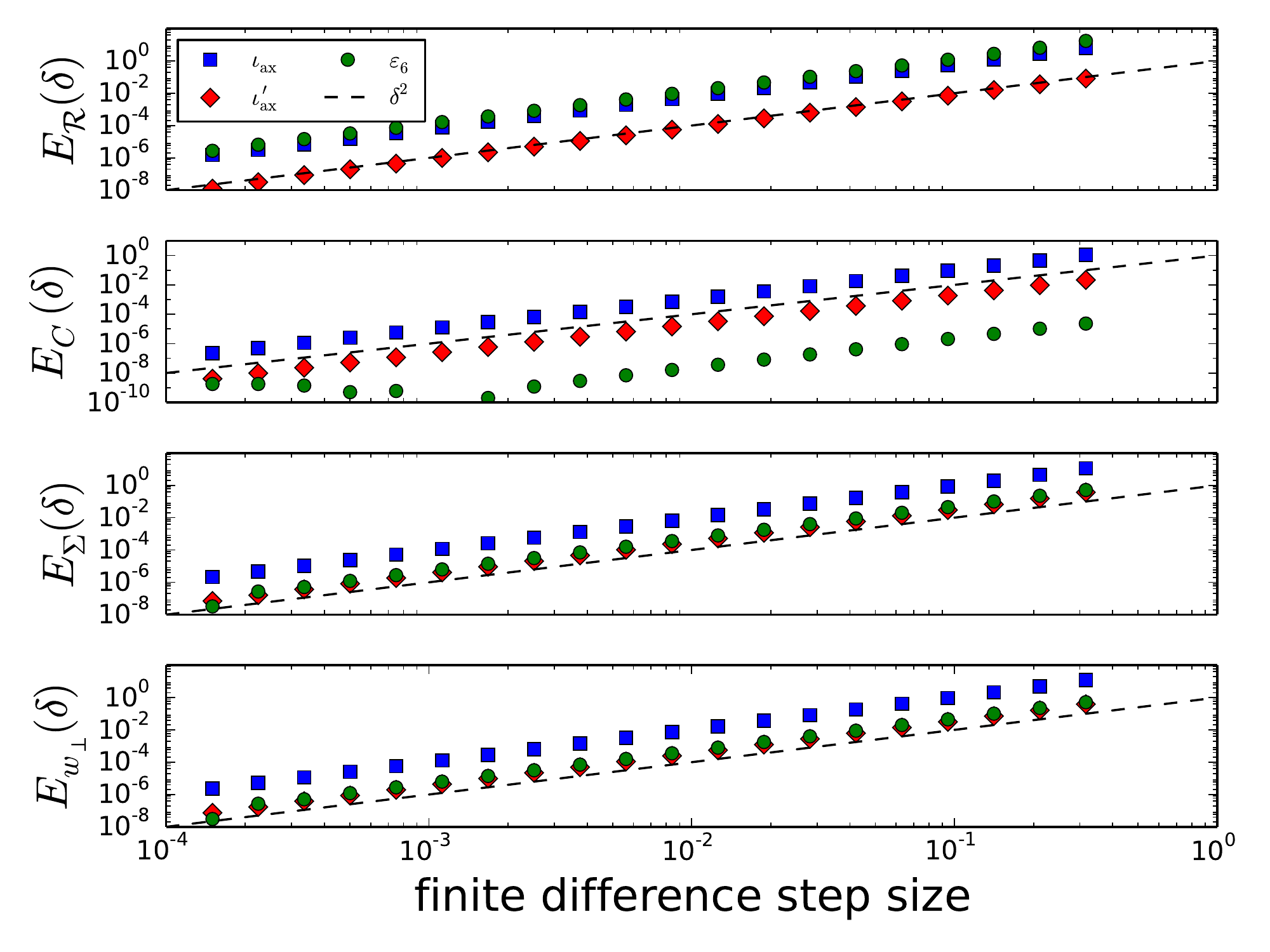}
\caption{Errors, relative to a centred difference approximation, in the gradients of residue $ \mathcal{R}$, circumference $C$, $\Sigma$ and island width $w_{\perp}$ calculated with respect to the on-axis rotational transform $\iota_{\rm ax}$, its first derivative $\iota_{\rm ax}'$ and the amplitude of the resonant perturbation $ \varepsilon_6 $ in the Reiman model. The errors $E(\delta)$, defined in equation (\ref{Error}), are shown as a function of a normalized finite difference step size $\mathcal{R}^{-1} \left( \partial \mathcal{R} / \partial \kappa \right) \delta $. The configuration parameters are $\iota_{\rm ax} = 0.15$, $\iota'_{\rm ax} = 0.38$, $\varepsilon_6 = 0.01$ and $\varepsilon_i = 0$ for $i\neq 6$. The resonant flux surface has rotational transform $\iota_{\rm res} = 1/6$. The dashed line is $E(\delta ) = \delta^2$. The toroidal angle resolution was $N_{\varphi} = 80$.}
\label{fig-gradwidtherror-Reiman}
\end{figure}

The results of equations (\ref{gradC}), (\ref{gradSigma}), (\ref{gradwidth}) and (\ref{gradRes}) can be compared with a numerical derivative of the quantities $C$, $\Sigma_k$, $w_{\perp, k}$ and $\mathcal{R}$ calculated using finite differences.
To this end, we calculate centred difference (CD) derivatives of the form
\begin{align} \label{gradf-centred}
\frac{\partial^{\rm CD} \ln f}{\partial \kappa} (\kappa, \delta )  = \frac{ \ln \left[f \left( \kappa  + \delta \right) \right] - \ln \left[ f \left( \kappa - \delta  \right) \right] }{ 2\delta } \rm .
\end{align}
The error with respect to the adjoint calculation of the derivative in equation (\ref{gradf-centred}) is
\begin{align} \label{Error}
E_{f} (\delta )  = \left|   \frac{\partial^{\rm CD} \ln f}{\partial \kappa} (\kappa, \delta )  -  \frac{ \partial \ln f }{ \partial \kappa } (\kappa ) \right|   
\end{align} 
In equations (\ref{gradf-centred})-(\ref{Error}), $f \in \lbrace \mathcal{R}, C, \Sigma_k, w_{\perp, k} \rbrace$.
We expect the centred difference derivative of the numerically calculated island width to approach the numerically calculated derivative as $\delta$ is decreased.
Having ignored the pieces of the derivative of $\Sigma_k$ that are small in $\hat{\epsilon}$ in equation (\ref{Deltaw-includingDeltaevecs}), one might expect that $E(\delta) \sim \delta^2$ for $\delta$ larger than a threshold value and $E(\delta ) \sim \hat{\epsilon}$ for smaller $\delta$.
However, comparing the Reiman model to the derivation of the island width, the quantity $\zeta_0 ( \psi )$ is independent of $\psi$ in the Reiman model. 
Thus,  the diagonal elements of the tangent map in (\ref{EOM-flux}), and correspondingly the off-diagonal elements of the scalar invariant matrix in (\ref{deltaK}), are exactly zero.
Hence, the eigenvectors $\uvec{e}_{\perp}$ and $\uvec{e}_{\parallel}$ are exactly aligned with $\nabla_{\vec X} \psi$ and $\nabla_{\vec X} \theta$ respectively, for any value of the parameters $\iota_{\rm ax}$, $\iota'_{\rm ax}$ and $\varepsilon_6$.
Thus, the terms small in $\hat{\epsilon}$ that were neglected in the derivation of the gradient of the island width are exactly zero in the Reiman model, and $E(\delta )  \sim \delta^2$ should hold for all values of $\delta$. 
In figure \ref{fig-gradwidtherror-Reiman} the quantity $E_{f} (\delta)$ for $f\in \left\lbrace \mathcal{R}, C, \Sigma, w_{\perp} \right\rbrace $ is shown for derivatives with respect to $\kappa \in (\iota_{\rm ax}, \iota'_{\rm ax}, \varepsilon_6 )$ 
for $\iota_{\rm ax} = 0.15$, $\iota'_{\rm ax} = 0.38$ and $\varepsilon_6 = 0.01$.
The proportionality $E_{f} (\delta) \propto  \delta^2$ for all $\kappa$ is strong evidence that the gradient calculation is accurate for all quantities. 
In addition, the gradient of the on-axis rotational transform $\bar{\iota}$ calculated from equation (\ref{gradiota}) is, as expected, unity for $\kappa = \iota_{\rm ax}$ and zero for $\kappa \in \left\lbrace \iota_{\rm ax}', \varepsilon_6 \right\rbrace$.

\subsection{Shape gradient calculation: explicit coils} \label{subsec-num-NCSX}

In this section, it will be useful to denote as $\vec R = (X, Y, Z)$ the position along a magnetic field line expressed in a set of right-handed Cartesian coordinates. 
The coordinate $Z$ is the same as in the two-dimensional vector $\vec{X} = (R, Z)$, while $X = R \cos \varphi$ and $Y = R \sin \varphi$.

The magnetic field produced by explicit coils is calculated using the Biot-Savart law.
The number of inputs required for an explicit coil calculation is just the number of field periods and a set of (sufficiently resolved) positions along the coils.
Since the coils producing the magnetic field are continuous (even though they are numerically approximated as discrete), the magnetic field is a functional of the continuous periodic function $\vec{r}_c (l_c ) = (x_c, y_c, z_c )$ specifying the coil shape.
Lowercase letters and a $c$ subscript distinguish coil positions from positions along a magnetic field line.
Here $l_c$ is the arc length along the coil measured from some reference point, $l_c \in [0, L_c)$, where $L_c$ is the total coil length.
The magnetic field is specified by the Biot-Savart law
\begin{align} \label{B-coils}
\vec{B}_{\rm coil} = - \sum_{c=1}^N \frac{\mu_0 I_c }{4\pi }  \int  dl_c  \frac{ \vec{R} - \vec{r}_c }{|\vec{R} - \vec{r}_c |^3}  \times \frac{d \vec{r}_c}{dl_c}  \text{.}
\end{align}
The poloidal gradient of the magnetic field is
\begin{align}
\nabla_{\vec X} \vec{B}_{\rm coil} = \sum_{c=1}^N \frac{\mu_0 I_c }{4\pi } \int dl_c \frac{1}{|\vec{R} - \vec{r}_c |^3 } \left[ -  \mat{I}_p    + 3  \frac{ (\vec{X} - \vec{x}_c) (\vec{R} - \vec{r}_c) }{ |\vec{R} - \vec{r}_c |^2 }  \right]  \times \frac{d \vec{r}_c}{dl_c}    \text{.}
\end{align}
where we introduced the poloidal identity matrix $\mat{I}_p = \uvec{e}_R \uvec{e}_R + \uvec{e}_Z \uvec{e}_Z $, with $\uvec{e}_Z = \nabla_{\vec X} Z$, and the position vector of the coil projected in the poloidal plane, $\vec{x}_c = \vec{r}_c \cdot \left( \uvec{e}_R \uvec{e}_R + \uvec{e}_Z \uvec{e}_Z \right)$.
The second derivative of the magnetic field is
\begin{align}
\nabla_{\vec X} \nabla_{\vec X} \vec{B}_{\rm coil} =  \sum_{c=1}^N \frac{\mu_0 I_c }{4\pi } \int dl_c \frac{1}{|\vec{R} - \vec{r}_c |^5 } \left[ 3 \left(  \mat{I}_p - 5  \frac{  (\vec{X} - \vec{x}_c) (\vec{X} - \vec{x}_c) }{ |\vec{R} - \vec{r}_c |^2 } \right)   \left( \vec{R} - \vec{r}_c \right)    \right. \nonumber  \\ \left. 
 + 3  \left( \vec{X} - \vec{x}_c  \right) \mat{I}_p + 3 \left(  \uvec{e}_R  \left( \vec{X} - \vec{x}_c  \right)\uvec{e}_R  +  \uvec{e}_Z  \left( \vec{X} - \vec{x}_c  \right)  \uvec{e}_Z   \right) 
   \right]  \times \frac{d \vec{r}_c}{dl_c}
    \text{.}
\end{align}

The variation $\Delta \vec{B}$ of a coil-produced magnetic field is conveniently expressed in terms of the shape gradient $\boldsymbol{\mathcal G}_{\vec{r}_c} \vec{B} $ of the magnetic field with respect to coils \citep{Landreman-Paul-2018}.
The shape gradient $\boldsymbol{\mathcal G}_{\vec{r}_c}$ is a vector operator which, like $\nabla_{\vec X}$, is meaningful only when acting on a (scalar or tensor) quantity to its right.
It is defined via
\begin{align} \label{shapegradient-def}
\Delta f = \sum_{c=1}^N \int dl_c \left(   \Delta \vec{r}_{c} \times \frac{d \vec{r}_c}{dl_c} \right) \cdot \boldsymbol{\mathcal G}_{\vec{r}_c} f \text{.}
\end{align}
The shape gradient of the magnetic field can be extracted from the Biot-Savart law and (\ref{shapegradient-def}) as shown in appendix \ref{app-shape-mag}, giving
\begin{align} \label{B-shape}
\boldsymbol{\mathcal G}_{\vec{r}_c} \vec{B}_{\rm coil}  = \frac{\mu_0 I_c }{4\pi |\vec{R} - \vec{r}_c |^3} \left[ \mat{I}  -  \frac{ 3  (\vec{R} - \vec{r}_c) \left(  \vec{R} - \vec{r}_c \right) }{ |\vec{R} - \vec{r}_c |^2 }   \right] \rm ,
\end{align}
where $\mat{I} = \uvec{e}_X \uvec{e}_X +  \uvec{e}_Y \uvec{e}_Y +  \uvec{e}_{Z} \uvec{e}_{Z} $.

\begin{figure}
\centering
\includegraphics[width=1.0\textwidth]{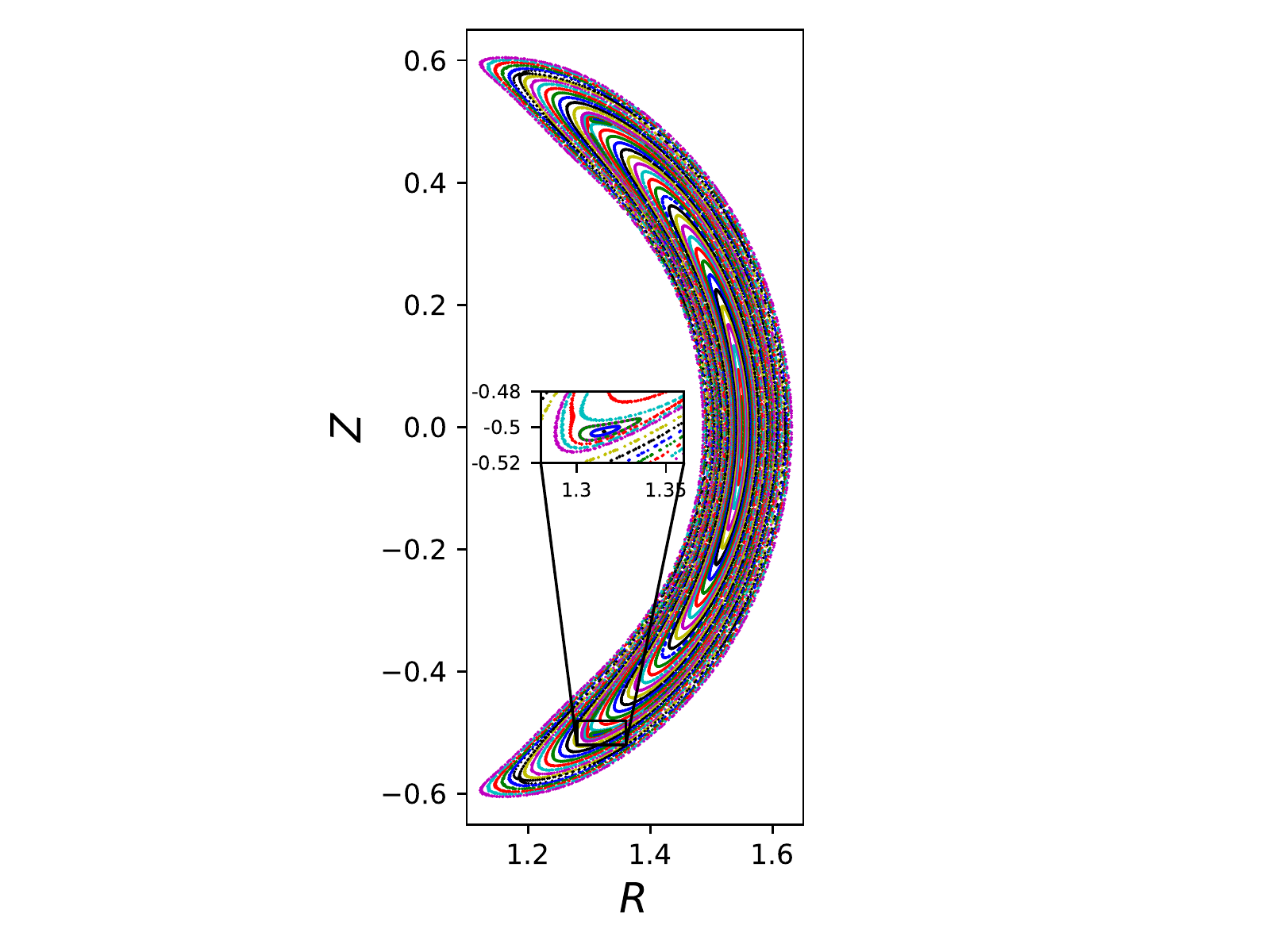}
\caption{Poincar\'e plot for the configuration produced by NCSX coils. The island for which the shape gradient of the width is calculated in figure~\ref{fig-NCSXgradient} is shown enlarged in the inset; its width using (\ref{width}) is $w_{\perp} = 0.0106$ ($\mathcal R = 0.0149$).} 
\label{fig-NCSXconfig}
\end{figure}
\begin{figure}
\includegraphics[width=\textwidth]{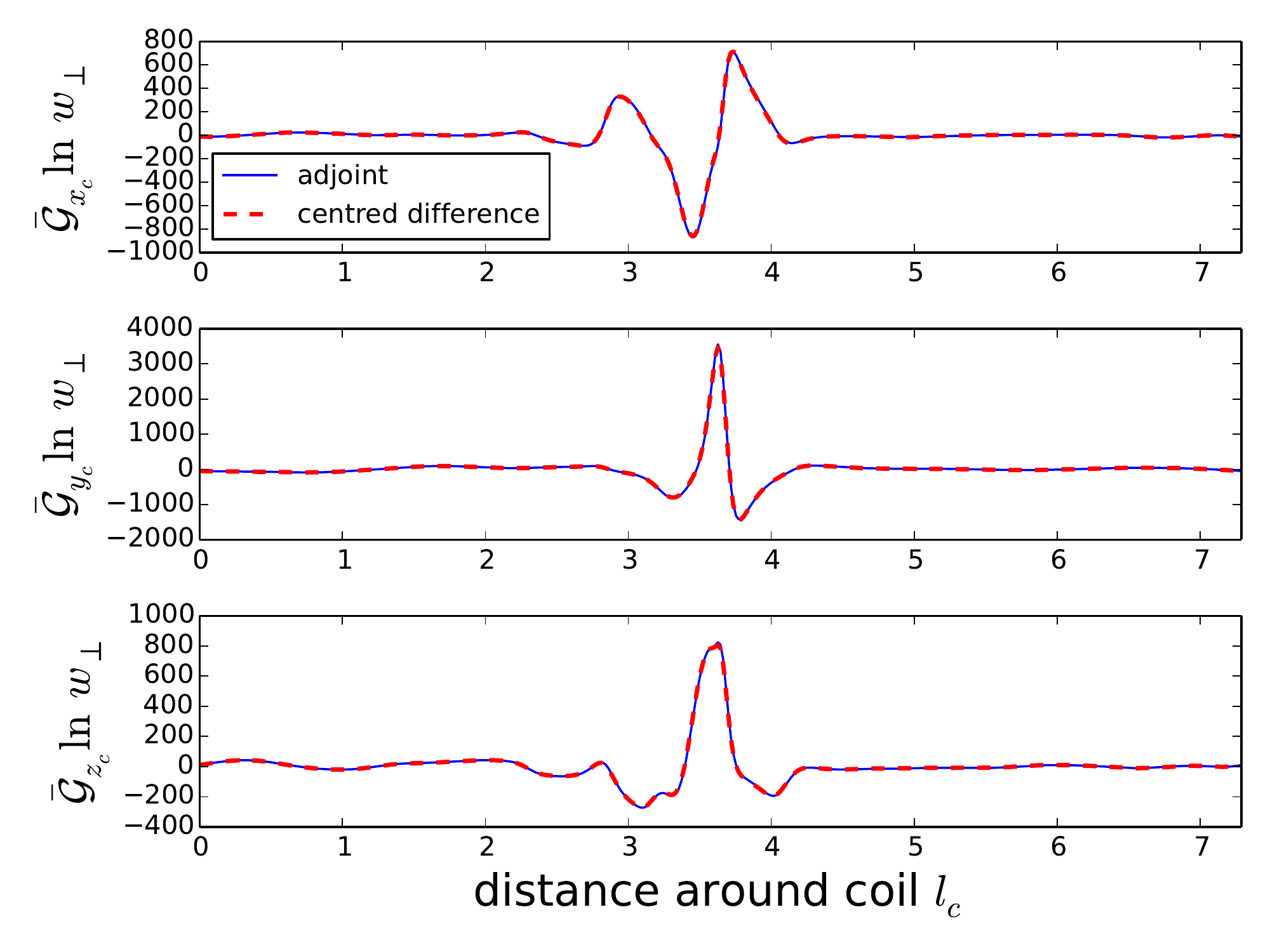}
\caption{Shape gradient of the width of one of the magnetic islands in the NCSX configuration (figure \ref{fig-NCSXconfig}) with respect to the positions along a type A modular coil (length $L_c = 7.29$), calculated using $N_{\varphi} = 30$ with: the adjoint method (solid lines); the centred difference scheme with $\delta \kappa_c = 10^{-4}$ (dashed lines). For each component, the mean residual between the two calculations is about $2\%$ of the mean absolute value. The adjoint calculation is over a hundred times faster.}
\label{fig-NCSXgradient}
\end{figure}

The shape gradient of the island width with respect to the coils producing the magnetic field is
\begin{align}
\boldsymbol{\mathcal G}_{\vec{r}_c} \ln w_k  = \boldsymbol{\mathcal G}_{\vec{r}_c} \ln C -  \boldsymbol{\mathcal G}_{\vec{r}_c} \ln \Sigma_k  \text{.}
\end{align}
Here $\boldsymbol{\mathcal{G}}_{\vec{r}_c} C$ is given by 
\begin{align} \label{C-shape}
\boldsymbol{\mathcal G}_{\vec{r}_c} C = - \int_0^{2\pi L / n_0} d\varphi \boldsymbol{ \mathcal G}_{\vec{r}_c} \vec{V} \cdot \boldsymbol{\lambda} \rm ,
\end{align}
with 
\begin{align}
\boldsymbol{ \mathcal G}_{\vec{r}_c} \vec{V} =  \frac{R \boldsymbol{\mathcal{G}}_{\vec{r}_c} \vec{B}_{\rm p} }{B_{\varphi}}  - \frac{R ( \boldsymbol{\mathcal{G}}_{\vec{r}_c} B_{\varphi} )  \vec{B}_{\rm p} }{B_{\varphi}^2 }   \text{,}
\end{align}
and $ \boldsymbol{\mathcal G}_{\vec{r}_c} \Sigma_k $ is given by 
\begin{align} \label{Sigma-shape}
\boldsymbol{\mathcal{G}}_{\vec{r}_c} \Sigma_k =  -  \sum_{Q=0}^{Q_0-1} \int_{2\pi LQ/n_0}^{2\pi L(Q+1)/n_0} d\varphi \left( \left( \boldsymbol{\mathcal{G}}_{\vec{r}_c}  \vec{V} \right) \cdot \boldsymbol{\lambda}_Q   +  \vec{s}_{\perp k} \cdot  \left( \nabla_{\vec X}   \boldsymbol{\mathcal{G}}_{\vec{r}_c} \vec{V} \right) \cdot \boldsymbol{\mu} \right)  \rm ,
\end{align}
with
\begin{align}
\nabla_{\vec X} \boldsymbol{\mathcal G}_{\vec{r}_c} \vec{V} = & \frac{\uvec{e}_R \boldsymbol{\mathcal G}_{\vec{r}_c} \vec{B}}{B_{\varphi}} 
- \frac{\uvec{e}_R \vec{B}}{B_{\varphi}^2} \boldsymbol{\mathcal G}_{\vec{r}_c} B_{\varphi} 
+ \frac{R \nabla_{\vec X} \boldsymbol{\mathcal G}_{\vec{r}_c} \vec{B} }{B_{\varphi}} 
- \frac{R (\nabla_{\vec X} \vec{B} ) \boldsymbol{\mathcal G}_{\vec{r}_c} B_{\varphi}}{B_{\varphi}^2}
 - \frac{R (\nabla_{\vec X} B_{\varphi}) \boldsymbol{\mathcal G}_{\vec{r}_c} \vec{B} }{B_{\varphi}^2}  \nonumber \\
&  
 - \frac{R (\nabla_{\vec X} \boldsymbol{\mathcal G}_{\vec{r}_c} B_{\varphi } ) \vec{B} }{B_{\varphi}^2} 
 +  \frac{2 R (\nabla_{\vec X} B_{\varphi } ) \vec{B} \boldsymbol{\mathcal G}_{\vec{r}_c} B_{\varphi} }{B_{\varphi}^3}  \rm .
\end{align}
The spatial derivative (in the poloidal plane) of the shape gradient of the magnetic field is
\begin{align} \label{gradB-shape}
\nabla_{\vec X} \boldsymbol{\mathcal G}_{\vec{r}_c} \vec{B} & = \frac{\mu_0 I_c }{4\pi |\vec{R} - \vec{r}_c |^5} \left[ 3\left( \vec{X} - \vec{x}_c\right) \left( \mat{I}  - 5  \frac{ \left( \vec{R} - \vec{r}_c \right) \left( \vec{R} - \vec{r}_c \right) }{ |\vec{R} - \vec{r}_c |^2 } \right) \right. \nonumber \\ & \left.  +  3\uvec{e}_R  \left( \vec{R} - \vec{r}_c \right) \uvec{e}_R + 3\uvec{e}_R \uvec{e}_R  \left( \vec{R} - \vec{r}_c \right)  
 + 3\uvec{e}_Z   \left( \vec{R} - \vec{r}_c \right) \uvec{e}_Z  + 3\uvec{e}_Z \uvec{e}_Z  \left( \vec{R} - \vec{r}_c \right)    \right] \rm .
\end{align}

Using these equations, a numerical approximation to the shape gradient of the width of an island in an NCSX vacuum configuration, shown in figure \ref{fig-NCSXconfig}, was calculated for a type A modular coil.
The magnetic field from the toroidal field coils and from most of the poloidal field coils is included.
The resonant rotational transform of the island chain is $\iota_{\rm res} = 1/4$.
The shape gradient calculation is compared in figure~\ref{fig-NCSXgradient} with a finite difference approximation calculated by perturbing the discrete coil positions at certain locations.
In figures \ref{fig-NCSXconfig} and \ref{fig-NCSXgradient}, all plotted quantities are in SI units.
For simplicity, in figure~\ref{fig-NCSXgradient} we have plotted an alternative definition of the shape gradient $ \bar{ \boldsymbol{ \mathcal G}_{c}} f = \left( d\vec{r}_c / dl_c \right) \times \boldsymbol{ \mathcal G}_{\vec{r}_c} f$, which satisfies 
\begin{align} \label{shapegradient-altdef}
\Delta f = \sum_{c=1}^N \int dl_c    \Delta \vec{r}_{c} \cdot \bar{\boldsymbol{\mathcal G}}_{\vec{r}_c} f \text{}
\end{align}
and thus has a more intuitive geometric interpretation. 
In the adjoint calculation, the shape gradient is calculated by evaluating $ \boldsymbol{ \mathcal G}_{\vec{r}_c} f$ using the adjoint expressions outlined in this section and a numerical approximation of $d\vec{r}_c /d l_c  $,
\begin{align} \label{shapegradient-adjoint-num}
\bar{\boldsymbol{\mathcal{G}}}_{\vec{r}_{\rm c}} f =  \frac{ \vec{r}_c ( l_c + \delta l_c ) -  \vec{r}_c ( l_c - \delta l_c ) }{ 2 \delta l_c } \times \boldsymbol{\mathcal G}_{\vec{r}_c} f \text{,}
\end{align}
where $\delta l_c$ is the separation between equally spaced positions on the coil. 
Using a centred difference approximation instead, the components of the shape gradient, $\bar{\mathcal G}^{\rm CD}_{\kappa_c} f$ for $\kappa_c \in \{ x_c, y_c, z_c \}$, are calculated from 
\begin{align} \label{shapegradient-finitediff}
 \bar{\mathcal G}^{\rm CD}_{\kappa_c} f    = \frac{  f(\kappa_c + \delta \kappa_c ) - f(\kappa - \delta \kappa_c ) }{2 \delta \kappa_c \delta l_c } \text{.}
\end{align}
For the coil considered in this study, the centred difference shape gradient taken with $\delta \kappa_c = 10^{-4}$ has a good overlap with the shape gradient calculated using the adjoint method, as shown in figure~\ref{fig-NCSXgradient}.
The direct adjoint calculation is, however, over a hundred times faster.
This is because the adjoint approach requires solving a few equations along the same periodic magnetic field line, so that the magnetic field and its gradients do not need to be re-evaluated when computing the shape gradient with respect to all coil positions.
Conversely, the finite difference calculation requires searching for a new periodic field line after \emph{each} finite difference step, which is expensive because it requires re-computing the magnetic field at new locations for each coil perturbation.

\subsection{Optimization of helical coils via gradient of residue} \label{subsec-num-heli}

We proceed to consider a model magnetic field studied by \cite{Hanson-Cary-1984} which consists of two pieces: a toroidal magnetic field generated by a long current-carrying wire passing through the centre of the torus, and  a magnetic field generated by a pair of helical coils with opposite current ($\pm I_{\rm heli}$). 
The helical coil positions are specified by the poloidal angle $\eta_\pm$ as a function of the toroidal angle $\varphi$,
\begin{align}
 \eta_\pm = \frac{ n_0 \varphi }{l_0}  + \sum_{k=0}^{k_{\rm max}} \left[ A_{\pm, k} \cos \left( \frac{ kn_0 \varphi}{l_0}  \right) + B_{\pm, k} \sin \left( \frac{ kn_0 \varphi}{l_0}  \right) \right] \rm .
\end{align}
The position of the coils $\vec{r}_\pm = (x_\pm , y_\pm, z_\pm )$ is then obtained using the equations
\begin{align}
x_{\pm} = \left( R_0 + r_0 \cos \eta_\pm \right) \cos \varphi \rm ,
\end{align}
\begin{align}
y_\pm = \left( R_0 + r_0 \cos \eta_\pm \right) \sin \varphi \rm ,
\end{align}
\begin{align}
z_\pm = - r_0 \sin \eta_\pm \rm .
\end{align}
Here we have introduced the major and minor radius of the toroidal surface in which the helical coils lie, $R_0$ and $r_0$ respectively.
The magnetic field configuration is obtained by adding the toroidal magnetic field $\uvec{e}_{\varphi} R_0 B_t / R $ to the magnetic field obtained by applying the Biot-Savart law to the helical coils, 
\begin{align}\label{B-heli}
\vec{B}_{\rm heli} (\vec{R}) = \frac{R_0 B_t }{ R } \uvec{e}_{\varphi}  +  \sum_{\pm} \frac{\pm \mu_0 I_{\rm heli} }{4\pi } \int d\varphi \frac{1}{ |\vec{R} - \vec{r}_{\pm} |^3} \frac{d\vec{r}_{\pm}}{d\varphi} \times (\vec{R} - \vec{r}_{\pm})  \rm .
\end{align}

Upon fixing $R_0 $, $r_0 $, $I_{\rm heli} $ and $B_t$, as done in \cite{Hanson-Cary-1984}, the continuous parameters that can be used to perturb the magnetic field configuration are $A_k$ for $0 \leqslant k \leqslant k_{\rm max}$ and $B_k$ for $1\leqslant k \leqslant k_{\rm max}$.
Perturbing any one parameter of any one coil (carrying a current $\pm I_{\rm heli}$), denoted $\kappa_{\pm}$, such that the variation in the coil position is $\vec{r}_{\pm} (\varphi ) \rightarrow \vec{r}_{\pm} (\varphi ) + \Delta \kappa_{\pm} \partial \vec{r}_{\pm} (\varphi )  / \partial \kappa_{\pm}$ causes the magnetic field to change such that $\vec{B}(\vec{R})  \rightarrow \vec{B}(\vec{R}) + \Delta \kappa_{\pm} \partial \vec{B}(\vec{R})   / \partial \kappa_{\pm} $.
From the expressions (\ref{shapegradient-def}) and (\ref{B-shape}), we get
\begin{align} \label{DeltaB}
\frac{\partial \vec{B}}{\partial \kappa_{\pm}} = \pm \frac{\mu_0 I_{\rm heli} }{4\pi} \int d\varphi  \frac{1}{|\vec{R} - \vec{r}_{\pm} |^3}  \left[ - \mat{I} + \frac{ 3  (\vec{R} - \vec{r}_{\pm}) \left(  \vec{R} - \vec{r}_{\pm} \right) }{ |\vec{R} - \vec{r}_{\pm} |^2 } \right] \cdot \left( \frac{\partial \vec{r}_{\pm} }{\partial \kappa } \times \frac{ d \vec{r}_{\pm} }{ d \varphi }  \right) \text{.}
\end{align}
The gradient of the magnetic field changes according to $\nabla_{\vec X} \vec{B}(\vec{R})  \rightarrow \nabla_{\vec X} \vec{B}(\vec{R}) + \Delta \kappa_{\pm} \partial \nabla_{\vec X} \vec{B}(\vec{R})   / \partial \kappa_{\pm} $; from (\ref{shapegradient-def}) and (\ref{gradB-shape}) we deduce 
\begin{align} \label{DeltanablaB}
\frac{\partial \nabla_{\vec X} \vec{B}}{\partial \kappa_{\pm}} = \pm \frac{\mu_0 I_{\rm heli} }{4\pi}  \int d\varphi  \frac{1}{|\vec{R} - \vec{r}_{\pm} |^5}  \left[ 3 (\vec{R} - \vec{r}_{\pm} ) \mat{I} - \frac{ 5 (\vec{R} - \vec{r}_{\pm} )  (\vec{R} - \vec{r}_{\pm}) \left(  \vec{R} - \vec{r}_{\pm} \right) }{ |\vec{R} - \vec{r}_{\pm} |^2 } \right. \nonumber \\ \left. +   \mat{I} \left(  \vec{R} - \vec{r}_{\pm} \right)  +   \uvec{e}_R  \left(  \vec{R} - \vec{r}_{\pm} \right) \uvec{e}_R  +  \uvec{e}_Z  \left(  \vec{R} - \vec{r}_{\pm} \right) \uvec{e}_Z \right] \cdot \left( \frac{\partial \vec{r}_{\pm} }{\partial \kappa } \times \frac{ d \vec{r}_{\pm} }{ d \varphi }  \right) \text{.}
\end{align}

A gradient-based optimization scheme to demonstrate an application of the adjoint gradient formulation was performed on the system with $R_0 = 1$, $r_0 = 0.3$, $\mu_0 I_{\rm heli} / 4\pi = 0.0307 $ and $B_t = 1$. 
This was previously optimized in \cite{Cary-Hanson-1986} using a derivative-free algorithm. 
The fixed parameters are $A_{+,0} = \pi$, $A_{-,0} = A_{-,1} = B_{+,1} = 0$, and $A_{\pm,k} = B_{\pm,k}  = 0$ for $k \geqslant 2$.
The vector $\vec{p} = ( A_{+,1}, B_{-,1} )$ is composed of the only parameters which are optimized.
Two fixed points are initially located and their residues are used in the optimization. 
A priori, an infinite choice of appropriate objective functions exist: the sensible properties are that they be functions of the residues of all the periodic field lines and that they be zero when all the residues are zero.
The island width can only be used in the objective function if it is known that the periodic field line is an O point instead of an X point.
However, periodic field lines may switch from O to X points during the optimization as $\vec{p}$ is varied.
Moreover, the measure of island width used in this work is only accurate if the island size is small.
For these reasons, using the residues in the optimization is preferable.
We choose a linear combination $P$ of the pairs of residues, 
\begin{align}
P = \frac{1}{2} \left( \mathcal{R}_1 + \mathcal{R}_2 \right) \rm .
\end{align}
One could in principle take the difference of the two residues or the mean square of the residues as the objective function:
the former converges to a non-optimal solution with $\mathcal{R}_1 =  \mathcal{R}_2$ where the objective function is zero yet the residues have not been reduced enough;
the latter has a much slower convergence.
The objective function was minimized using gradient descent with a line search, with more details discussed in appendix \ref{app-op}.

\begin{figure}
\includegraphics[width=\textwidth]{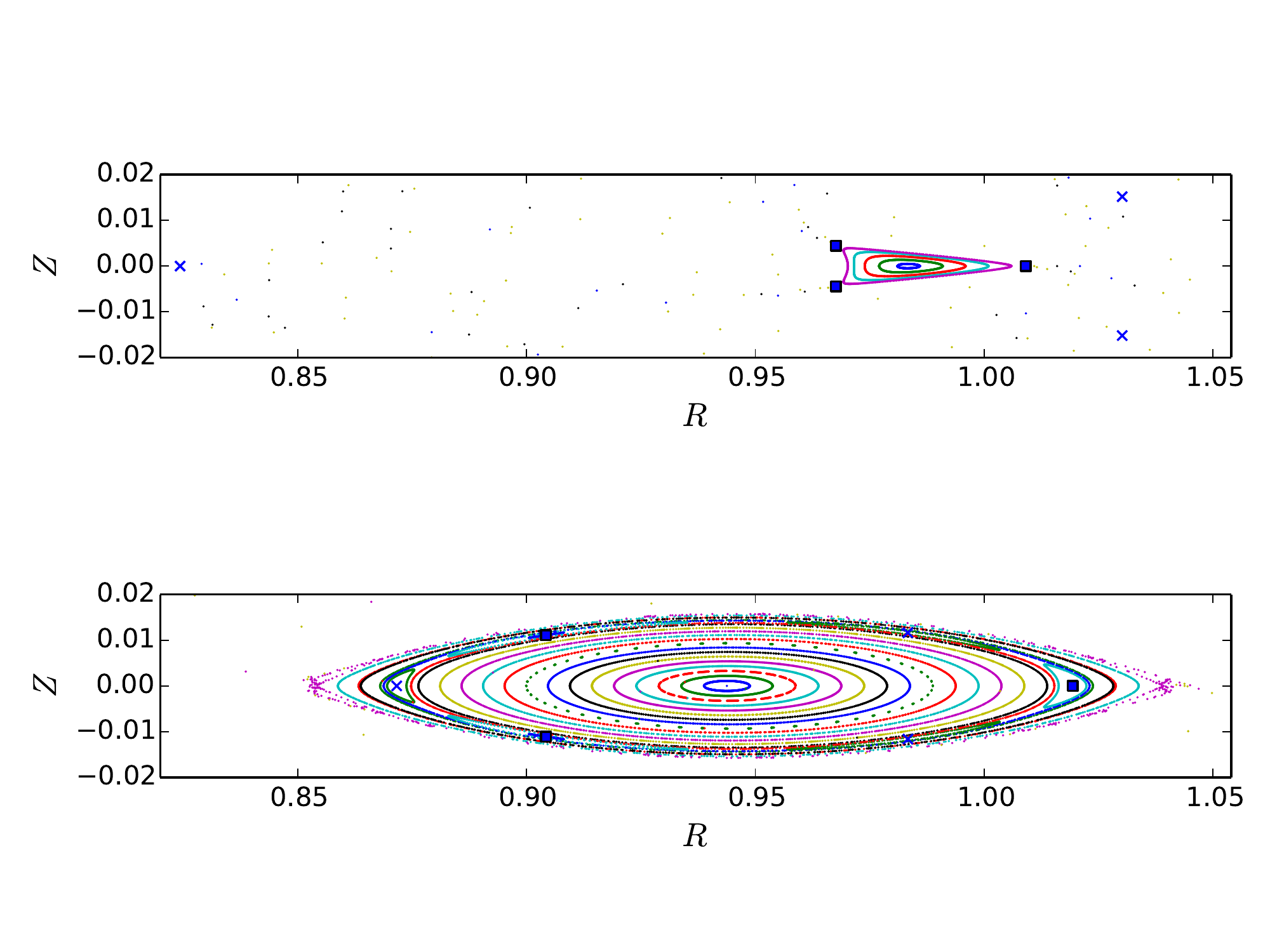}
\caption{Poincar\'e plots of unoptimized (top) and optimized (bottom) helical coil configuration. The squares indicate fixed points with residue $\mathcal{R}_1 $, while the crosses are those with residue $\mathcal{R}_2 $ }
\label{fig-heli}
\end{figure}

\begin{figure}
\includegraphics[width=\textwidth]{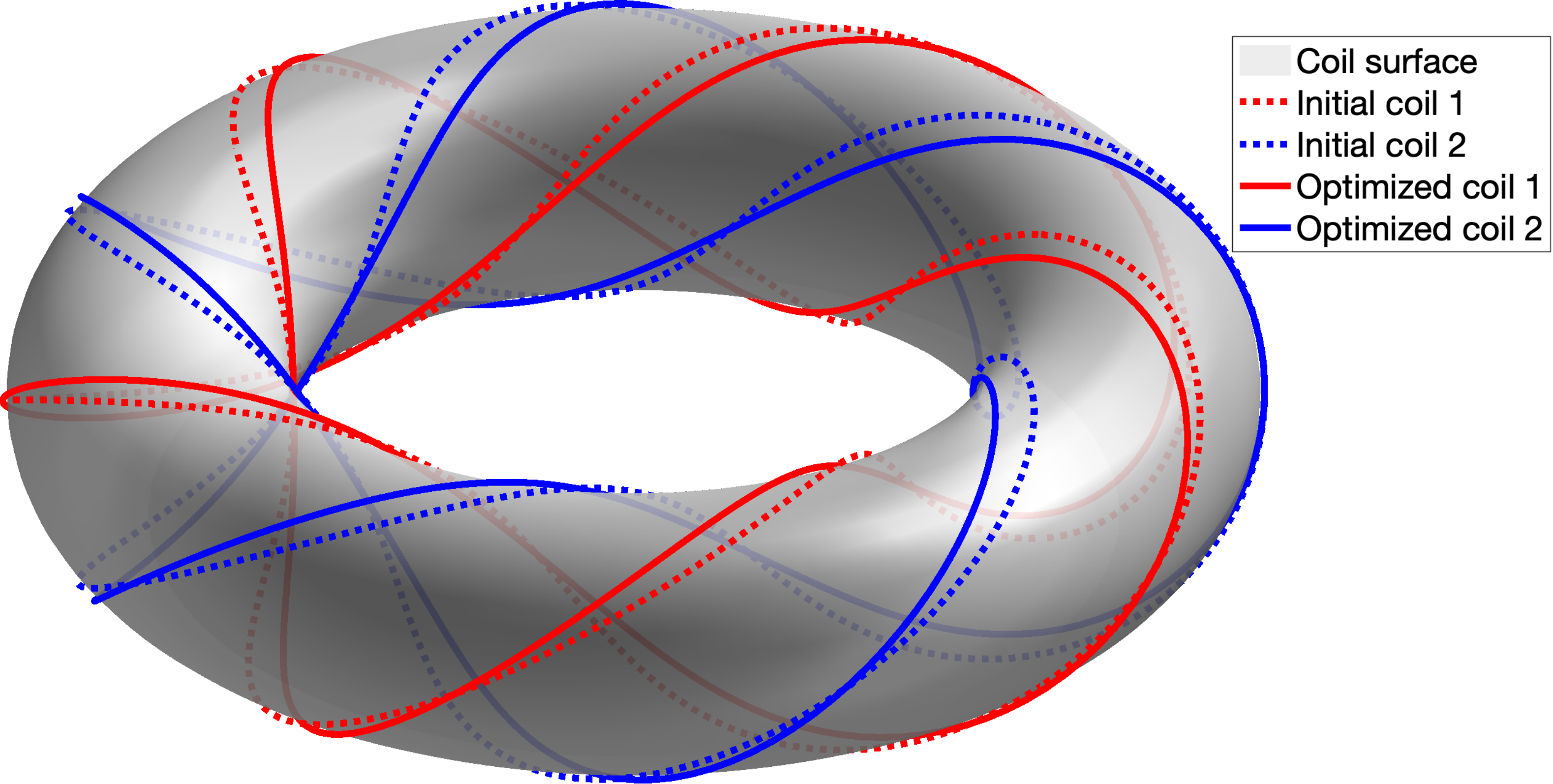}
\caption{Helical coils corresponding to the initial unoptimized configuration (dotted line) and to the optimized configuration (solid line). Coil $1$ (red) carries the negative current ($-I_{\rm heli}$).}
\label{fig-opcoils}
\end{figure}

\begin{figure}
\includegraphics[width=1.0\textwidth]{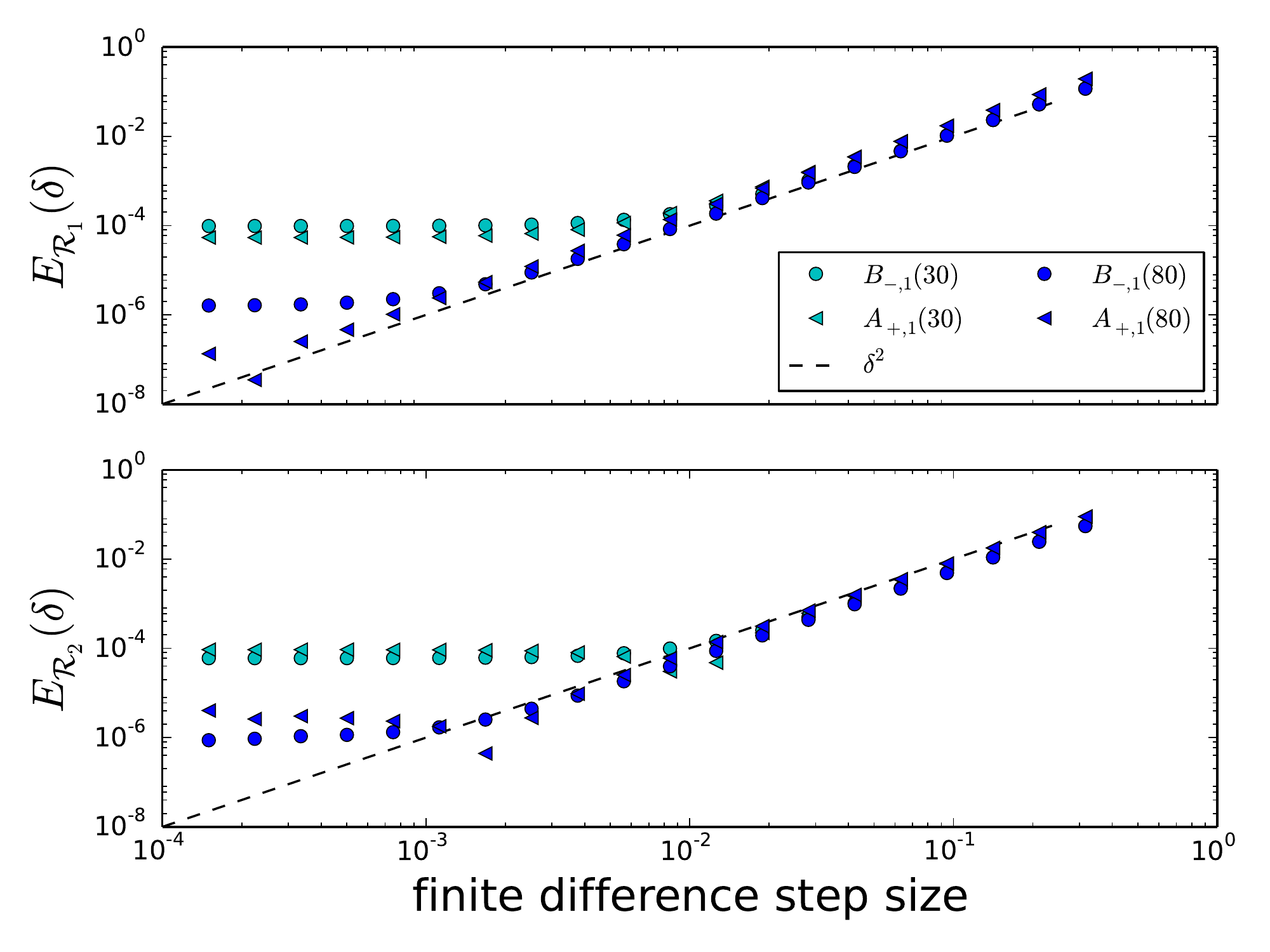}
\caption{Error convergence of the centred difference approximation of the gradient of the two residues to the adjoint calculation when $B_{-,1} = 0.18079$ and $A_{+,1} = 0.25268$. 
The finite difference step size on the horizontal axis is normalized to $\mathcal{R}^{-1} \left( \partial \mathcal{R} / \partial \kappa \right) \delta $ for $\kappa \in \lbrace A_{+,1}, B_{-,1} \rbrace$.
The legend labels which parameter the gradient was taken with respect to for the different symbols, and in brackets the value of $N_{\varphi}$.
The higher resolution, $N_{\varphi} = 80$, has an increased accuracy, although the optimization was carried out at $N_{\varphi} = 30$ to make it faster.
The decrease of the converged error for larger $N_{\varphi}$ indicates discretization error in the adjoint-based approach.}
\label{fig-heli-errors}
\end{figure}

The initial configuration with $\vec{p}_0 = (0.0, 0.0)$ has residues $\mathcal{R}_1 = -0.143$ and $\mathcal R_2 =  79.9$.
Applying the optimization scheme described above results in $\vec{p} = (0.3414, 0.3066)$, with $\mathcal{R}_1 = 0.0160$ and $\mathcal R_2 = 0.0128$.
The values of the parameters are similar to the ones obtained by \cite{Cary-Hanson-1986}, but give a slightly more optimized configuration: the small difference may be caused by different numerical resolutions in the Biot-Savart integral around the helical coils.
The optimized and unoptimized configurations are shown in figure~\ref{fig-heli}, and the helical coils that produce them are shown in figure \ref{fig-opcoils}.
The quadratic convergence of the errors of the gradients of the residue for an intermediate step during the iteration is shown in figure~\ref{fig-heli-errors}.


\section{Conclusion} \label{sec-conclusion}

In this paper we have derived the equations for the gradients of several quantities related to magnetic islands, including the island width and the residue of the periodic field line, using an adjoint method.
The residue is a quantity that, when small and positive, is strongly correlated with island size but that can be calculated for any periodic field line, even X points in stochastic regions.
Thus, although it does not quantify the physical width of an island, it is more versatile and can be used to minimize stochasticity in magnetic configurations. 
The gradient of the island width was obtained by differentiating the measure of island width introduced by \cite{Cary-Hanson-1991}.
Although the island width calculation is only accurate for small islands, optimized configurations usually have small islands.
Our adjoint approach thus provides an efficient and reliable method to compute the sensitivity of the island size in optimized or near-optimized stellarators.

We have performed and verified numerical gradient calculations on different magnetic field configurations.
The analytical configuration of \cite{Reiman-1986}, shown in figure~\ref{fig-surfaces-Reiman} provides the ideal system for efficiently and accurately verifying the island width calculation and the gradient calculations.
Figure~\ref{fig-width-Reiman} shows a comparison of different measures of island width for this system.
Figure \ref{fig-gradwidtherror-Reiman} demonstrates the correct quadratic convergence of centred difference gradient approximations to the adjoint gradient calculations as the centred difference step is decreased. 
In figure \ref{fig-NCSXgradient} we show shape gradients of the island width with respect to a discrete set of positions along a type A modular coil in an NCSX configuration, highlighting the overlap between the adjoint and the centred difference calculation.
For NCSX, where the magnetic field evaluation is computationally expensive, a factor of over one hundred in computation time is gained by calculating the shape gradient using the adjoint method.
The gradient calculation using adjoints has been successfully implemented also to a system composed of a toroidal field coil and a pair of helical coils, previously optimized without derivatives in \cite{Hanson-Cary-1984} and \cite{Cary-Hanson-1986}.
For one such configuration, the helical coils were optimized using the adjoint calculation of the gradient of the residue, with a result consistent with Cary and Hanson's. 
The result of the optimization is shown in figure~\ref{fig-heli}. 

The tool we have developed in this work can be applied in several areas of stellarator design.
Firstly, shape gradient calculations using the adjoint approach can be used to efficiently calculate coil tolerances with respect to island size in stellarator configurations. 
But they can also be used in optimization, as demonstrated in section \ref{subsec-num-heli}.
Gradient-based minimization of stochasticity and island size, as well as optimization of island width sensitivity, are possible applications of this work.

~~ 

The authors would like to thank Antoine Cerfon for suggesting to use of the Reiman model as a test magnetic field configuration.
AG would like like to thank Chris Smiet and Mohsen Sadr for interesting and stimulating conversations.
This work was supported by the US Department of Energy through grant DE-FG02-93ER-54197.

\appendix 

\section{Magnetic field line trajectory as a Hamiltonian system} \label{app-Ham}

A fundamental property of the magnetic field, namely that it is divergenceless, is imposed by defining $\vec{B}$ as the curl of a vector potential,
\begin{align}\label{B-A}
\vec{B} = \nabla \times \vec{A} \rm .
\end{align}
Note that $\vec{A}$ is not uniquely defined: adding the gradient of a scalar function, $\nabla g$, called a gauge transformation, leaves $\vec{B}$ unchanged.
The form of $\vec{B}$ we have adopted in (\ref{B-flux}) corresponds to the vector potential
\begin{align} \label{A-flux}
\vec{A} = \psi \nabla \theta - \chi (\psi, \theta, \varphi ) \nabla \varphi \rm ,
\end{align}
where we have chosen a particular gauge. 

Considering $ \vec{R} (\varphi)$ as the position vector following a field line, equation (\ref{B-A}) can be derived by extremizing the action 
\begin{align} \label{S-action-1}
S = \int_{\varphi_1}^{\varphi_2} \mathcal{L} \left( \vec{R}(\varphi), \varphi \right) d\varphi \rm 
\end{align} 
with respect to $\vec{R}(\varphi)$, where the end-points along the path, $\vec{R} ( \varphi_1 )$ and $\vec{R} ( \varphi_2 )$, are fixed, and the Lagrangian $\mathcal{L}$ is
\begin{align} \label{L-def}
\mathcal{L} = \vec{A} (\vec{R}) \cdot \frac{d\vec{R}}{d\varphi} \rm .
\end{align}
The choice (\ref{L-def}) can be verified using the Euler-Lagrange equation resulting from the extremization of the action,
\begin{align}
\frac{d}{d\varphi} \left( \frac{d\mathcal{L}}{\partial \dot{\vec{R}}} \right) = \frac{d\mathcal{L}}{\partial \vec{R}} \rm ,
\end{align}
which leads to the equation $\vec{B} \times d\vec{R} / d\varphi = 0$, implying that $d\vec{R} / d\varphi $ is always parallel to $\vec{B}$.
Upon inserting equation (\ref{A-flux}) into (\ref{L-def}), the action $S$ in equation (\ref{S-action-1}) becomes
\begin{align} \label{S-action}
S = \int_{\varphi_1}^{\varphi_2} \left( \psi  \frac{d \theta}{d\varphi} - \chi (\psi, \theta, \varphi)  \right) d\varphi \rm .
\end{align} 
The action expressed in the form (\ref{S-action}) can be directly compared with the standard form for the action of a Hamiltonian system,
\begin{align} \label{action-Ham}
S = \int_{t_1}^{t_2} \left( p  \frac{dq}{dt} - H (q, p, t)  \right) dt \rm .
\end{align} 
Hence, it follows that the trajectory of a magnetic field line as a function of toroidal angle constitutes a Hamiltonian system where the canonical co-ordinate $q$ is $\theta$, the canonical momentum $p$ is $\psi$, the Hamiltonian $H$ is $\chi$ and the time $t$ is $\varphi$.

\subsection{Hamiltonian after replacing $\theta$ with $\Theta$} \label{subapp-Ham}

We proceed to obtain the Hamiltonian after the change of variables $(\theta , \psi ) \rightarrow (\Theta , \psi )$, where $\Theta = \theta - \iota_0(\psi_0) \varphi$ is the poloidal angle relative to the unperturbed magnetic field line at the flux surface $\psi = \psi_0$ crossing $\varphi = \theta = 0$.
To this end, we re-express $d\theta / d\varphi$ by applying the chain rule to the equation $\theta = \Theta + \iota_0 (\psi_0) \varphi$, 
\begin{align} \label{dthetadphi-chain}
\frac{d\theta}{d\varphi} = \left( \frac{\partial \theta }{\partial \Theta } \right)_{\varphi} \frac{d\Theta}{d\varphi} + \left( \frac{\partial \theta}{\partial \varphi} \right)_{\Theta} = \frac{d\Theta}{d\varphi} + \iota_0 ( \psi_0 ) \rm .
\end{align}
Here, a subscript to the right of the parentheses indicates what variable is being kept constant in the partial differentiation within the parentheses.
By using (\ref{dthetadphi-chain}), the action in equation (\ref{S-action}) becomes 
\begin{align} \label{S-action-final}
S = \int_{\varphi_1}^{\varphi_2}  \left( \psi \frac{d\Theta}{ d\varphi } +  \psi \iota_0 (\psi_0) - \chi (\psi, \Theta ) \right) d\varphi\rm .
\end{align}
Therefore, the Hamiltonian $K$ in the new variables is given by equation (\ref{K-Ham}).

\section{Dominance of resonant perturbation} \label{app-resonant}

The equations of the magnetic field line trajectory including the perturbation are
\begin{align} \label{thetadot-pert}
\frac{d\theta }{d\varphi} = \iota_0(\psi) +   \sum_{m,n} \chi_{m,n}'(\psi) \exp \left( im\theta - in \varphi \right) \text{,}
\end{align}
and
\begin{align} \label{psidot-pert}
\frac{d\psi }{d\varphi} = - \sum_{m,n} \chi_{m,n}(\psi)  im\exp \left( im\theta - in \varphi \right) \text{.}
\end{align}
The solutions of equations (\ref{thetadot-pert}) and (\ref{psidot-pert}) are assumed to be approximately given by the functions obtained from equations (\ref{dpsidvarphi-lowest})-(\ref{dthetadvarphi-lowest}),
\begin{align}
\psi \simeq \psi_0 + \psi_1 (\varphi) \text{,}
\end{align}
and
\begin{align}
\theta \simeq \theta_{\text{i}} + \iota_0 (\psi_0) \varphi +  \theta_1 (\varphi) \text{,}
\end{align}
where $\psi_1$ and $\theta_1$ are additional functions --- assumed small --- that depend on the perturbation, and $\theta_{\rm i}$ is the value of $\theta$ at $\varphi = 0$.
Since $|\chi_1 | \ll \chi_0$, we assume that $\psi \simeq \psi_0$ so that $\psi_1$ and $\theta_1$ are small corrections, and calculate $\theta_1$ by neglecting it in the phase of the perturbation,
\begin{align} \label{dudvarphi-1}
\frac{d\theta_1 }{d\varphi} \simeq  \sum_{m,n} \chi_{m,n}'(\psi_0)  \exp \left( im \theta_{\text{i}}  + im \iota_0 \varphi - in \varphi \right) \text{.}
\end{align}
This gives
 \begin{align} \label{theta1-div}
 \theta_1 \simeq \sum_{m,n} \frac{ \chi_{m,n}'(\psi_0) }{ im \iota_0 - i n} \exp \left( im\theta_{\text{i}} + im \iota_0 \varphi - in \varphi \right) \text{,}
\end{align}
which is divergent when $ n/m = \iota_0 $.
Hence, we deduce that the effect of the perturbation is dominated by the Fourier modes that coincide (resonate) with the rotational transform of the unperturbed flux surface, $ n/m = \iota_0 $.

\section{Variation of periodic field line position} \label{app-DeltaX}

The periodic field line position changes due to the variation of the magnetic field configuration, affecting the circumference and in turn the island width as discussed in section \ref{sec-width}.
However, the gradient of the poloidal position vector $\vecbar X_k$ is itself useful in order to search faster and more reliably for the new position of the periodic field line after a small change in parameters, e.g. during the optimization considered in section \ref{subsec-num-heli}. 
We thus proceed to derive the variation $\Delta \vecbar X_k$.

We introduce the vector Lagrangian
\begin{align}
\mathcal{L}_{\vecbar{X}} = \vecbar{X}_k + \int_{2\pi k/n_0}^{2\pi (k+L)/n_0}  \left( \frac{d\vec{X}}{d\varphi} - \vec{V} \right) \cdot \mu_{\vecbar{X}_k} d\varphi \text{.}
\end{align}
In (\ref{LRes}) we have introduced a constraint using the adjoint variable $\mat{\mu}_{\vecbar{X}_k}$ (a matrix).
Extremization with respect to variations in $\vec{X} (\varphi )$ gives
\begin{align}
\Delta \mathcal{L}_{\vecbar{X}} = \Delta \vecbar{X}_k + \int_{2\pi k/n_0}^{2\pi (k+L)/n_0} \Delta \vec{X} \cdot \left( -\frac{d\mat{\mu}_{\vec{X}_k}}{d\varphi} - \nabla_{\vec X} \vec{V} \cdot \mat{\mu}_{\vecbar{X}_k} \right) d\varphi  \\
+  \Delta \vec{X} \left( 2\pi (k+L)/n_0 \right)  \cdot  \mat{\mu}_{\vecbar{X}_k} \left( 2\pi (k+L)/n_0 \right) - \Delta \vec{X} \left( 2\pi k / n_0 \right)  \cdot    \mat{\mu}_{\vecbar{X}_k} \left( 2\pi k / n_0 ) \right)  \text{.}
\end{align}
Considering $ \Delta \vec{X} \left( 2\pi (k+L)/n_0 \right) =  \Delta \vec{X} \left( 2\pi k/n_0 \right) = \Delta \vecbar{X}_k$,  the adjoint equation is
\begin{align} \label{muX}
\frac{d\mat{\mu}_{\vecbar{X}_k}}{d\varphi} = - \nabla_{\vecbar X} \vec{V} \cdot \mat{\mu}_{\vecbar{X}_k} \rm ,
\end{align}
and must be solved with the boundary condition $ \mat{\mu}_{\vecbar X_k} \left( 2\pi (k+L)/n_0 \right) =  \mat{\mu}_{\vecbar X_k} \left( 2\pi k/ n_0 \right) - \mat{I}$. 

The boundary condition can be simplified as follows.
The transpose of (\ref{EOM-U}) is 
\begin{align}
\frac{d\mat{S}_k^\intercal}{d\varphi} =  \mat{S}_k^\intercal  \cdot   \nabla_{\vec X} \vec{V} \rm .
\end{align}
Premultiplying and postmultiplying by the adjoint of $\mat S_k$, $\mat{S}_k^{\dagger} = \left( \mat{S}_k^\intercal \right)^{-1} $, and noting that $\mat{S}_k^{\dagger} \cdot d\mat{S}_k^{\intercal} / d\varphi = - \left( d\mat{S}_k^{\dagger} / d\varphi \right) \cdot \mat{S}_k^\intercal$ (from the chain rule and $\mat{S}_k^{\dagger} \cdot \mat{S}_k^\intercal = \mat{I}$) results in 
\begin{align} \label{adjoint-eq}
\frac{d \mat{S}_k^{\dagger}}{d\varphi} = -  \nabla_{\vec X} \vec{V}   \cdot  \mat{S}_k^{\dagger} \rm .
\end{align}
Comparing equations (\ref{muX}) and (\ref{adjoint-eq}) and using that $\mat S_k \left( 2\pi (k + L) / n_0 \right)= \mat M_k $ with $\mat S_k \left( 2\pi k / n_0 \right)= \mat  I $ gives 
\begin{align}
\mat{\mu}_{\vecbar X_k}  \left( 2\pi (k + L) /n_0 \right) = \mat{M}_k^{\dagger} \cdot \mat{\mu}_{\vecbar X} \left( 2\pi k/n_0 \right) \rm .
\end{align}
Hence, the appropriate boundary condition for equation (\ref{muX}) is
\begin{align}
\mat{\mu}_{\vecbar{X}_k}(2\pi k / n_0) = \left[ \mat{I} - \mat{M}_k^{\dagger} \right]^{-1} \rm .
\end{align}
The variation of $\vecbar{X}_k$ is then given by
\begin{align} \label{DeltaX}
\Delta \vecbar{X}_k   =-  \int_{2\pi k/n_0}^{2\pi (k+L)/n_0} \Delta \vec{V} \cdot \mat{\mu}_{\vecbar{X}_k} d\varphi \rm .
\end{align}

\section{Formulas for the Reiman magnetic field configuration} \label{app-Reiman}

The analytical magnetic field configuration is adapted from a model magnetic field used in \cite{Reiman-1986}.
The magnetic field components are, from equations (\ref{B-flux}) and (\ref{psi-Reiman})-(\ref{theta-Reiman}),
\begin{gather}
B_R = \frac{Z}{R} D_0 + \frac{R-R_0}{ R} D_1 \text{,}   \label{BR-Reiman} \\ 
B_Z = - \frac{R-R_0}{R} D_0 + \frac{Z}{ R} D_1 \text{,}  \\
B_{\varphi} = - 1 \text{,}   \label{Bphi-Reiman}
\end{gather}
where 
\begin{gather}
D_0 =  \iota_{\rm ax} + \iota'_{\rm ax} r^2 - \sum_{k=1}^{k_{\rm max}} k \varepsilon_k r^{k-2} \cos \left( k \theta - \varphi \right) \text{,} \\
D_1 =  \sum_{k=1}^{k_{\rm max}} k \varepsilon_k r^{k-2} \sin \left( k \theta - \varphi \right) \text{.} 
\end{gather}
From equations (\ref{BR-Reiman})-(\ref{Bphi-Reiman}), the gradients of the magnetic field components with respect to the poloidal coordinates $R$ and $Z$ are
\begin{gather}
\partial_R B_R  = - \frac{Z}{R^2} D_0 + \frac{Z}{R} \partial_R D_0 + \frac{R_0}{ R^2} D_1  + \frac{R-R_0}{ R} \partial_R D_1 \text{,}  \label{dRBR-Reiman}  \\
\partial_Z B_R  = \frac{1}{R} D_0 + \frac{Z}{R} \partial_Z D_0 + \frac{R-R_0}{ R} \partial_Z D_1 \text{,} \\
\partial_R B_Z  = - \frac{R_0}{R^2} D_0 - \frac{R-R_0}{R} \partial_R D_0 -  \frac{Z}{ R^2} D_1 + \frac{Z}{ R} \partial_R D_1 \text{,}  \\
\partial_Z B_Z  =  - \frac{R-R_0}{R} \partial_Z D_0 +  \frac{1}{ R} D_1 + \frac{Z}{ R} \partial_Z D_1 \text{,}   \\
\partial_R B_{\varphi}  = \partial_Z B_{\varphi} = 0  \label{dRBphi-dZBphiReiman} \text{,}
\end{gather}
where
\begin{gather}
\partial_R D_0 = 2 \iota'_{\rm ax} \left( R - R_0 \right) -  \sum_{k=1}^{k_{\rm max}} k  \varepsilon_k r^{k-4} \left( \left( k - 2 \right) \left(R-R_0 \right) \cos \left( k \theta - \varphi \right)  +  kZ \sin \left( k \theta - \varphi \right)  \right) \text{,} \\ 
\partial_R D_1 =  \sum_{k=1}^{k_{\rm max}} k  \varepsilon_k r^{k-4} \left( \left( k - 2 \right) \left( R - R_0 \right) \sin \left( k \theta - \varphi \right)  -  kZ \cos \left( k \theta - \varphi \right)  \right)  \text{,} \\
\partial_Z D_0 =  2 \iota'_{\rm ax} Z -  \sum_{k=1}^{k_{\rm max}} k  \varepsilon_k r^{k-4} \left( \left( k - 2 \right) Z \cos \left( k \theta - \varphi \right)  -  k\left( R-R_0 \right) \sin \left( k \theta - \varphi \right)  \right) \text{,} \\
\partial_Z D_1 =  \sum_{k=1}^{k_{\rm max}} k  \varepsilon_k r^{k-4} \left( \left( k - 2 \right) Z \sin \left( k \theta - \varphi \right)  +  k \left(R-R_0 \right) \cos \left( k \theta - \varphi \right)  \right) \text{.} 
\end{gather}
From equations (\ref{dRBR-Reiman})-(\ref{dRBphi-dZBphiReiman}), the second derivatives of the magnetic field components with respect to $R$ and $Z$ are
\begin{gather}
\partial_{RR} B_R  =  \frac{2Z}{R^3} D_0 - \frac{2R_0}{ R^3} D_1 - \frac{2Z}{R^2} \partial_R D_0 + \frac{2R_0}{ R^2} \partial_R D_1  + \frac{Z}{R} \partial_{RR}D_0 + \frac{R-R_0}{ R} \partial_{RR} D_1 \text{,} \\
\partial_{RZ} B_R  =  - \frac{1}{R^2} D_0 - \frac{Z}{R^2} \partial_Z D_0 +  \frac{1}{ R} \partial_R D_0  + \frac{R_0}{R^2} \partial_Z D_1 + \frac{Z}{ R} \partial_{RZ} D_0 + \frac{R-R_0}{R} \partial_{RZ} D_1 \text{,}  \\
\partial_{ZZ} B_R =  \frac{2}{R} \partial_{Z} D_0 +  \frac{Z}{R} \partial_{ZZ} D_0 +  \frac{R-R_0}{R} \partial_{ZZ} D_1\text{,} \\
\partial_{RR} B_Z  = \frac{2R_0}{R^3} D_0 +  \frac{2Z}{R^3} D_1 - \frac{2R_0}{R^2} \partial_R D_0 - \frac{2Z}{R^2} \partial_R D_1   - \frac{R-R_0}{R} \partial_{RR} D_0 + \frac{Z}{ R} \partial_{RR} D_1 \text{,} \\
\partial_{RZ} B_Z  =  - \frac{1}{R^2} D_1 - \frac{R_0}{R^2} \partial_Z D_0  - \frac{Z}{ R^2} \partial_Z D_1 + \frac{1}{R} \partial_R D_1 -  \frac{R-R_0}{ R} \partial_{RZ} D_0 + \frac{Z}{ R} \partial_{RZ} D_1 \text{,}  \\
\partial_{ZZ} B_Z = \frac{2}{R} \partial_Z D_1 -  \frac{R-R_0}{ R} \partial_{ZZ} D_0 + \frac{Z}{ R} \partial_{ZZ} D_1 \text{,} \\
\partial_{RR} B_{\varphi}  =  \partial_{RZ} B_{\varphi}   = \partial_{ZZ} B_{\varphi}  = 0  \text{,} 
\end{gather}
where $\partial_{RZ} = \partial_{ZR}$ and
\begin{gather}
\partial_{RR} D_0 =  2 \iota'_{\rm ax} -  \sum_{k=1}^{k_{\rm max}} k  \varepsilon_k r^{k-6} \left[  \left( \left( k - 2 \right) \left( r^2 + (k-4)(R-R_0)^2 \right) - k^2 Z^2 \right) \cos \left( k \theta - \varphi \right) \right. \nonumber \\ \left. +  kZ (R - R_0 ) (2k - 6) \sin \left( k \theta - \varphi  \right)  \right] \text{,} \\  
\partial_{RZ} D_0 =  -  \sum_{k=1}^{k_{\rm max}} k  \varepsilon_k r^{k-6} \left[  k \left( r^2 - (k-2)(R-R_0)^2 + (k-4)Z^2 \right)  \sin \left( k \theta - \varphi \right) \right. \nonumber \\ \left. +  kZ (R - R_0 ) (2k^2 - 6k + 8) \cos \left( k \theta - \varphi  \right)  \right] \text{,} \\  
\partial_{ZZ} D_0 =  2  \iota'_{\rm ax} -  \sum_{k=1}^{k_{\rm max}} k  \varepsilon_k r^{k-6} \left[  \left( \left( k - 2 \right) \left( r^2 + (k-4)Z^2 \right) - k^2 (R-R_0)^2 \right) \cos \left( k \theta - \varphi \right) \right. \nonumber \\ \left. -  kZ (R - R_0 ) (2k - 6) \sin \left( k \theta - \varphi  \right)  \right] \text{,} \\  
\partial_{RR} D_1 =    \sum_{k=1}^{k_{\rm max}} k  \varepsilon_k r^{k-6} \left[  \left( \left( k - 2 \right) \left( r^2 + (k-4)(R-R_0)^2 \right) - k^2 Z^2 \right) \sin \left( k \theta - \varphi \right) \right. \nonumber \\ \left. -  kZ(R - R_0 ) \left( 2k-6 \right)  \cos \left( k \theta - \varphi  \right)  \right] \text{,} \\  
\partial_{RZ} D_1 =   \sum_{k=1}^{k_{\rm max}} k  \varepsilon_k r^{k-6} \left[  k \left( - r^2 + (k-2)(R-R_0)^2 - (k-4)Z^2 \right)  \cos \left( k \theta - \varphi \right) \right. \nonumber \\ \left. +  kZ (R - R_0 ) (2k^2 - 6k + 8) \sin \left( k \theta - \varphi  \right)  \right] \text{,} \\  
\partial_{ZZ} D_1 =   \sum_{k=1}^{k_{\rm max}} k  \varepsilon_k r^{k-6} \left[  \left( \left( k - 2 \right) \left( r^2 + (k-4)Z^2 \right) - k^2 (R-R_0)^2 \right) \sin \left( k \theta - \varphi \right) \right. \nonumber \\ \left. +  kZ (R - R_0 ) (2k - 6) \cos \left( k \theta - \varphi  \right)  \right] \text{.} 
\end{gather}

All the equations in this appendix are linear in the parameters $\kappa \in (\iota_{\rm ax}, \iota'_{\rm ax}, \varepsilon )$.
Hence, the gradients of the magnetic field and its first poloidal derivatives can be straightforwardly extracted.

\section{Shape gradient of coil-produced magnetic field} \label{app-shape-mag}

The magnetic field produced by a set of $N$ coils, indexed $c$, is given by the Biot-Savart law (\ref{B-coils}).
Perturbing the coils such that $\vec{r}_c (l_c ) \rightarrow \vec{r}_c (l_c ) + \Delta \vec{r}_c (l_c ) $ changes the magnetic field such that $\vec{B}(\vec{R}) \rightarrow \vec{B}(\vec{R}) + \Delta  \vec{B}(\vec{R}) $, where the change in magnetic field is given by
\begin{align} \label{DeltaB-init}
\Delta \vec{B}_{\rm coil} =  \sum_{c=1}^N \frac{\mu_0 I_c }{4\pi} \oint dl_c \left( \Delta \vec{r}_c \cdot  \nabla_{\vec{r}_c} \left( \frac{\vec{R} - \vec{r}_c}{ | \vec{R} - \vec{r}_c |^2 } \right) \times \frac{d\vec{r}_c}{dl_c} + \left( \frac{\vec{R} - \vec{r}_c}{ | \vec{R} - \vec{r}_c |^2 } \right) \times \frac{d\Delta \vec{r}_c}{dl_c} \right) \text{.}
\end{align}
where 
\begin{align} \label{traceless}
 \nabla_{\vec{r}_c} \left( \frac{\vec{R} - \vec{r}_c}{ | \vec{R} - \vec{r}_c |^2 } \right) = \frac{1}{ | \vec{R} - \vec{r}_c |^2 } \left( - \mat{I} + \frac{ \left( \vec{R} - \vec{r}_c \right) \left( \vec{R} - \vec{r}_c \right) }{ | \vec{R} - \vec{r}_c |^2 } \right) \text{.}
\end{align}
An important property of (\ref{traceless}) is that 
\begin{align} \label{traceless-exp}
\rm Tr \left[  \nabla_{\vec{r}_c} \left( \frac{\vec{R} - \vec{r}_c}{ | \vec{R} - \vec{r}_c |^2 } \right) \right] = 0 \rm .
\end{align} 
The second term in (\ref{DeltaB-init}) can be re-expressed using integration by parts as
\begin{align}
 \oint dl_c  \left( \frac{\vec{R} - \vec{r}_c}{ | \vec{R} - \vec{r}_c |^2 } \right) \times \frac{d \Delta \vec{r}_c}{dl_c}  = -  \oint dl_c  \frac{d\vec{r}_c}{dl_c} \cdot  \nabla_{\vec{r}_c} \left( \frac{\vec{R} - \vec{r}_c}{ | \vec{R} - \vec{r}_c |^2 } \right) \times \Delta \vec{r}_c  \text{,}
\end{align}
where the term involving an integral of a total derivative has vanished due to the integral being periodic.
Inserting this into the previous equation gives
\begin{align} \label{DeltaB-int2}
\Delta \vec{B}_{\rm coil} =  \sum_{c=1}^N \frac{\mu_0 I_c }{4\pi} \oint dl_c \left( \Delta \vec{r}_c \cdot  \nabla_{\vec{r}_c} \left( \frac{\vec{R} - \vec{r}_c}{ | \vec{R} - \vec{r}_c |^2 } \right) \times \frac{d \vec{r}_c}{dl_c} - \frac{d \vec{r}_c}{dl_c}  \cdot  \nabla_{\vec{r}_c} \left( \frac{\vec{R} - \vec{r}_c}{ | \vec{R} - \vec{r}_c |^2 } \right) \times \Delta \vec{r}_c \right) \text{.}
\end{align}
It is an exercise in vector identities to show that
\begin{align} \label{after-identities}
\Delta \vec{r}_c \cdot  & \nabla_{\vec{r}_c} \left( \frac{\vec{R} - \vec{r}_c}{ | \vec{R} - \vec{r}_c |^2 } \right)  \times \frac{d \vec{r}_c}{dl_c}  - \frac{d \vec{r}_c}{dl_c}  \cdot \nabla_{\vec{r}_c} \left( \frac{\vec{R} - \vec{r}_c}{ | \vec{R} - \vec{r}_c |^2 } \right)  \times \Delta \vec{r}_c \nonumber \\  & = \rm Tr \left[  \nabla_{\vec{r}_c} \left( \frac{\vec{R} - \vec{r}_c}{ | \vec{R} - \vec{r}_c |^2 } \right) \right] \Delta \vec{r}_c \times \frac{d \vec{r}_c}{dl_c}  -  \nabla_{\vec{r}_c} \left( \frac{\vec{R} - \vec{r}_c}{ | \vec{R} - \vec{r}_c |^2 } \right)  \cdot \Delta \vec{r}_c \times  \frac{d \vec{r}_c}{dl_c}  \rm .
\end{align}
Using (\ref{traceless-exp}), the first term on the right hand side of (\ref{after-identities}) vanishes.
Inserting equations (\ref{traceless}), (\ref{traceless-exp}) and (\ref{after-identities}) into equation (\ref{DeltaB-int2}) gives equation (\ref{B-shape}) for the shape gradient.

\section{Optimization scheme} \label{app-op}

The iteration scheme used to search for solutions of $P = 0$ exploits the direction vector $\vec{d}_n$ obtained from the gradient of $P_n$ with respect to the two variable parameters,
\begin{align}
\vec{d}_{n} = \left(  \frac{\partial P_n}{ \partial A_{-,2} }  , \frac{ \partial P_n}{  \partial B_{+,1} }  \right) \rm .
\end{align}
Here, a subscipt $n$ denotes the value of the quantity at the $n$th step in the iteration.
Starting from an initial guess $\vec{p}_0 = (0.0, 0.0)$, the next step in the iteration is tentatively specified by
\begin{align} \label{move}
\vec{p}_{n+1} = \vec{p}_n + \vec{p}_{n, \rm step}   \rm ,
\end{align}
where 
\begin{align}
 \vec{p}_{n, \rm step} = - \frac{g_{n}}{2^r} \frac{ P_{n} \vec{d}_n}{|\vec{d}_n|^2} \rm ,
\end{align}
and where $g_{n} > 0$ is a number chosen at each iteration and $r$ is the number of times the move (\ref{move}) is rejected at a particular iteration.
The reference value of $g_n$ is denoted $\bar{g}_n$ and is specified as follows.
Initially we set $\bar{g}_0 = 1$.
Then, if at any particular iteration the number of times $r$ the move (\ref{move}) gets rejected before being accepted exceeds a threshold value $r_{\rm limit} = 3$, we set $\bar{g}_{n+1} = \bar{g}_{n} / 2$, otherwise we keep the reference value unchanged, $\bar{g}_{n+1} = \bar{g}_{n} $.
The move (\ref{move}) is rejected if the step in parameter space leads to an increase in the objective function, $P_{n+1} \geqslant P_n$.

There is an additional rejection criterion that is linked to a maximum value of the step in the fixed point position, $| \vecbar{X}^{n+1} - \vecbar{X}^{n} |$ (the subscript labelling the fixed point is omitted to avoid clutter, and the superscript is the iteration step number).
This is necessary in order to ensure that the iteration proceeds successfully.
At the iteration step $n+1$  the scheme to search for a periodic field line is applied with the initial guess equal to $  \vecbar{X}^{n} - \vec p_{n,\rm step}  \cdot \nabla_{\vec{p}} \vecbar{X}^n   $.
The gradient of the fixed point position with respect to the parameters, $ \nabla_{\vec{p}} \vecbar{X}^n = \partial \vecbar{X}^n / \partial \vec{p}$, is calculated from the equations of Appendix \ref{app-DeltaX}.
A common way for the search for the new fixed point position $ \vec{X}^{n+1}$ to fail is by returning the magnetic axis (which is always a periodic solution of the magnetic field line equations for any toroidal interval which is a multiple of the the field periodicity).
In order to avoid this, the maximum step size is set to be a fraction $u$ of the smallest distance between the fixed point and the magnetic axis $ \left| \vecbar{X}^{n} - \vecbar{X}_{\rm axis}^{n} \right| $. The value $u = 0.3 $ was used.

To summarize, the criterion for accepting or rejecting a step (\ref{move}) is:
\begin{align} \label{accept-reject}
& \text{accept move if } P_{n+1} < P_n \text{ and }  |\vecbar{X}^{n+1} - \vecbar{X}^{n} | < u \left| \vecbar{X}^{ n} - \vecbar{X}_{\rm axis}^n \right|  \text{;} \nonumber \\
& \text{reject move if } P_{n+1} \geqslant P_n \text{ or } |\vecbar{X}^{n+1} - \vecbar{X}^{n} | \geqslant u \left| \vecbar{X}^{ n} - \vecbar{X}_{\rm axis}^n  \right|       \text{.}
\end{align}
Once a move (\ref{move}) is accepted, it becomes permanent and the following move $ \vec{p}_{n+1, \rm step}$ is calculated.
The optimization is stopped when $| P_{n-1}  - P_n | < 10^{-12}$.

\bibliographystyle{jpp}

\bibliography{islands}

\end{document}